%% file: MAIN-sigconf-authordraft.tex
\documentclass[sigconf, final]{acmart}

\AtBeginDocument{%
  }

\usepackage{xcolor}
\usepackage{tabularx}  
\usepackage{multirow}   
\usepackage{array} 

\copyrightyear{2026}
\acmYear{2026}
\setcopyright{cc}
\setcctype{by}
\acmConference[CHI '26]{Proceedings of the 2026 CHI Conference on Human Factors in Computing Systems}{April 13--17, 2026}{Barcelona, Spain}
\acmBooktitle{Proceedings of the 2026 CHI Conference on Human Factors in Computing Systems (CHI '26), April 13--17, 2026, Barcelona, Spain}
\acmPrice{}
\acmDOI{10.1145/3772318.3791301}
\acmISBN{979-8-4007-2278-3/2026/04}

\begin{document}

% \title{Exploring AI Opportunities for Healthcare Providers to Support the Workflow of Conducting Serious Illness Conversations with Older Adults in the Emergency Department}
\title[Balancing Efficiency and Empathy]{%Why Doctors Do Not Want to Have the Hard Conversation in the Emergency Department, and Is AI a Silver Bullet?
%"It Would Be Helpful To Have Help Like If The AI Model Was Listening": Explore Challenges And Needs Healthcare Providers Have In Their Workflow Of Conducting Serious Illness Conversations In The Emergency Department
Balancing Efficiency and Empathy: Healthcare Providers' Perspectives on AI-Supported Workflows for Serious Illness Conversations in the Emergency Department}

\author{Menglin Zhao}
\affiliation{%
  \institution{Northeastern University}
  \city{Boston}
  \state{Massachusetts}
  \country{USA}}
  \email{zhao.mengl@northeastern.edu}

\author{Zhuorui Yong}
\affiliation{%
  \institution{Northeastern University}
  \city{Boston}
  \state{Massachusetts}
  \country{USA}}
\email{yong.zh@northeastern.edu}

\author{Ruijia Guan}
\affiliation{%
  \institution{Northeastern University}
  \city{Boston}
  \state{Massachusetts}
  \country{USA}}
  \email{r.guan@northeastern.edu}

\author{Kai-Wei Chang}
\affiliation{%
 \institution{University of California, Los Angeles}
 \city{Los Angeles}
 \state{California}
 \country{USA}}
 \email{kw@kwchang.net}

\author{Adrian Haimovich}
\affiliation{%
  \institution{Beth Israel Deaconess Medical Center}
  \city{Boston}
  \state{Massachusetts}
  \country{USA}}
  \email{ahaimovi@bidmc.harvard.edu}

%\author{Kei Ouchi}
%\email{kouchi@bwh.harvard.edu}

%\affiliation{
%  \institution{Harvard Medical School}
%  \city{Boston}
%  \state{Massachusetts}
%  \country{USA}
%}

%\affiliation{
%  \institution{Dana-Farber Cancer Institute}
%  \city{Boston}
%  \state{Massachusetts}
%  \country{USA}
%}

%\affiliation{
%  \institution{Brigham and Women’s Hospital}
%  \city{Boston}
%  \state{Massachusetts}
%  \country{USA}
%}

\author{Kei Ouchi}
\affiliation{
  \institution{Harvard Medical School, Dana-Farber Cancer Institute, Brigham and Women’s Hospital}
  \city{Boston}
  \state{Massachusetts}
  \country{USA}}
\email{kouchi@bwh.harvard.edu}

\author{Timothy Bickmore}
\affiliation{%
  \institution{Northeastern University}
  \city{Boston}
  \state{Massachusetts}
  \country{USA}}
\email{t.bickmore@northeastern.edu}

\author{Zhan Zhang}
\affiliation{%
  \institution{Pace University}
  \city{New York City}
  \state{New York State}
  \country{USA}}
\email{zzhang@pace.edu}

\author{Bingsheng Yao}
\affiliation{%
  \institution{Northeastern University}
  \city{Boston}
  \state{Massachusetts}
  \country{USA}}
\email{b.yao@northeastern.edu}

\author{Dakuo Wang}
\affiliation{%
  \institution{Northeastern University}
  \city{Boston}
  \state{Massachusetts}
  \country{USA}}
\email{d.wang@northeastern.edu}

\author{Smit Desai}
\authornote{Corresponding author}
\affiliation{
  \institution{Northeastern University}
  \city{Boston}
  \state{Massachusetts}
  \country{USA}}
\email{sm.desai@northeastern.edu}

\renewcommand{\shortauthors}{Zhao et al.}

\begin{abstract}

Serious Illness Conversations (SICs)—discussions about values and care preferences for patients with life-threatening illness—rarely occur in Emergency Departments (EDs), despite evidence that early conversations improve care alignment and reduce unnecessary interventions. We interviewed 11 ED providers to identify challenges in SICs and opportunities for technology support, with a focus on AI. Our analysis revealed a four-stage SIC workflow (identification, preparation, conduction, documentation) and barriers at each stage, including fragmented patient information, limited time and space, lack of conversational guidance, and burdensome documentation. Providers expressed interest in AI systems for synthesizing information, supporting real-time conversations, and automating documentation, but emphasized concerns about preserving human connection and clinical autonomy. This tension highlights the need for technologies that enhance efficiency without undermining the interpersonal nature of SICs. We propose design guidelines for ambient and peripheral AI systems to support providers while preserving the essential humanity of these conversations.

\end{abstract}



\keywords{Serious illness conversations, Emergency department, Older adults, Clinical workflows, Artificial intelligence, Healthcare communication, Palliative care}
 %%Description{Enjoying the baseball game from the third-base
 %% seats. Ichiro Suzuki preparing to bat.}
 %% \label{fig:teaser}

%   \centering
%   \includegraphics[width=\textwidth]
%   {Figures/tease.jpg}
%   \caption{We systematically examined ED providers' current workflow of conducting SICs in EDs, which revealed four stages (Identification, Preparation, Conduction, and Documentation). Meanwhile, we uncovered specific challenges that ED providers encounter at each stage, as well as limitations in post-SIC that hinder the delivery of future SICs. Building on the insights, we identified how AI technologies could help address the challenges while being smoothly integrated into the clinical workflow. The role of AI technologies is designed to be ambient and peripheral, which allows ED providers to use them when needed.}

\maketitle

\input{sections/1introduction_v3}

\input{sections/2related_work_revised}

\input{sections/3interviews_method_and_participants}

\input{sections/4interviews_findings}

\input{sections/8discussion}

\input{sections/10conclusion}

\begin{acks}
We thank the members of the Conversational Human-AI Interactions (CHAI) Lab and the Northeastern University Human-Centered AI (NEU HAI) Lab for their support and feedback. We also thank the emergency department physicians and nurses who participated in this study for their time. This research was supported in part by the National Cancer Institute of the National Institutes of Health under Award Number R01CA301579 and by a Northeastern University Tier-1 Research Seed Grant. The content is solely the responsibility of the authors and does not necessarily represent the official views of the National Institutes of Health. Finally, we acknowledge the use of GPT-5 for creating  the initial graphics for Figure 1, which were further revised and edited manually using Figma. 
\end{acks}

\bibliographystyle{ACM-Reference-Format}
\bibliography{sample-base}

\end{document}

%% file: sections/1introduction_v3.tex
\section{Introduction}

\begin{figure*}[htbp]
  \includegraphics[draft=false,width=\textwidth]{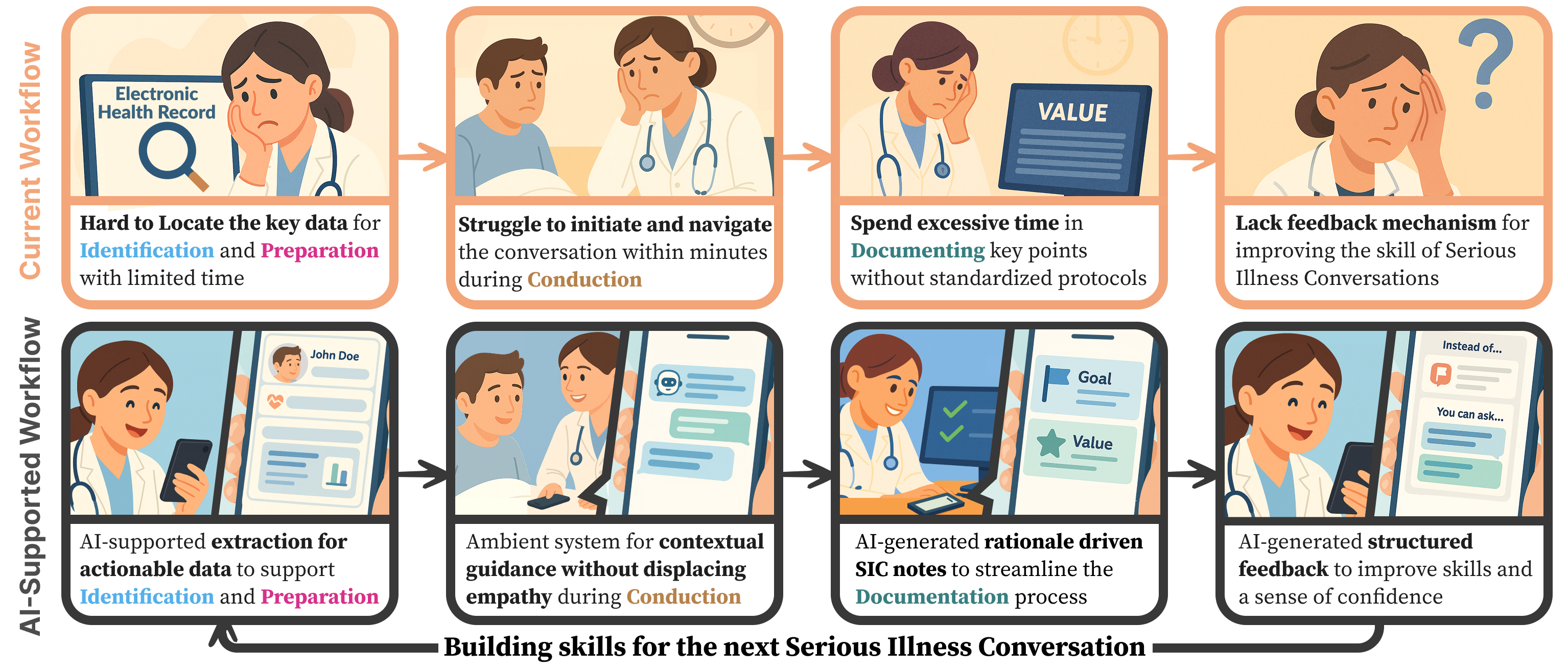}
  \caption{We systematically examined ED providers' current workflow of conducting SICs in EDs, which revealed four stages (Identification, Preparation, Conduction, and Documentation). Meanwhile, we uncovered specific challenges that ED providers encounter at each stage, as well as limitations in post-SIC that hinder the delivery of future SICs. Building on the insights, we identified how AI technologies could help address the challenges while being smoothly integrated into the clinical workflow. The role of AI technologies is designed to be ambient and peripheral, which allows ED providers to use them when needed.}
  \label{fig:teaser}
   \Description{Fig.1: The figure contrasts the current workflow of SICs with a proposed AI-assisted workflow. In the current process, providers must manually search the EHR to gather patient information, struggle to navigate sensitive topics during the conversation, spend excessive time manually documenting lengthy discussions, and lack structured feedback to improve their SIC skills. In contrast, the AI-assisted workflow introduces automated summarization of patient information to support identification and preparation, personalized suggestions from ambient systems to guide providers during conversations, AI-generated notes to streamline documentation, and personalized feedback to enhance skills for future SICs. Together, these AI supports aim to reduce manual burdens, bridge critical gaps in the current process, and enable providers to build their capacity for more effective and sustainable SICs over time.}
\end{figure*}

{Consider a patient with end-stage heart failure who arrives in the Emergency Department (ED) in severe respiratory distress. Within minutes, the healthcare provider must decide whether to intubate—a step that may offer temporary physiologic stability but could also initiate a prolonged and invasive course—or to pursue comfort-focused care that prioritizes relief in what may be the patient's final hours. In these moments, the provider must quickly try to understand the patient's values, goals, and care preferences so that the decision reflects the patient's wishes. Yet in the ED, where these conversations are the most needed, they are rarely conducted \cite{ouchi2019goals}.}

{These critical conversations are known as \emph{Serious Illness Conversations (SICs)}~\cite{ouchi2017preparing, peloquin2025discussions}, structured discussions where providers explore what matters most to patients facing life-threatening conditions.} When treating life-threatening illnesses such as heart failure and advanced cancer, providers need to understand patients' care preferences to guide critical decisions—{decisions that, as illustrated above, often arise with little warning and require immediate action}~\cite{wright2008associations}. These decisions typically require input from the patient or their family members to ensure alignment with the patient's values and wishes. {Through SICs}, providers {work to uncover} patients' care preferences by discussing expected disease trajectory~\cite{ouchi2017preparing}, uncovering fundamental values~\cite{naik2016health}, clarifying goals of care, and identifying specific treatment preferences~\cite{shilling2024let}. By doing so, providers are able to deliver care that minimizes suffering while offering emotional support and preserving quality of life~\cite{prachanukool2022emergency, ouchi2017preparing, geerse2019qualitative}.

{However, the unique constraints of the ED create significant challenges for conducting these conversations. Unlike routine clinical environments (e.g., primary care) where SICs can be planned and unhurried, ED providers must navigate them under time pressure, with unfamiliar patients in crisis, and often without access to prior documentation of patient wishes.} Currently, SICs often take place in ICUs, inpatient wards, and post-acute settings (e.g., long-term care) \cite{ouchi2019goals, johnson2025benefits}, where patients have already been admitted and may have undergone treatments such as ventilation and intubation \cite{ouchi2019goals}. Despite EDs serving as the primary entry point into the hospital system and the site where initial treatment plans are formulated~\cite{ouchi2019goals,wilson2020end}, SICs are infrequently conducted in these settings~\cite{ouchi2019goals, murray2025goc}. For example, older adults—who account for up to one-quarter of all ED visits~\cite{samaras2010older, galvin2017adverse} and often present with life-threatening illnesses requiring urgent decisions~\cite{smith2012half}—receive SICs in fewer than one-fifth of cases in EDs~\cite{johnson2025benefits}, leaving many vulnerable to life-prolonging interventions that may result in unnecessary suffering and reduced quality of life~\cite{wright2008associations}. Evidence demonstrates that ED-based SICs can provide substantial benefits~\cite{prachanukool2022emergency}, including reduced hospital stay and better utilization of healthcare resources~\cite{johnson2025benefits} and greater alignment of care with patients' wishes~\cite{ouchi2017preparing}. Consequently, a growing number of practitioners advocate for initiating SICs in the ED~\cite{prachanukool2022emergency, ouchi2019goals}.

%% NEW PARAGRAPH 3 - Study Focus and Research Questions

Although prior research has identified barriers to SICs (e.g., difficulty in determining which patients need them~\cite{shilling2024let}, hesitation to initiate SICs~\cite{ouchi2019goals}, lack of accessible documentation for providers to reference~\cite{ouchi2017preparing}, and concern about triggering patient distress through poor prognostic disclosure~\cite{shilling2024let}), most of these studies were conducted outside EDs. As a result, there is limited understanding of the feasibility and challenges of conducting SICs in the ED, one of the most time-critical medical settings where providers (e.g., physicians) are often overwhelmed. To address this gap, our study asks the first research question (RQ) (\textbf{RQ1:}) \textbf{How do healthcare providers approach conducting SICs with patients or their family members in EDs, and what barriers do they face?}

Leveraging technological solutions to support patient–provider communication has long been a central focus of the HCI community (e.g., \cite{ramesh2024data, patel2013visual, jang2014bodydiagrams, ni2011anatonme, bascom2024designing, jo2024exploring, liaqat2024promoting, wang2024commsense, berry2019supporting, yang2024talk2care, chen2025designing}). For example, one study explored tools for eliciting patients' values from caregivers \cite{foong2024designing}. However, limited research has examined technological solutions to support SICs from the healthcare provider's perspective. Recent advances in artificial intelligence (AI) have shown promise in supporting the conduction of SICs \cite{manz2020effect,chua2022enhancing}—for example, by helping providers identify patients who would most benefit from SICs through AI-based mortality risk prediction models \cite{kwon2019artificial, pourhomayoun2020predicting, avati2017improving, wang2019development}, or by generating contextually appropriate, emotionally sensitive, and condition-specific questions to guide conversations between providers and patients or caregivers~\cite{maity2024future, yang2024talk2care,zheng2025customizing, seo2025enhancing, kumar2023exploring}. Recognizing these research gaps and the potential benefits of emerging technologies, we pose the second research question (\textbf{RQ2:}) \textbf{How do healthcare providers perceive the opportunities and barriers of technologies such as AI to support SIC workflows in EDs?}

%\vspace{\baselineskip}

To explore these research questions, semi-structured interviews were conducted with 11 ED providers. Given the absence of a standardized, evidence-based SIC workflow in this setting, we examined how they conduct SICs in practice. Our analysis revealed a four-phase ED-specific workflow for conducting SICs: identification, preparation, conduction, and documentation. We further investigated providers' specific needs and challenges at each phase, as well as their perceptions of how emerging technologies, such as AI, could support SICs in the fast-paced, time-constrained ED environment. Overall, our work makes the following key contributions: %%1) A detailed characterization of care providers' workflow for conducting SICs in the ED. 2) An empirical analysis of providers' needs, challenges, and opportunities across each phase of the SIC workflow. 3) Proposed AI-assisted SIC workflow (as shown in Fig.~\ref{fig:teaser}) and design guidelines for developing AI systems that support SICs in emergency care.

\begin{itemize}
\item {A novel four-phase workflow for conducting SICs in Emergency Departments, developed from providers’ real-world practices and tailored to the constraints of ED settings.}
\item {An empirical analysis of the needs and challenges providers face at each phase of this ED-specific SIC workflow.}
\item {Recognition of opportunities and concerns for integrating AI systems into each workflow phase (as shown in Fig. \ref{fig:teaser}), along with design guidelines that outline how AI can support SICs in EDs while balancing efficiency and empathy.}
\end{itemize}

%% file: sections/2related_work_revised.tex
\section{Related Work}

\subsection{Serious Illness Conversations in Clinical Settings}

Life-threatening illness refers to a health condition that carries a high risk of mortality and either negatively impacts a person’s daily function or quality of life, or excessively strains their caregivers ~\cite{kelley2018identifying}. 
When patients arrive at EDs with life-threatening illnesses, ED providers must engage with the patients and/or their family members to have SICs to guide their clinical decision-making.
Through SICs, healthcare providers systematically explore patient preferences: they discuss expected disease trajectory~\cite{ouchi2017preparing}, uncover fundamental values (e.g., the desire to be alive for grandchildren~\cite{naik2016health}), clarify goals of care (e.g., curative treatment versus comfort care), and identify specific treatment preferences~\cite{shilling2024let}.
In one study of 340 older adult patients with life-threatening illnesses, 64\% preferred not to be connected to machines, and 93\% prioritized being free from pain~\cite{steinhauser2000factors}.
%yet patients may receive care that conflicts with their goals~\cite{teno2002medical}, resulting in unnecessary suffering~\cite{wright2008associations}.
SICs allow providers to deliver care that minimizes suffering, offers emotional support, and preserves quality of life~\cite{prachanukool2022emergency, ouchi2017preparing, geerse2019qualitative}.
%Through SICs, patients and family members can articulate their values, goals, and care preferences, allowing clinicians to develop treatment plans that align with their wishes~\cite{houben2014efficacy,silveira2010advance} and can reduce both the frequency and duration of unnecessary hospitalizations~\cite{klingler2016does}. 
However, many may not have participated in an SIC~\cite{prachanukool2022emergency}. 
Even among those who had SICs, their care preferences can evolve over time~\cite{kim2016natural}, creating ongoing gaps in care alignment. 

As the primary entry point into the hospital system~\cite{ouchi2019goals}, the ED is where initial treatment plans are formulated. This makes it a particularly critical setting for SICs~\cite{prachanukool2022emergency}, as the decisions made during this first encounter often set the entire trajectory for a patient's care.
For patients with life-threatening illnesses, SICs can reduce aggressive interventions, decrease surrogate stress, and improve care alignment with patient goals and wishes~\cite{ouchi2017preparing}. Despite this critical need, SICs are infrequently conducted in EDs~\cite{ouchi2019goals, murray2025goc}. 
Previous studies have identified some barriers faced by healthcare providers during SIC, including difficulty in determining which patients need them~\cite{shilling2024let}, hesitation to initiate SICs~\cite{ouchi2019goals}, lack of accessible documentation to reference~\cite{ouchi2017preparing}, and concern about triggering patient distress through poor prognostic disclosure~\cite{shilling2024let}.
%including severe time constraints (averaging only 16 minutes of direct patient-provider contact \cite{Reznek_Mangolds_Kotkowski_Samadian_Joseph_Larkin_2023}) and concerns about triggering negative emotional responses~\cite{shilling2024let}. 
%Additionally, prognostic uncertainty often leads to clinician overoptimism~\cite{christakis2000extent}, which patients may internalize, resulting in preferences for overly aggressive care.
However, many of these studies were conducted in non-ED settings. Furthermore, many of these studies are descriptive, documenting observations without investigating the underlying causes.

A few guidelines and frameworks have been developed to support healthcare providers in conducting SICs. 
For example, the Serious Illness Conversation Guide (SICG)~\cite{bernacki2015development} is a validated framework that offers a structured sequence of open-ended questions.  
Moreover, the Serious Illness Care Program (SICP) adds training and workflow reminders~\cite{paladino2020training, paladino2023improving}.   However, both frameworks can face limitations in EDs. SICG takes more than 20 minutes to complete~\cite{mandel2023pilot}, and the fixed questions are not adaptable to varying clinical scenarios.  SICP does not fully alleviate the emotional burden clinicians face~\cite{paladino2023improving}.  
These limitations further highlight the need for novel approaches tailored to ED workflows. However, limited research has examined the specific workflow of conducting SICs in EDs and identified the unique needs and challenges healthcare providers face in this time-sensitive and high-stake clinical environment. 
These critical gaps create the urgent need to understand ED-specific SIC workflows and requirements, which form the foundation for RQ1.

\subsubsection{{Differences Between Serious Illness Conversations in Emergency Departments and Routine Clinical Settings}}
{The role of SICs in EDs differs fundamentally from those in routine clinical settings, such as outpatient oncology or primary care. In outpatient settings, SICs are typically pre-scheduled, longitudinal engagements~\cite{lakin2017systematic, curtis2018effect}. For instance, trials of the SICP in oncology clinics treat SICs as a planned, multi-component process involving patient preparation, structured guides, and follow-up sessions, which aims to improve care alignment over weeks or months~\cite{bernacki2015development}. Healthcare providers in these settings often have established relationships with patients, access to medical histories, and protected time for SICs.}

{In contrast, SICs in the ED differ markedly from those in specialized or outpatient settings. As \citet{ouchi2019goals} notes, ED SICs occur at clinical “turning points,” where patients experience unexpected decline and an urgent need for values-based decision-making under severe time pressure. Unlike specialty contexts—such as oncology—where SICs may unfold along a more predictable disease trajectory, ED clinicians must conduct SICs across a wide range of acute presentations, disease types, and patient populations. Moreover, ED providers face substantially higher workload demands than outpatient physicians~\cite{morley2018emergency}, often managing several simultaneous patients~\cite{chisholm2001work}. Prior work shows that ED physicians spend 37.5 minutes per hour responsible for three or more concurrent patients, compared to just 0.9 minutes for primary care providers~\cite{chisholm2001work}. Despite these challenges, ED-based SICs offer unique advantages. Evidence shows that conducting SICs in the ED—rather than waiting for inpatient admission—can meaningfully alter downstream care, reducing hospital length of stay, ICU overuse, and increasing hospice utilization~\cite{johnson2025benefits}. However, we still lack a detailed understanding of how ED clinicians actually conduct SICs and the breakdowns they encounter in this high-pressure environment.}

%%The role of SICs in EDs differs fundamentally from those in routine clinical settings, such as outpatient oncology or primary care. 
%%In outpatient contexts, SICs are typically pre-scheduled, longitudinal engagements. For instance, trials of the SICP in oncology clinics treat SICs as a planned, multi-component process involving patient preparation for SICs, and structured guides, which aims at improving care alignment~\cite{bernacki2015development}.
%%In contrast, ED providers require a distinct approach for SICs. 
%%For example, ~\citet{ouchi2019goals} characterizes SICs in EDs as occurring at critical "turning points" of acute deterioration.
%%Accordingly, ED providers have to make decisions under extreme time pressure. 
%%Moreover, a study suggested that initiating SICs in EDs, as opposed to waiting for inpatient admission, significantly alters the care trajectory, which results in shorter hospital stays, reduced ICU overuse, and increased hospice utilization~\cite{johnson2025benefits}. 
%%Ultimately, whereas SICs in outpatient settings favor comprehensive, iterative planning based on longitudinal conversations, SICs in EDs are brief, high-stakes interactions focused on immediate decision-making.
\subsection{Technology-Supported Patient-Provider Communications}

Effective patient-provider communication is fundamental to patient-centered care~\cite{sunnerhagen2013enhancing}, improving patient engagement~\cite{butz2007shared}, reinforcing positive expectations~\cite{wright2004doctors}, and enhancing mental health outcomes~\cite{wong2019associations}. The HCI community has increasingly recognized communication as a critical design space for healthcare technologies. Various technological interventions have been developed to support patient-provider communication~\cite{ryu2023you, ramesh2024data, seljelid2021digital, ryu2023you}.  Notable examples include BodyDiagrams, which lets a patient annotate pain intensity and timing on a human body map~\cite{jang2014bodydiagrams}; recovery dashboards facilitate joint data interpretation between stroke patients and clinicians~\cite{ramesh2024data}; and mobile applications display family-reported needs in ICUs to prompt palliative care consultations~\cite{cox2025mobile}. 
Additionally, researchers have developed communication feedback tools to help clinicians recognize and address implicit bias in real-time interactions~\cite{bascom2024designing}. A subset of these technologies specifically targets the elicitation of patient values and care preferences~\cite{sudore2017effect, berry2021supporting}. One example, Living Voice~\cite{vitaldecisions_mylivingvoice}, allows patients to document treatment wishes for sharing with family or providers, while Foong et al.~\cite{foong2024designing} proposed five value-elicitation probes focused on caregiver input. Curtis et al.~\cite{curtis2023intervention} introduced the Jumpstart Guide, which automatically extracts EHR data and generates conversation templates.

However, these tools are predominantly designed for low-pressure care settings where patients have time to reflect, and clinicians can engage in unhurried conversations. The ED presents fundamentally different challenges: high-pressure situations, time constraints (typically four-hour stays~\cite{ouchi2019goals}), patients in crisis, and limited prior rapport.  Existing tools may not address the urgency, emotional intensity, or rapid decision-making demands inherent in ED-based SICs.
Given that SICs are fundamentally conversational in nature, involving complex dialogue between providers and patients about sensitive topics, AI-powered Conversational Agents (CAs) represent a potential solution to some of these communication challenges, as they have shown promise in various clinical contexts \cite{chua2022enhancing}. Yang et al.~\cite{yang2024talk2care} developed Talk2Care, an LLM-powered telehealth system that enables patients to share health information through voice interfaces while providing clinicians with AI-generated conversation summaries. Similarly, Li et al.~\cite{li2024beyond} created a GPT-4 chatbot that streamlines pre-visit workflows by collecting patient information and generating interactive summaries for healthcare providers. Another line of research focuses on the use of AI to streamline communication and share decisions in clinical settings~\cite{bartle2022second}. ~\citet{hao2024advancing} built an interactive AI system that supports cancer patients to collaborate with their cancer care clinicians during consultations. 
~\citet{seo2025enhancing} designed an AI chatbot to support communication among pediatric patients, parents, and providers during clinical visits.

Building on these broader AI communication applications, researchers have also begun exploring AI solutions specifically designed for SICs. One line of work focuses on patient identification and workflow integration. For example, AI algorithms have been developed to predict patient mortality rates, helping clinicians efficiently identify high-risk patients who would benefit from SICs~\cite{avati2017improving}. Manz et al.~\cite{manz2020effect} demonstrated that combining AI mortality predictions with automated reminders increased SIC frequency among oncology clinicians.  

A second area of research addresses SIC documentation and analysis. Natural language processing (NLP) techniques have been developed to identify previous SIC documentation in EHRs~\cite{lindvall2022natural} and to use NLP to extract SIC content from clinical notes~\cite{lee2021identifying}. Hachem et al.~\cite{hachem2025electronic} created dashboards to track clinician SIC training progress and facilitate conversation documentation. Moreover, ~\cite{chua2022enhancing} proposed an AI-human collaborative workflow for conducting SICs. They suggested using conversational agents to collect relevant patients' information before healthcare providers have SICs with patients to improve the efficiency of SICs. After SICs, they highlighted the potential of using NLP to improve both the efficiency and quality of SIC documentation. Given AI's potential to evaluate communication skills in clinical settings~\cite{ryan2019using}, they proposed leveraging AI to analyze SICs and provide feedback to healthcare providers.

While this body of work demonstrates AI's potential in healthcare communication, significant gaps remain in understanding how clinicians want AI support for SICs, particularly in high-stakes ED environments. Existing research has primarily focused on technical feasibility rather than user-centered design considerations. From an HCI perspective, it remains unclear what AI capabilities would be most valuable to ED clinicians, how such tools should integrate with existing workflows, and what design principles should guide AI-supported SIC tools for time-constrained, emotionally intense clinical encounters. 
These research gaps inform our RQ2.

%% file: sections/3interviews_method_and_participants.tex
\section{Methods}

\begin{figure*}[htbp]
  \includegraphics[draft=false,width=\textwidth]{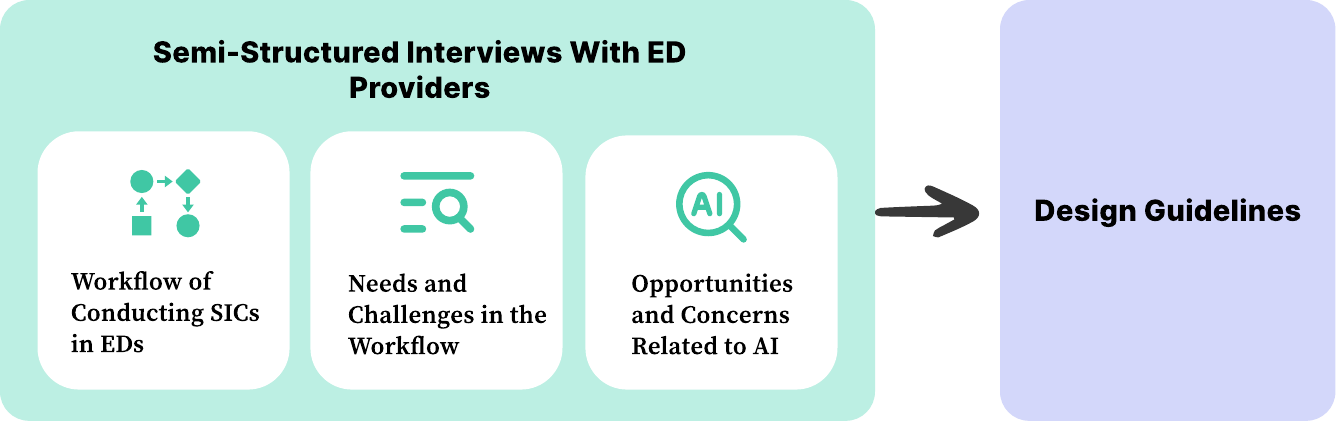}
  \caption{In this paper, we conducted semi-structured interviews with 11 ED providers to systematically examine their workflow of conducting SICs and challenges they face during their workflow. We also explored the AI opportunities and concerns in supporting ED-specific SIC workflow with our participants. Finally, we derived design guidelines grounded in our findings.}
  \label{fig:study_procedure}
   \Description{Fig.2: The figure shows the procedure of our study. The interviews focused on three areas: understanding the workflow of conducting SICs in EDs, identifying needs and challenges within that workflow, and exploring opportunities as well as concerns related to the use of AI. Findings from these three areas were then synthesized to inform a set of design guidelines aimed at improving SIC practices in emergency care contexts.}
\end{figure*}

To address our research questions, we conducted semi-structured interviews with 11 healthcare providers working in EDs, including both nurses and physicians.
Since there is a lack of evidence-based SIC workflow in EDs, and the workflow may vary among individual ED providers, we first needed to understand their range of practices in order to synthesize and derive their workflow. 
The workflow helped us identify key moments and decision points that ED providers face when conducting SICs in EDs. 
Meanwhile, we examined the challenges providers encounter throughout the SIC workflow, including barriers that prevent them from initiating SICs and obstacles that arise while conducting them.
Moreover, we explored their perspectives on how technologies, such as AI, might support or hinder their practices.
Based on these findings, we developed design guidelines for AI-assisted SICs that overcome workflow barriers.
The summary of our study procedure is shown in Fig.~\ref{fig:study_procedure}.

\subsection{Study Participants}

To recruit the participants, we first used convenience sampling~\cite{sedgwick2013convenience} to recruit two ED providers who were readily accessible to us, and then employed snowball sampling~\cite{goodman1961snowball} by asking them to recommend additional participants.
In total, we recruited nine physicians and two nurses working in EDs to capture a holistic understanding of the interprofessional workflow and communication dynamics involved in SICs.
{All participants were based in the United States and worked in tertiary academic medical centers~\cite{TertiaryCareCentersMeSH}, which provide advanced specialty and subspecialty care and serve as training sites for residents and fellows. Among the 11 providers in our sample, three were female, and eight were male. A recent national analysis reports a similar gender distribution among ED providers, with about 70 percent male and 30 percent female~\cite{bennett2020national}, so our sample reflects this broader pattern.
Regarding AI exposure among our participants when conducting SICs, the adoption of automated technologies was minimal. Only P2 and P8 reported using tools to automatically transcribe SICs for retrospective review. Beyond these specific instances, we found no evidence of participants utilizing computational tools in their workflows other than the EHR system, such as Epic~\cite{chishtie2023use}.}
After the 11th provider interview, we stopped participant recruitment because data saturation was achieved~\cite{guest2006many}.
All participants had experience conducting SICs.
Among the participants, two (P1 and P2) are both practitioners as well as researchers focusing on SICs in EDs.
Table~\ref{tab:demographics2} shows our participant demographics, department, roles, and years of practice.

\begin{table*}
  \caption{Demographics of Participants in Semi-Structured Interview}
  \label{tab:demographics2}
  \begin{tabular}{cclcc} 
    \toprule
    P\# & Gender & Department & Job Title & Year of Practice \\ 
    \midrule
    P1 & Male & Emergency Department & Emergency Medicine Physician & 7 years \\ 
    P2 & Male & Emergency Department & Emergency Medicine Physician & 11 years \\
    P3 & Male & Emergency Department & Emergency Medicine Physician and Toxicologist & 30 years \\ 
    P4 & Female & Emergency Department & Nurse & 42 years \\
    P5 & Male & Emergency Department & Emergency Medicine Physician & 10 years \\ 
    P6 & Female & Emergency Department & Emergency Medicine Physician & 4 years \\
    P7 & Male & Emergency Department & Emergency Medicine Physician & 7 years \\ 
    P8 & Female & Emergency Department & Nurse & 20 years \\
    P9 & Male & Emergency Department & Emergency Medicine Physician and Intensivist & 20 years \\ 
    P10 & Male & Emergency Department & Emergency Medicine and Palliative Care Physician & 5 years \\
    P11 & Male & Emergency Department & Emergency and Critical Care Medicine Physician & 5 years \\
    \bottomrule
  \end{tabular}
\end{table*}

\subsection{Data Collection}

The interview protocol was developed iteratively. 
In the initial stage, our questions focused broadly on dimensions such as how clinicians make decisions during SICs. 
After the first two interviews, we found that ED providers encountered greater challenges in other steps, such as preparing for SICs or documenting them. 
Based on these insights, we refined the protocol to include these critical areas. Before conducting interviews, the interviewers familiarized themselves with key terminology related to SICs, as well as the goals and content of SICs, by reviewing relevant literature. {The interviews were conducted via Microsoft Teams and audio-recorded. At the start of each session, participants were asked not to share any personally identifiable information (e.g., names, locations). After recording, the audio was transcribed, and the research team manually reviewed each transcript, cleaned the data, such as transcription errors, for data accuracy. Meanwhile, the research team removed any remaining identifying details, including names of participants (P1–P11), other care providers, and the specific institutions.}

{At the beginning of the interviews,} participants were invited to describe a recent experience conducting a SIC in EDs. Based on their response, we asked follow-up questions about how they identified patients who needed SICs, how they elicited patients’ goals and values, how those informed treatment decisions, and what tools or documentation practices they currently used. Clarifying questions were asked when medical terminology arose, and clinician co-authors subsequently reviewed transcripts and interpretations to ensure accuracy. Finally, we explored participants’ perspectives regarding potential roles for AI technologies in supporting SICs as well as concerns. Each interview lasted between 30 and 60 minutes. 
Participants were compensated \$20. This study was approved by the first author's university Institutional Review Board.

\subsection{Data Analysis}

After the interviews, two researchers analyzed all transcripts using thematic analysis~\cite{fereday2006demonstrating}. Our first goal was to identify the workflow that ED providers follow when conducting SICs. Through participants’ descriptions of their current practices, we identified multiple procedural steps and consolidated related steps into broader categories. {For example, we grouped “explore care goals” and “address family concerns” into a single stage (conduction) because these overlapping activities naturally occur together rather than unfolding as separate, sequential tasks.} This process resulted in four workflow stages: identification, preparation, conduction, and documentation.

{Next, we examined the challenges and needs associated with each workflow stage. Two researchers independently performed open coding on a subset of transcripts (\textit{n}=2 each) to generate initial codes, compared their results, reconciled discrepancies, and developed a preliminary codebook. Using affinity diagramming~\cite{hartson2012ux}, the team clustered these early codes into high-level thematic groups. The codebook was then iteratively refined and applied to the remaining transcripts through both inductive analysis (to capture emergent needs or challenges) and deductive analysis (to categorize relevant excerpts within the developing structure). The coding team met weekly to compare interpretations, resolve disagreements through negotiated agreement~\cite{campbell2013coding}, and update the codebook. This approach aligns with prior HCI qualitative studies~\cite{desai2023painless, garg2021understanding}}

{To ensure rigor, we followed the trustworthiness criteria described by Nowell et al.~\cite{Nowell_Norris_White_Moules_2017}, including iterative codebook development, collaborative resolution of discrepancies, and transparent reporting of analytic decisions. Consistent with these criteria, we report theme prevalence using established qualitative terminology: “a few’’ refers to up to 20\% of participants, “some’’ to 21–50\%, “most’’ to 51–80\%, and “nearly all’’ to more than 80\%.} %%Importantly, the severity or significance of a challenge was not determined by frequency alone. Themes were prioritized when they reflected substantial emotional or logistical burden for providers, even if mentioned by fewer participants.}

%% file: sections/4interviews_findings.tex
\section{Findings}
\label{sec:finding}
In this section, {we first present the overview of the SIC workflow in EDs (Section 4.1). The subsequent subsections (Sections \ref{sec:identification}–\ref{sec:documentation}) elaborate on each phase in more detail, and its challenges and needs associated with each phase, addressing RQ1. Our identification of primary challenges was based on their impact on SICs rather than on frequency alone. Specifically, we prioritized challenges that (1) significantly disrupted or impeded the SIC process, or (2) elicited strong negative emotional responses from participants.}
Next, we discuss AI opportunities and concerns to support their SIC workflow in EDs, addressing RQ2 (Section \ref{sec:ai}).
Key findings are summarized and visualized in Fig. \ref{fig:key_findings}.

\subsection{Workflow of SICs and Related Challenges in EDs (RQ1)}

\begin{figure*}[htbp]
  \includegraphics[draft=false,width=\textwidth]{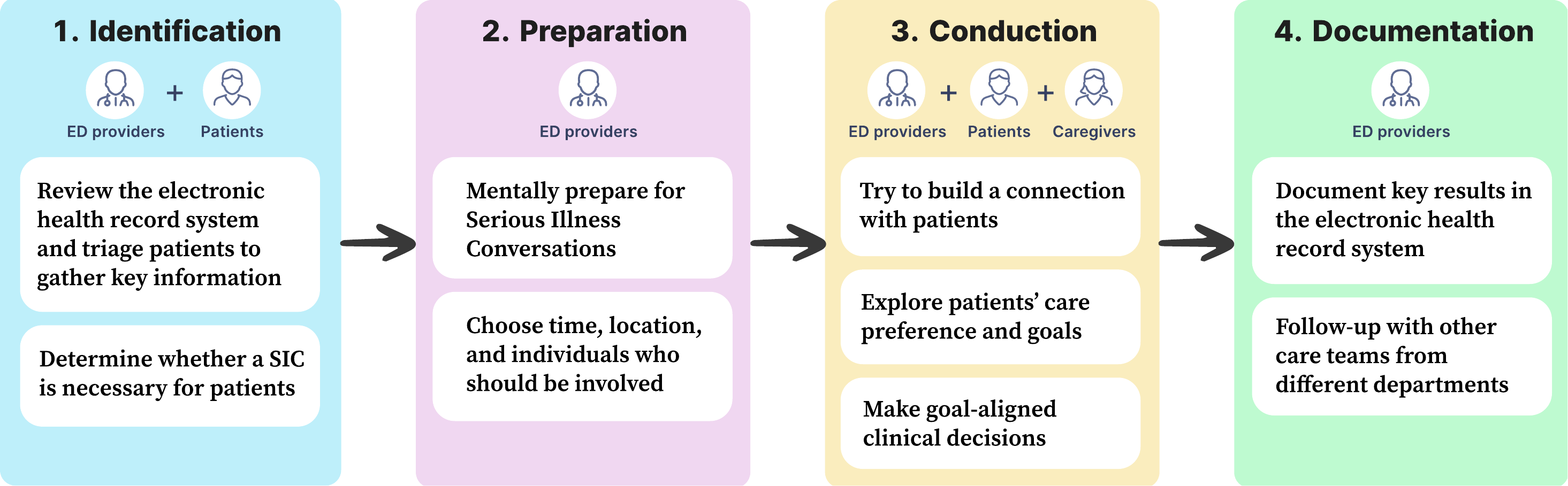}
  \caption{Four phases in ED provider’s workflow for conducting SICs: (1) identify patients who need SICs, (2) prepare for SICs mentally and logistically, (3) conduct SICs, (4) document the key information from SICs. We also include who is involved in each phase, including ED Providers (emergency physicians or nurses), Patients, and Caregivers (family members or surrogates)}
  \label{fig:workflow_ui}
   \Description{Fig3: This figure maps the workflow of conducting SICs against the challenges providers face, potential AI opportunities, and resulting design guidelines. The workflow is broken into five stages: (1) Identification, (2) Preparation, (3) Conduction, (4) Documentation. At each stage, we showed the specific barriers providers face. Specifically, fragmented information in the EHR during identification, lack of patient information and guidelines during preparation, difficulties sustaining smooth dialogue and missing key questions during conduction, inconsistent formats and missing details in documentation. Moreover, we identify the need for Feedback Mechanism to improve SIC skills. Corresponding AI opportunities include extracting and summarizing SIC-related data during identification, generating personalized conversation starters during preparation, offering real-time conversational support through ambient listening during conduction, automatically generating structured notes during documentation, and analyzing past recordings to provide personalized feedback. The figure also shows the design guidelines corresponding to each stage: synthesize patient information and provide SIC starters, support conversational flow with ambient technologies, generate notes while preserving clinical autonomy, and deliver feedback that fosters continuous skill development. Together, the figure illustrates a pathway from workflow challenges to AI-enabled interventions and design recommendations.}
\end{figure*}

{
We identified and synthesized four core phases in the workflow of conducting SICs in EDs: \textcolor{cyan}{\textbf{Identification}}, \textbf{\color{magenta}{Preparation}}, \textbf{\color{brown}{Conduction}}, and \textbf{\color{teal}{Documentation}} (Figure~\ref{fig:workflow_ui}). 
The process begins the moment a patient arrives at the ED with a life-threatening condition, prompting providers to enter the \textcolor{cyan}{\textbf{Identification}} phase to determine whether a SIC is needed. 
Once the need is confirmed, they move into the \textbf{\color{magenta}{Preparation}} phase to plan logistics (for example, where and with whom to talk) and to prepare themselves mentally for a difficult conversation.
Preparation then flows into the \textbf{\color{brown}{Conduction}} phase, the core interaction where providers explore the patient’s values and goals for treatment, often beginning with small talk to build rapport. 
Finally, providers enter the \textbf{\color{teal}{Documentation}} phase, recording key outcomes (for example, care preferences) in the EHR. 
\textbf{\color{teal}{Documentation}} also serves as a starting point for other clinical teams to continue SICs and deliver goal-concordant care as the patient moves through the system.
}

{
We use consistent colors for the four phases in both the figures and the text to help readers connect the narrative description with the visual model (Figure~\ref{fig:workflow_ui}).
Below, we describe each phase as it unfolds in ED practice and then surface the main challenges providers face in that phase.
}

%Understanding patients health conditions and previous SICs using EHR under time-pressure
\subsubsection{{\textcolor{cyan}{\textbf{Identification}} Phase: Understanding Patients' Conditions and Previous SICs Using the EHR Under Minutes}}
\label{sec:identification}

When patients arrive at EDs with life-threatening conditions (e.g., brain bleeding) requiring aggressive treatment, healthcare providers enter the identification phase to determine whether SICs are necessary.
Healthcare providers typically begin this process by consulting the EHR to locate documentation from treatment notes and previous SICs, which may contain critical information, such as code status, care preferences, treatment goals, and patient values.
{This information is essential for helping ED providers make difficult, time-pressured decisions: whether to pursue aggressive measures that may prolong life but increase suffering, or comfort care that prioritizes pain relief over a cure.}

{
However, this information is often scattered in the EHR, requiring providers to hunt for it. 
\textbf{This task becomes a critical barrier in \textcolor{cyan}{Identification}, as ED providers rarely have the time for such retrieval.} Unlike in other clinical settings, healthcare providers may have protected time to thoroughly review patient records and resolve ambiguities, often by asking patients detailed questions. In EDs, however, this is a ``\textit{luxury}'' (P2) that providers rarely have. Many times, especially for senior providers, they must remain in central treatment areas to constantly perform the treatment. This supervisory role physically tethers them to the patient's bedside. As a result, the physical location and high-priority duty in EDs lead to fewer opportunities to step away to retrieve the needed information from the EHR.
}

\begin{quote}
\textit{``If you open up a patient's chart on the left hand, most column where their demographic information is if they've previously had, but it's really limited information that's captured [in a patient's chart]. So it's usually just like a DNR DNI status to not resuscitate, do not intubate......If patients have had more extensive goals of care discussion, it's not always easily accessible. ''} (P6)
\end{quote}

{
These challenges persist even when using keyword searches, which often yield incomplete results. 
For instance, when P10 searched for SIC-related keywords, the EHR system returned "not addressed," which failed to provide the essential information needed to support informed clinical decision-making.}
The key factor underlying these challenges of information access is \textbf{inconsistent \textcolor{teal}{Documentation} practices for where to record SICs}, which results in SIC-related information being scattered across multiple locations within patient records. 
As P11 explained, some healthcare providers document SICs by creating separate ED progress notes, others include information in ED final notes, and some use advanced care planning (ACP) notes.
When SIC-related information is documented outside the standardized ACP note format, it becomes effectively invisible to other providers to locate this critical information.
The inconsistent documentation practices explain why healthcare providers frequently cannot access patients' previously documented code status, goals, and values precisely when this information is most needed for time-sensitive treatment decisions.

\begin{quote}
\textit{``I've seen variability where I've seen some people document it as a separate ED progress note. I've seen some people document as an ACP note. I've also seen some people just document it as part of the ED course that then populates into the final ED provider note.''} (P11)
\end{quote}

{Even when providers successfully locate the necessary information, \textbf{there is often a discrepancy between a patient’s documented treatment intent and their code status.} For example, P1 noted a case where the treatment intent prioritized pain management, yet the code status indicated intubation and resuscitation. This mismatch often signals a need for SICs to clarify treatment preferences. One reason for this discrepancy can be that patient preferences evolve over time without corresponding updates to medical records (P5, P9). Consequently, P5 and P9 emphasized that access to up-to-date patient information is essential for effective clinical decision-making.}

\begin{quote}
    \textit{``The information documented in the chart may not be accurate, the goals of care may have changed, and haven't been updated. Sometimes there's conflicting notes, and understanding what's right and current can be helpful.''} (P5)
\end{quote}

{When healthcare providers cannot locate SIC documentation in the EHR, or there is a mismatch, they resort to directly asking patients whether SICs have occurred previously.
However, some participants (P7, P8, P9, P10) discovered that many patients had never engaged in SICs at all. 
\textbf{This gap can stem from a systematic perspective where the EHR lacks standardized protocols for identifying patients who would benefit from SICs early in their illness trajectory, before acute crises develop.}
Without early identification, SICs become reactive conversations triggered by emergencies rather than proactive discussions integrated throughout the patient's care journey.}

%%Therefore, establishing early identification protocols is crucial not only for ensuring patients receive care aligned with their goals and values from the outset of their illness but also for equipping ED providers with the essential information needed to make rapid, patient-centered decisions during critical moments.
%%When ED providers cannot access a patient's previous treatment notes or SIC documentation in the EHR, or when they encounter discrepancies within these records, they have to initiate preparation for an SIC.}

{
\begin{quote}
\textit{``I find a lot of oncologists do not have these discussions with patients about advanced care and we are having to do that (SIC) in real time with limited information in the ED....Our team would greatly appreciate that these discussions (SICs) are made with family and the patient when they're alert and together and can think through the process... Victims of cancer and long-term disease have not had these conversations, or at least they're not documented in their chart, which makes it challenging for us in the ER (emergency room) to make these decisions.''} (P3)
\end{quote}}

\subsubsection{{\textbf{\color{magenta}{Preparation}} Phase: Psychological and Logistical Strains on ED Providers}}
\label{sec:preparation}

When patients are identified as needing an SIC, healthcare providers move to the preparation phase.
{Although often brief in EDs, this phase lays the foundation for a productive SIC.
There are two parts in this phase: ED providers' mental readiness and logistical planning. 
Mental readiness refers to the emotional and cognitive preparation that healthcare providers must undertake to engage in what are often sensitive, complex SICs.  
Logistical planning is to decide when and where the conversation should take place and who should be present. 
This dual preparation helps ensure that healthcare providers can navigate the complex terrain of an SIC with appropriate sensitivity and clinical grounding.}

{
\textbf{However, it can be hard to proceed with the \textcolor{magenta}{Preparation} phase if the patient is too ill to communicate and caregivers are unavailable, especially in the middle of the night.}
Given the continuous around-the-clock nature of EDs, some patients may arrive in EDs in the middle of the night and are too sick to participate in an SIC. 
Moreover, their family members may not be physically present.
To address this, P2 attempts to find a proxy by calling patients' families or even neighbors ``\textit{10 times, and be like, wake up}''. 
However, it is hard for people to pick up a call in the middle of the night. 
Consequently, healthcare providers are unable to conduct SICs and must make treatment decisions unilaterally, which increases the risk of providing care that conflicts with the patient's wishes.
Even when patients are able to participate in SICs, the preparation can be difficult.}
While professional training may equip healthcare providers to handle emotionally intense situations, our interviews revealed that \textbf{personal experiences can still significantly complicate the \textcolor{magenta}{Preparation}}, particularly in the ED (P3, P7).
For example, P7 described how his father's sudden death left emotional trauma that now makes the preparation more difficult.
He found these conversations more emotionally charged and noted becoming visibly affected, tearing up during conversations where he previously maintained composure. 
%This finding reveals that even highly trained healthcare providers are not immune to the psychological effects of personal experience when preparing for SICs, particularly when the patient's condition mirrors that of the loved one they lost.

\begin{quote}
\textit{``And then even the, the provider, the clinician comes with a lot of their own baggage. So, for example, you know my father passed away in 2021 from COVID. That next year or two, I was emotionally different than I was before. These conversations were more emotional for me. I was more connected to the conversation as I just lost my father. Even I might tear up during these conversations when historically throughout my career, I never would cry or show emotion like that.''} (P7)
\end{quote}

{
This psychological stress is compounded by \textbf{the sudden, emergent nature of SICs without support.} As P2 noted, the need for an SIC is often unknown until ED providers physically enter the room and realize how critically ill the patient is, which triggers the need for an SIC.
This sudden need for an SIC can induce what P2 described as "panic." 
The underlying reason for this panic is not only the limited time to prepare, but also the fact that ED providers are not specifically trained to conduct SICs.
As P1 said that they only received the training for one or several sessions once a year, and the training stops after they become an independent practicing ED provider.
Thus, some participants (P1, P3, P8, P9) expressed the need for SIC training.
Because of the limited training and practice, a few participants (P1, P2, P4) desired support from specialists who conduct SICs more frequently, such as the patient's oncologist.
However, such support is often inaccessible, as other providers have limited time and are unavailable during night hours.
Therefore, when ED providers are preparing for SICs, they have to tell themselves to "bite the bullet" without any support. }
{
\begin{quote}
\textit{``So if anybody's available, I am happy like I will be ecstatic to involve them (healthcare providers in other departments) in this decision making. 95 percent of the time, nobody's available. It's just me. At 3:00 in the morning when someone's about to die...I am panicked. I want to hit this panic button and say I need help.''} (P2)
\end{quote}}

Beyond personal experience, {\textbf{logistical constraints (time and space) further compound ED providers' psychological burden of \textcolor{magenta}{preparing} for SICs.}
Although time pressure and limited space for patients are well-known challenges in EDs, we identified their negative effect on SIC preparation, as well as the relation between these logistical constraints with the psychological burden of preparing for SICs.}
First, some participants (P6, P9, P10, P11) described challenges in finding an available time slot.
The high patient volume in the ED forces ED providers to manage multiple cases simultaneously, creating constant interruptions and time pressure.
Therefore, this environment increases cognitive load and emotional strain during preparation, particularly when SICs require emotional presence, careful communication, and active listening. Moreover, space limitations create additional preparation challenges.
Some participants (P5, P6, P8, P10, P11) noted that having access to a private and quiet room is important not only for helping patients and their family members remain calm, but also for supporting ED providers emotionally in conducting emotionally charged SICs.
However, the ED's overcrowded and chaotic environment makes such spaces scarce (P10).
For example, some patients wait 10-15 hours in crowded rooms, and some are placed in noisy hallways, which leaves virtually no private spaces available for sensitive conversations.
Together, insufficient time and inadequate space intensify ED providers' psychological burden to prepare SICs and thus make SICs hard to conduct. 
\begin{comment}
\begin{quote}
\textit{``I think in the emergency setting, it is usually [challenging] like physical, like [finding] time and space [to conduct SICs]. So like I said, you know, even finding private care space sometimes is challenging. Having the time to have these conversations is really challenging.''} (P6)
\end{quote}

\end{comment}

\subsubsection{{\textbf{\color{brown}{Conduction}} Phase: Conducting SICs with Unprepared Patients and Families Without Support}}
\label{sec:conduction}

%This decision often involves a stark dispositional choice: discharge to home, admit to the hospital, or transition to palliative care. Furthermore, if admitted, the ED team must determine the level of intervention: aggressive, life-sustaining treatment, which may increase suffering, or a focus on symptom management and comfort. 

{Whether or not ED providers feel prepared, they must proceed with SICs, which represent the core interaction where values and goals are explored. 
As described by P2, this phase typically begins with an opening question to the patient or family that introduces the topic to set the expectation and initiates SICs sensitively. 
ED providers first assess what the patient and family understand about the current health condition and treatment, addressing any confusion or misinformation. 
The conversation then shifts to the heart of SICs: exploring the patient's treatment preference that needs to be conducted in EDs, and any care trade-offs.
The overarching aim of this phase is to elicit patients' care preferences to guide ED providers' decision-making (e.g., aggressive treatment or comfort care).
As a result, ED providers propose a plan of care that authentically aligns with what the patient values most.}
{
\begin{quote}
\textit{``I do the same steps, which include asking about patients' baseline function, and quality of life, like how do they perceive their life ... I ask about tough questions about values and goals, such as what kind of conditions you would consider worse than dying.''} (P2)
\end{quote}}

{However, the conduction of SICs can be extremely difficult in EDs.
Specifically, ED providers always \textbf{struggle to initiate and proceed SICs with patients and caregivers who are not mentally prepared for such discussions within extremely short timeframes (e.g., a matter of minutes).}}
For patients, many of them are cognitively overwhelmed, emotionally unprepared, and, in some cases, unreceptive to discussing their care preferences and deciding between “live” (aggressive treatment, but can cause irreversible damage) or “die” (reduce their pain but not prolong their lives) in such a short period. 
Especially for patients who have never had SIC before, they haven't had the chance to think of their treatment trajectory toward the end of life. 
As a result, SICs could be longer and complex in EDs (P5, P8).
For their caregivers, a few participants (P4, P11) mentioned that some caregivers often hold unrealistic hopes that the patient will return to full health, which hinders these family members from engaging in SICs to talk about the “crucial reality” of the possibility of losing their loved ones. 

\begin{quote}
\textit{``You know you're talking to a family member of a young child, very difficult conversation, very difficult. The moment you add a child into the, as the patient, it becomes, it's hard to have these conversations. The more sudden and unexpected the emergency is, the more difficult the conversation is. The family or the patient, they're just not prepared. You know, like to tell a, a family member that, you know, that their 50-year-old husband who was just mowing the lawn and playing soccer with their son, now just had a massive heart attack...It's a different conversation.''} (P7)
\end{quote}

{Although} prior studies note that patients and family members often struggle to accept prognostic information {outside of EDs}~\cite{you2017barriers}, {ED providers still have time to guide the conversation and gradually break the bad news in a gentle way.
In stark contrast, the goal of an SIC in EDs is immediate, emotionally charging, and decisional. 
\textbf{ED providers must initiate the SIC, break the bad news, be an emotional container to deal with negative reactions, and ask treatment preference within minutes.}
As P6 mentioned that ``\textit{in the ER, we are very blunt with what we ask...It's not a gentle conversation'}'. }
These emotionally charged dynamics complicate SIC delivery while intensifying psychological burden on providers, particularly those already struggling with preparation.
{As P1 expressed, what makes patients uncomfortable also makes ED providers uncomfortable, which creates a vicious cycle and increases the difficulty of moving SICs forward.}
As a result, ED providers' capacity to manage and respond to others' emotional distress becomes compromised, creating a downward spiral that hinders effective SICs.

\begin{comment}
The primary reason for this challenge is that patients and their caregivers often lack a clear understanding of the patients' current medical condition or the potential poor prognosis of their illness. 
{For example, P2 noted that some patients still believe they are healthy and will remain so for a long time. For these patients with unrealistic optimism, it becomes extremely difficult for ED providers to help them grasp their actual health status, which is far from what they expect, and then ask them to decide their treatment preferences within a few minutes in EDs.}
This reflects a gap in patient-provider communication prior to their arrival at the ED, where few, if any, discussions about the progression of their illness have taken place.
{
\begin{quote}
\textit{``She (the patient) was doing great until a couple days ago. And this is like a huge shock......That's in one case, they're (the patient and her caregivers) shocked, like they're sort of reality has been destroyed.''} (P10)
\end{quote}
}
\end{comment}
{
To navigate these challenges, participants (P4, P7, P8, P11) emphasized the importance of establishing a personal connection with the patient. 
Building this connection fosters trust, which makes patients more open to receiving bad news and more willing to engage in SICs.}
However, a few participants (P4, P7, P10) face challenges in building the connection.
For example, P7 noted that, unlike in primary care, where providers have time to get to know patients and {pave the way to build connections, \textbf{providers in EDs often meet patients for the first time under urgent and stressful conditions}. 
Moreover, patients' acute clinical deterioration compresses this entire SIC timeline of deliberation into a few critical moments with limited time to know patients and build a connection.}
To address this challenge, supplying patient-specific information (e.g., family support, personal interests such as favorite sports teams) before SICs is essential for helping ED providers build personalized strategies to {build the connection in a short period of time.
As P2 suggested, breaking the ice by finding shared interests, such as sports teams or children, is a good way to quickly build a connection with patients in EDs.}

\begin{quote}
\textit{``We're at a significant disadvantage in the emergency department as I am not their primary care doctor. I have not cared for them for a decade and earned all their trust and faith. They are meeting me at point blank, and they have to develop faith and trust.''} (P7)
\end{quote}

{
Our findings indicate a critical systems-level limitation is that \textbf{current training and guidelines fall short in supporting ED providers in \textcolor{brown}{conducting} SICs}.
%Unlike their counterparts in oncology or palliative care, who have more experience and training, ED providers have to conduct SICs with limited training and support.
Although there are guidelines for supporting SICs, they are primarily designed for clinical settings where the patient’s condition is stable, and there is enough time and cognitive capacity to conduct SICs.}
%Moreover, current guidelines lack support to personalize icebreakers that could help providers initiate SICs with patients they first meet, and to {break the bad news} in a way that minimizes additional cognitive and emotional strain on patients and their family members.
Additionally, there is no clear support for helping providers de-escalate heightened emotions and move the conversation forward in a limited timeframe. 
%These limitations point to the need for providing patient-specific information and contextualized support that are specifically designed for time-constrained and high-pressure EDs.
%{
%\begin{quote}
%\textit{``So the guides are there for too big and too cumbersome and too long when sometimes what we need is just like a little bit, just a a small unit.''} (P1)
%\end{quote}}

\subsubsection{{\textbf{\color{teal}{Documentation}} Phase: Creating Time-Consuming Notes with Fragmented Practices}}
\label{sec:documentation}

When healthcare providers successfully overcome the numerous challenges of conducting SICs, they enter the documentation phase, where they must document each SIC as a note in the EHR system, {including the patient's goals, values, and care preferences.}
However, this phase presents its own significant obstacles.
As participants (P7, P10) described, the documentation process is a time-consuming and burdensome task, particularly for healthcare providers working in EDs. 
Given the high workload in the ED, providers have limited or no time for documentation during their scheduled shifts.  
{Consequently, many providers must stay late or use unpaid personal time to complete this task; time that could otherwise be spent caring for additional patients.
}

When we examined why the process of documenting SICs is time-consuming and burdensome, we identified two primary factors: {\textbf{inadequate training and the absence of standardized protocols for SIC \textcolor{teal}{Documentation}}}.
First, current training emphasizes how to conduct SICs but overlooks the crucial step of completing SIC documentation. 
Without training, ED providers often document SICs in different locations, which makes it difficult for other healthcare providers to locate previously documented patient goals and values where they expect them to be, as we mentioned in Section \ref{sec:identification}.
Consequently, this inconsistency can undermine continuity of care.
Moreover, {the challenge is compounded by the absence of effective documentation methods. 
As P10 noted, no standardized approach exists for documenting SICs, further complicating an already difficult process.}
{
\begin{quote}
\textit{``A lot of what you do is looking through a ton of data, trying to synthesize it, seeing the patient, addressing too many things in too little time, and then trying to document it and then do it, you know, 30 or 40 more times in a shift. And so many of us end up staying late or working unpaid hours, finishing charts because we care, and the system takes advantage of us.''} (P10)
\end{quote}
}
{
The absence of systematic protocols leads to two key problems. First, ED providers document based on subjective definitions of importance, which means the content may not align with what other providers are looking for. Second, ED providers may record SICs using inconsistent structures and store them in disparate EHR locations. Consequently, there is a need for a standardized documentation protocol to enhance the efficiency of the SIC workflow while reducing the documentation burden.}

{
Importantly, although documentation represents the final step in the SIC workflow (individual level), P10 and P11 emphasized that it also serves as the starting point and foundation for subsequent SICs with other healthcare providers (organizational level). %As noted in Section 4.1.2, many patients arriving in EDs have neither previously participated in nor documented SICs. Therefore, the first time for them to have an SIC is in EDs. 
%However, due to severe time constraints, ED providers can only elicit information, such as patients' treatment preferences, to guide their decision-making, instead of the overall rationale of goals of care.
This initial documentation in EDs, therefore, becomes essential but incomplete groundwork that requires future, more comprehensive SICs as patients move through the healthcare system. }

\begin{quote}
\textit{``Really having a clear sense as you go into the conversation about what your goal is, I think that can be challenging because ultimately I think in the arc of a patient's illness, we might have the goal of really getting a clear understanding of their functional status of their preferences, and then making a recommendation. But oftentimes, doing all of that, that conversation, if you have it all at once, can easily be a 45-minute to 90-minute conversation. And that is not available in the emergency department. And so you have to basically decide, I can't do all of that.''} (P11)
\end{quote}

To complete the broader spectrum of SICs, collaborative efforts across multiple healthcare providers are required to build up a holistic picture of patient values, goals, and care preferences. 
{However, the current EHR system has \textbf{limited support for the collaborative effort for conducting SICs across provider teams}.}
One example is that when P3 wanted to find other providers to discuss patient treatment plans, the EHR system displays the contact information for multiple clinical care team members (e.g., attending physicians, residents, nurse practitioners), which makes it unclear whom he should contact.
{As a result, patients may experience inconsistent, misaligned care, as each point of care and SIC conducted across different departments remains disconnected.
Therefore, a systematic protocol not only needs to support SIC documentation but also be able to support information sharing and communication about SIC between different clinical departments.}

\begin{quote}
\textit{``You'll open up a patient's chart, and there'll be so many healthcare providers, physicians, nurse practitioners, physician assistants, residents...You don't know who to call, and if you do, they're often not on calls.''} (P3)
\end{quote}

\subsection{Opportunities and Concerns of AI Integration in SIC Workflows in EDs (RQ2)}
\label{sec:ai}
{Building on the workflow and challenges described above, we asked providers how AI technologies might support or complicate their work regarding SIC. 
Participants expressed a strong interest in AI assistance that has the potential to address the pain points in each phase of the SIC workflow, while also raising concerns about how these tools might affect the humane nature of SICs. 
In this section, we organize AI opportunities and concerns by workflow phase, from \textcolor{cyan}{\textbf{Identification}} through \textbf{\color{teal}{Documentation}}.
}
{
\subsubsection{AI Support for \textcolor{cyan}{\textbf{Identification}}: Prioritizing Actionable Patients Information}
In the Section \ref{sec:identification}, we identified that participants find it hard to look for the patients' key information (e.g., previous treatment plans and SIC documentation) in the EHR with limited time.
To address this, the first opportunity involves AI-supported identification and preparation for SICs.} 
Some participants (P5, P6, P9, P10) suggested that they wish they could have an AI system to extract and summarize key information from extensive health records to save a large amount of time. 
{Importantly, this process of extraction and summarization should not just passively list all the prior SICs and health records, but actively prioritize the patient's information in a way that can help ED providers quickly consume the information that is useful at that point.
For example, P2 mentioned that knowing the major issue among multiple diseases, that is, "trying to kill the patients," is important to know how severe patients are. 
Another benefit is that ED providers can use extracted data to 'confirm with patients,' and verify whether the information is up to date, which could address a challenge with information accuracy in the EHR (Section \ref{sec:identification}).
Collectively, AI-supported extraction could provide ED providers with actionable information.}
As a result, they could make decisions about whether they need to conduct SICs based on patients' information in a short period of time.

\begin{quote}
\textit{``Anyone who's previously documented, like if there was a goal of care button, I could click and it would go through everything that's been documented on a patient's chart. Because maybe they had a conversation with a social worker a few weeks ago or their primary care doctor, and I just don't have the time to click through every note.''} (P6)
\end{quote}

{
\subsubsection{AI Support for Rapid SIC \textbf{\color{magenta}{Preparation}}: Synthesizing a Coherent Patient Narrative}
ED providers often enter SICs underprepared because they lack the time to piece together a coherent "patient story" from fragmented medical records. 
This fragmentation prevents them from identifying the most appropriate entry point for the discussion. To better prepare SICs, a few participants (P6, P10, P11) expressed a strong desire for a tool that automatically synthesizes relevant records, such as prior SICs, recent ICU admissions, and key care team contacts, into a rapid summary. 
For instance, P1 requested a summary distinguishing "what’s done" from "what’s missing," enabling providers to bypass covered ground and focus immediately on unresolved issues. 
Beyond clinical efficiency, P11 emphasized that technology should surface patient-specific details to demonstrate that the provider is familiar with the patient’s history, thereby establishing trust. Together, these insights highlight the need for a centralized, skimmable preparation space that supports both efficient review and personalized SIC planning.
}
{
\begin{quote}
\textit{``So if there was a way for AI to sort of screen and say: 'hey, this really important thing, came back, you should probably take a look'...it lets me provide better care, but it also builds trust because they say: 'hey, this doctor is on top of it.' Even though he is taking care of 30 other people.''} (P11)
\end{quote}
}

{
\subsubsection{AI Support for \textbf{\color{brown}{Conduction}}: Real-Time, Contextual Guidance Without Displacing Empathy}
\label{sec:ai_conduction}
Next opportunity for AI support lies in challenges in smoothly keeping the conversation forward during SIC conduction (identified in Section \ref{sec:conduction}).}
To deal with this challenge, some participants (P6, P7, P8) proposed  AI systems that can passively listen to conversations (with patient and family consent), dynamically tailored to the ongoing dialogue, and generate real-time, contextual question suggestions that can be used or ignored by the provider based on their needs. 
{Importantly, P2 noted that the suggested questions need to be personalized to the patient's belief system. For instance, offering a "more spiritual perspective" if the patients are spiritual. }
With the support, ED providers may be able to maintain meaningful SICs while alleviating their psychological stress.
{As P8 explained, AI-generated suggestions could help ED providers feel less uncomfortable and keep the SICs moving forward.
While the promise of AI-generated questions and responses is promising, something to keep in mind is that the support must fit a rapid timeframe. Conducting SICs in EDs is not one-hour, thoroughly planned SICs, but conversations that happen "within minutes." The goal, therefore, is not comprehensive information gathering. Instead, the goal is to help ED providers be able to make a specific clinical decision at that point. 
To support this objective in EDs, the role of AI may not be as an autonomous agent, but as a support tool for moment-to-moment conversational repair to maintain a smooth, empathetic, and less-stressed SICs.}

\begin{quote}
\textit{``I think honestly, that would probably be so helpful because again, I think that these are hard conversations and put us in uncomfortable territory. And that would be a good, like, I don't, I'm, I'm trying to think of the right word of, like guide to like keep the conversation flowing, to keep things moving.''} (P8)
\end{quote}

While participants expressed enthusiasm for AI-supported opportunities in conduction, they articulated important concerns.
First, some participants (P3, P5, P7, P8, P11) were concerned that the AI system might diminish the deeply human aspect of SICs. 
The interaction between patients and providers during SICs represents more than structured information exchanges; they are highly personal moments characterized by emotional depth. 
{Moreover, AI might not be able to have an empathetic response when dealing with patients' emotions (P2).}
Consequently, introducing the AI system could potentially reduce the genuine emotional engagement and connection that patients value.
As P5 expressed that, patients could perceive AI-supported conversations as mechanical rather than sincere, thereby further hindering the process of SICs.
Moreover, some participants (P3, P6, P7, P8, P11) worried that AI integration might create distracting and disrespectful interactions when they attempt to build connections with patients. 
They emphasized that frequent or noticeable interaction with computer screens could disrespect the gravity and intimacy of these sensitive SICs.
To preserve human connection during SICs, P7 suggested that AI systems should remain ambient.
So that providers can use it only when they need it without disrupting their natural workflow.

\begin{quote}
\textit{``if there's a perception that the physician is just using an electronic tool to ask [the] right questions, that's not really sort of coming from the heart, that could be negatively perceived.''} (P5)
\end{quote}

\subsubsection{AI Support for \textbf{\color{teal}{Documentation}}: Rationale-Driven Notes and Reduced Charting Burden}

In addition, many participants (P3, P5, P6, P7, P9, P10) identified a significant opportunity in AI-generated notes to address the substantial documentation burden. 
{However, this is not a simple automated extraction or summarization task. Some participants (P3, P4, P7, P9) noted that the decision of what to condense and what to keep in detail needs to align with what ED providers perceive as clinically important. In terms of what is important in the documentation, P2 emphasized that knowing only the result of an SIC, such as the patient's final treatment choice, is insufficient. It is far more important to capture the rationale behind that result, such as a patient's "core values" that shape their preference. Understanding this "why" may be beneficial for other provider teams to better understand the patient and make future clinical decisions that genuinely reflect the patient's values.}
While the need for rationale-driven notes was clear, participants had differing preferences on how this information should be presented. For example, P5 suggested the AI could filter unrelated conversation details and highlight only patient goals and care preferences. In contrast, P11 preferred viewing the full conversation text with key information highlighted, while P3 wanted structured templates or word prompts to guide their own note creation. 
{The different insights highlight a key design challenge in balancing AI-generated notes with provider-specific needs for detail.}
Regardless, the potential impact of the AI-generated notes for documentation can be significant. P10 estimated that an AI system automating 80\% of this work could "cut post-shift charting time in half," which frees up substantial time for direct patient care.

\begin{quote}
\textit{``But if AI could offload even say 80 percent of that (documentation), then the other 20 percent you're tweaking and verifying the transcript or the output, that would save close to half of my working hours that I could actually be spending with seeing more patients.''} (P10)
\end{quote}

%Beyond this patient-facing identification, P1 pointed out that the challenge is not just identifying the right patient, but also the right healthcare provider at the right moment. Given the provider coordination challenges under time pressure in EDs, he suggested that AI could identify available healthcare providers suitable for the SIC and send a notification. This would allow the notified ED providers, if they have time, to "check out" the patient and confirm the necessity of the conversation. Following this confirmation, ED providers could then decide whether they want themselves to do that or whether they think there are more suitable health care providers to do this, which helps manage the in-the-moment workflow. This insight indicates that SIC identification is not just a patient-facing problem; it exposes a hidden layer of provider-side triage and delegation that current systems might ignore.

%\begin{quote}
%\textit{``Once you (AI) find the right patient and make sure I'm in the right time of my day to do something. You can ask me if I agree with the premise that something that this should happen.
%You can ask me if I would do it or you (other ED providers)... the purpose of that is to say then the flag can be passed to someone else in the pipeline who's not me. But with my support for... If I say I will do it, and then you (AI) can show me briefly what I might do, and by saying patient has talked to so and so on this date, and said this it would be useful to know this next thing.''} (P1)
% \end{quote}

However, P11 and P6 expressed a concern that AI might overemphasize irrelevant content while failing to adequately highlight the information most relevant to clinical care teams. 
Specifically, P4 provided an example of how AI might mistakenly capture minor patient comments, such as a casual mention of "chest pain", as clinically significant events, which potentially leads to unnecessary medical interventions and resource usage. 
Avoiding irrelevant content also underscores the importance of highlighting critical information.
P11 mentioned an important information that may not be directly related to key elements in SICs (i.e., values, goals, or care preferences), yet it still plays a critical role in decision-making and care continuity. 
\begin{comment}
One particularly important information is the patient’s baseline functional status, including whether the patient lives with severe functional limitations but nonetheless perceives their quality of life as good. 
Capturing such information is crucial: if a patient considers life with significant limitations still meaningful, this shapes how providers interpret what outcomes the patient may accept or even desire. 
Furthermore, such information supports other providers to understand the definition of patients' preferences and deliver goal-aligned care when patients survive an ED encounter and transition to another department.

\begin{quote}
\textit{``...but despite those limitations of function, this person still also believes that they have an excellent quality of life. Because then, you know, then you have to say, OK, well, they that means that this individual accepts these limitations and is if we can return them to something near dysfunctional status, that might actually be something that the patient would find acceptable and even desired.''} (P11)
\end{quote}
\end{comment}

{
Moreover, as detailed in Section \ref{sec:conduction}, ED providers report significant difficulty initiating SICs, building the connection, breaking bad news, and extracting critical information within minutes. 
A primary contributor to this struggle is the limited SIC training available in EDs. 
Without this training, ED providers often lack the benchmarks to determine whether they handled a delicate interaction effectively.
To address this uncertainty, a few participants (P2, P5) expressed a need for immediate AI-generated feedback following an SIC. 
For example, P3 specifically wished to know what information was missed and how they might have phrased key points differently. 
}
{
Furthermore, feedback could interrupt the "negative loop" where a stressful SIC experience leads to a dread of future SICs. 
\textbf{By objectively highlighting strengths and necessary corrections, the AI-generated feedback could help rebuild confidence.}
This approach extends existing HCI work on clinical communication support by closing the learning loop for providers' professional learning afterward. 
Moreover, the participants' need reveals a gap in feedback loop in high-stakes clinical settings, suggesting a new design space for real-time, post-hoc feedback systems.
}
{
\begin{quote}
\textit{``...I just finished having this conversation (the SIC) and then I leave the room... the phone to text, just like ping me and say, 'hey doctor, you did this really well, and this was really great. One thing you might consider is when the daughter said this, you might consider saying this next time.' That would really help me get better every time.''} (P2)
\end{quote}
}

%\subsubsection{SIC Progress Tracking Across Clinical Departments For Future SICs}
%{
%A final opportunity area identified in this study is AI-assisted progress tracking, which reframes the SIC as a longitudinal process rather than a singular event. As P5 noted that a complete SIC may consist of several components, such as treatment preferences, values, preferred lifestyle, and minimum acceptable lifestyle. However, an ED provider might only accomplish one or two under the time pressure (a challenge we mentioned in Section \ref{sec:documentation}). To address this challenge, P7 proposed an AI system that could track the patient's SIC progress along this route. The system would then notify the next provider team, potentially in a different department, of how far the patient progressed and prompt them to address the remaining parts. The purpose is to prepare and inform future SICs, creating a connected SIC ecosystem across departments so that care remains aligned with the patient’s core values throughout the entire treatment trajectory.
%}
%{
%\begin{quote}
%\textit{``I can only maybe accomplish one or two of them (components in SICs). The AI recognizes it and is able to inform the next provider: 'Hey, [the patient] was only able to get through one and two, you know, consider focusing on three, four and five.'''} (P7)
%\end{quote}}

{
\begin{figure*}[htbp]
  \includegraphics[draft=false,width=\textwidth]{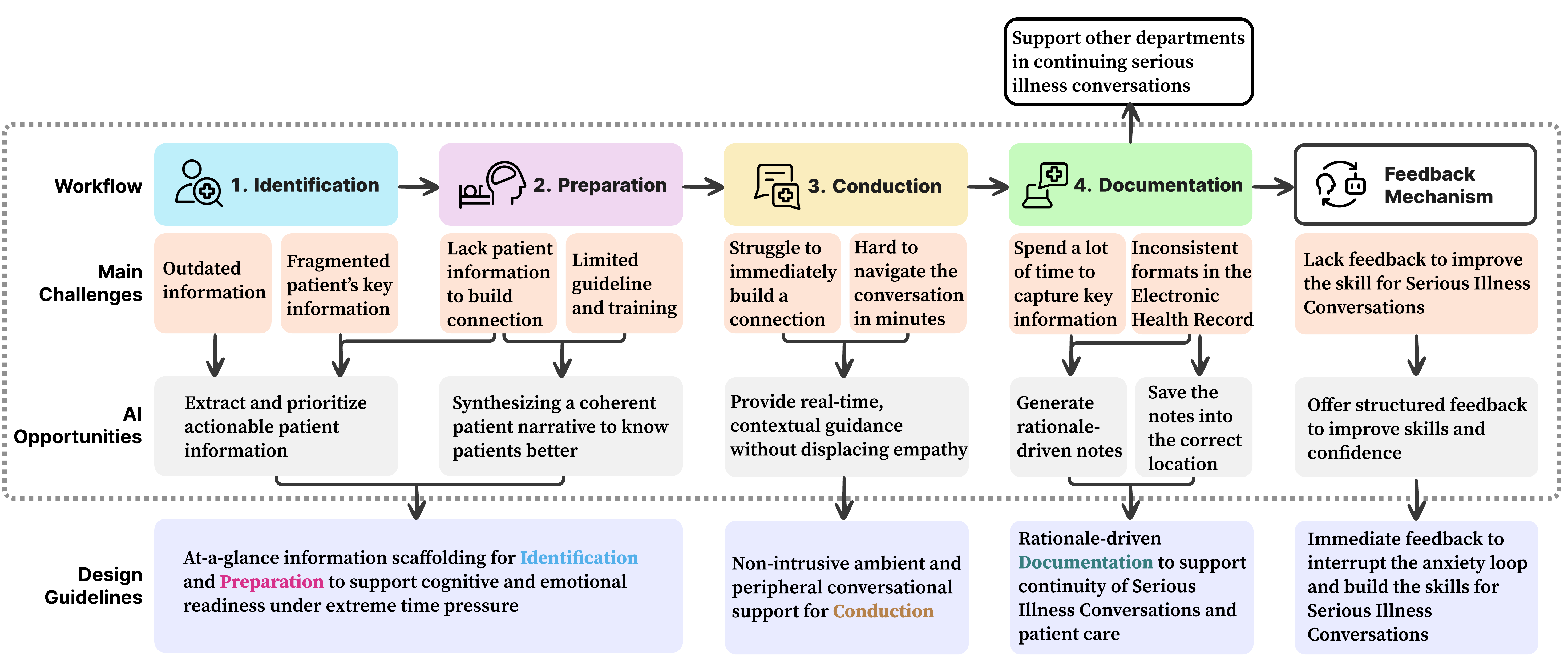}
  \caption{This figure summarizes the key challenges ED providers face at each phase of their workflow of SICs. The rationale for identifying the key challenges was based on their impact on SICs rather than their frequency of mention. Specifically, we prioritized challenges that (1) most significantly impede or lead to a breakdown of the SIC process in EDs, or (2) elicit the most significant negative emotional responses from participants. Moreover, the figure maps the AI opportunities identified to address these challenges across each phase. Besides, the figure presents four design guidelines for SICs grounded in the unique challenges and needs of EDs.}
  \label{fig:key_findings}
   \Description{Fig.4: The figure illustrates a four-stage workflow for conducting SICs. The Identification stage involves reviewing the EHR and conducting triage to gather patients’ key information, followed by determining whether an SIC is necessary. The Preparation stage requires providers to mentally prepare for the conversation and to select the appropriate time, setting, and participants. The Conduction stage focuses on clarifying the patient’s understanding of their health conditions, exploring their goals, values, and tradeoffs, and jointly developing a goal-aligned care plan. Finally, the Documentation stage involves recording key information from the SIC in the EHR system and coordinating follow-up with other care teams.
}
\end{figure*}
}

%% file: sections/8discussion.tex
\section{Discussion}

{In this section, we begin by unpacking the Efficiency–Empathy Paradox—a central tension surfaced in our findings that serves as a design lens for systems supporting SICs in the ED.}  {Building on this, we present four design guidelines that articulate what an AI system—conceived as a coordinated, multi-phase solution spanning \textbf{\textcolor{cyan}{Identification}}, \textbf{\textcolor{magenta}{Preparation}}, \textbf{\textcolor{brown}{Conduction}}, and \textbf{\textcolor{teal}{Documentation}}—must enable in practice. Rather than enumerating isolated features, these guidelines outline how an integrated AI system can fit seamlessly into ED providers' time-pressured workflows while safeguarding the human connection at the heart of SICs.}

%%we derive design guidelines for AI-supported SICs. 
%%It is important to note that our participants’ current SIC workflows lack AI support. The following discussion focuses on how our proposed AI systems can be designed to bridge these gaps and integrate seamlessly into clinical practice.}

\subsection{Navigating the Efficiency-Empathy Paradox in AI-Supported SICs in EDs}

Building on HCI research on human–AI collaboration, we show that healthcare providers want the AI system to serve in an assistive role during the four stages of conducting SICs, with humans remaining in control. This preference aligns with prior work advocating for AI as an assistant rather than a decision maker~\cite{zhang2024rethinking}. Concerns that AI could erode the “human touch” in healthcare (e.g., worries about diminished empathy or reduced patient voice; cf. \citet{kong2024envisioning}) are well documented, but our findings extend this line of work by identifying a deeper tension. Providers fear that technology may distance them from patients, yet simultaneously request AI features that would reduce their cognitive load, preserve their presence, and allow them to engage more directly with patients. Their descriptions reveal a paradox: automation that appears to threaten empathy can also create the cognitive conditions in which empathy becomes possible. This echoes the automation paradox literature~\cite{Bainbridge_1983}, where automation reshapes rather than removes human work.
{In our context, we observe an “Empathy–Efficiency Paradox”: providers need efficiency support to sustain empathetic relational work, yet remain cautious of tools that appear unempathetic or overly controlling.}

Drawing from calm computing principles~\cite{Weiser1996THECA} and peripheral interaction research~\cite{Bakker_van_den_Hoven_Eggen_2015}, providers consistently expressed a desire for AI that operates at the “periphery” of attention, handling information processing while leaving relational work in the foreground.
{We advance this paradox as a design lens rather than a descriptive tension. Because cognitive labor and emotional labor are tightly intertwined in emergency SICs, efficiency interventions must be evaluated in terms of how they redistribute the emotional work ED providers must perform. This framing offers designers a structured way to reason about when AI should accelerate workflow, when it should remain unobtrusive, and how it can create space for human connection rather than inadvertently diminishing it.} {This reading of the paradox also connects to long-standing CSCW and HCI work that foregrounds emotional labor as an essential, though often invisible, component of professional practice. Studies of invisible work highlight how technologies reshape affective and relational responsibilities rather than simply automating tasks~\cite{Fox_Shorey_Kang_Montiel_Valle_Rodriguez_2023, Star_Strauss_1999, Mazmanian_Orlikowski_Yates_2013}. In health and care contexts in particular, emotional and logistical work are deeply entangled~\cite{Smriti_Wang_Huh-Yoo_2024, Verdezoto_Bagalkot_Akbar_Sharma_Mackintosh_Harrington_Griffiths_2021}. Our findings extend this literature by showing how, in emergency SICs, ED providers’ emotional labor is constrained by the cognitive burden of assembling fragmented information under time pressure. AI support that reduces cognitive load therefore indirectly sustains the emotional labor on which empathic communication depends.}

{Understanding the empathy–efficiency paradox as a matter of emotional labor clarifies why it recurs across all four SIC workflow stages.}
{During \textbf{{\textcolor{cyan}{Identification}}}, ED providers wanted help noticing subtle cues (e.g., medical history patterns or repeated ED visits) while retaining moral judgment about whether a SIC is appropriate. AI support thus must remain peripheral and contextual rather than directive.}
During \textbf{\textcolor{magenta}{\textbf{Preparation}}}, surfacing consented, humanizing patient context (family relationships, past care preferences, or small personal details) helps ED providers enter the conversation with rapport already forming. For instance, the system might remind a provider that “Mr. Johnson is a lifelong Red Sox fan who watches games with his grandson” or suggest opening with “I understand your daughter Maria has been helping coordinate your care; would you like her to join us for this conversation?”
During \textbf{\textbf{\color{brown}{Conduction}}}, ED providers envisioned the AI as a quiet “listener”, providing peripheral conversational support such as tracking questions or flagging emotional cues without interjecting or taking the conversational floor. When a patient says, “I just want to be comfortable,” the system might gently note: “Patient may be expressing preference for comfort care. Consider exploring what ‘comfort’ means to them.”
Finally, during {\textbf{\textcolor{teal}{Documentation}}, providers saw opportunities for AI to offload note-taking while preserving the ED provider’s authorship of the narrative, ensuring that patient values are recorded faithfully rather than flattened into generic templates.}
{Across these stages, the empathy–efficiency paradox acts as a guidepost: design decisions must respect how emotional labor is distributed and how AI’s presence alters the relational texture of SICs.}

The empathy dimension of this paradox also requires reframing how we think about AI’s role in healthcare communication. Prior work has explored how AI systems can be designed to display empathy, while others have highlighted the risks of these attempts. For instance, \citet{cuadra2024illusion} cautioned that empathetic conversational agents may generate an “illusion” of understanding without genuinely grasping a user’s situation. Related CHI work shows that empathetic chatbot responses can reduce perceived authenticity even when they appear emotionally attuned~\cite{seitz2024artificial}. In our setting, we see a different model: instead of asking AI to perform empathy, ED providers want AI to create the cognitive space in which human empathy can occur. When AI handles tasks such as information synthesis or documentation, providers gain time and mental clarity for the emotional labor that genuine empathy requires. AI-generated suggestions become materials for ED providers to interpret, adapt, or disregard based on their understanding of the patient.
{This shifts the central question from whether AI can act empathetically to how AI might sustain the conditions necessary for authentic human empathy under severe time pressure.}

Moreover, our findings have implications for how cognitive empathy can be theorized in high-stakes, time-constrained healthcare settings. As Alam and Mueller argue, cognitive empathy requires AI systems to use patient reasoning, perspectives, and information to create shared understanding~\cite{alam2023cognitive}. In EDs, this requirement becomes especially urgent because ED providers have only minutes, not hours, to build the mental models that SICs demand. The AI capabilities requested by providers map directly onto these cognitive empathy requirements. Patient history synthesis supports understanding of patients’ reasoning by revealing patterns in past medical decisions and treatment preferences. Conversation tracking supports perspective-taking by identifying recurring concerns or emotional cues that providers might miss during multitasking. Real-time prompting facilitates information integration by signaling when to explore unstated values or family dynamics that shape patient preferences. Together, these capabilities help ED providers construct the cognitive models that make empathetic engagement possible. This reconceptualization complements ongoing HCI work on empathetic interfaces~\cite{chen2025patient} while responding to long-standing concerns in medicine that technology may dehumanize care.

\subsection{Design Guidelines}

Based on our study, we identified major challenges and unmet needs across the full workflow of SICs in EDs. 
These findings reveal opportunities where AI could offer support as well as critical constraints that design efforts must carefully consider. {They build on ~\citet{jacobs2021designing}'s emphasis on embedding AI tools into high-pressure ED environments and are grounded in prior HCI work on human–AI collaboration, automation, and emotional labor.} 
Together, the guidelines aim to help the design of AI systems that reduce cognitive and logistical burden while protecting the relational and empathic dimensions that providers view as essential to effective SICs. 

{
\textbf{DG1: Provide At-a-Glance Information Scaffolding for SIC \textcolor{cyan}{\textbf{Identification}} and \textbf{\color{magenta}{Preparation}} to Support Cognitive and Emotional Readiness Under Extreme Time Pressure.}} As identified in Section \ref{sec:identification} and \ref{sec:preparation}, ED providers struggle to efficiently access fragmented patient information scattered across EHR systems. This fragmentation creates significant barriers during the time-critical identification and preparation phases, where providers must rapidly determine whether a SIC is necessary and simultaneously gather enough context to enter a highly sensitive conversation with confidence. To address this challenge, AI systems should automatically extract, synthesize, and summarize key patient information identified in our study—{including code status, primary serious illness, previously expressed goals and values, and even mortality prediction}—in easily consumable formats. This information should be organized by clinical relevance rather than strict chronology, with direct links back to source documentation for verification.

{
Critically, such synthesis must do more than surface what is known. It must also identify what is missing—for example, whether a patient has never had an SIC—so that providers do not waste scarce minutes searching for documentation that does not exist. Highlighting gaps is especially important in ED SICs, where many patients arrive without any prior engagement in goals-of-care discussions. The core purpose of this guideline is to support both cognitive triage and emotional readiness in the first few minutes of an ED encounter. Unlike in inpatient or outpatient settings, ED providers often meet patients for the first time under crisis conditions, must decide whether an SIC is warranted almost immediately, and lack the emotional runway that typically precedes these conversations. An AI system should therefore provide “at-a-glance” scaffolding that helps reconstruct a usable patient narrative under extreme time pressure, reducing the sense of panic and isolation ED providers described when preparing to enter SICs.
}

{
In addition to information organization, the system should generate context-specific entry points into SICs—brief cues or scripts grounded in the patient’s medical history and caregiving context—to help providers initiate these conversations smoothly. Such supports function as a digital form of emotional backup, particularly for ED providers who receive limited SIC-specific training.
}

Prior research has demonstrated the feasibility of such approaches, including AI-driven mortality prediction systems that help identify high-risk patients~\cite{avati2017improving} and interventions using predictions to prompt SICs~\cite{manz2020effect}. {However, prediction alone is insufficient in ED SICs. Drawing on critiques of prediction-centric design~\cite{zhang2024rethinking}, we argue that an AI system in this context must support the entire process leading into an SIC—organizing conflicting records, surfacing gaps, scaffolding emotional readiness, and preparing providers with meaningful conversational entry points—rather than functioning as an oracle.} {This aligns with and extends Human–AI collaboration frameworks~\cite{lee2021human} by emphasizing that in ED SICs, AI must serve as a partner in reconstructing actionable understanding, not merely a predictor of risk.}
\begin{comment}
{
Taken together, DG1 articulates a design principle unique to the ED SIC context: that under conditions of extreme time pressure, AI must simultaneously support cognitive sensemaking and emotional readiness. The system’s value lies not in comprehensiveness but in rapid relevance to show what matters now, what is missing, and how to begin.}
\end{comment}

{
\textbf{DG2: Provide Non-Intrusive Ambient and Peripheral Conversational Support During SIC \textcolor{brown}{Conduction} to Support Relational and Emotional Stability}.} 
Based on Section \ref{sec:conduction}, ED providers, particularly those with limited experience, struggle to navigate these sensitive conversations smoothly with patients never met before, {while conducting SICs within minutes of a life-altering crisis.} 
However, as noted in Section \ref{sec:ai_conduction}, ED providers strongly emphasized that technology must not diminish the essential humanity of these deeply personal conversations. 
To address this tension, AI systems should offer conversational support that explicitly preserves intimate dialogue by operating through peripheral, non-intrusive interfaces. Ambient listening capabilities (with explicit patient consent) should passively monitor conversations to provide context-aware support without creating technological barriers between provider and patient. This support should appear on peripheral displays—secondary screens or subtle visual cues positioned outside the direct line of sight between provider and patient—allowing providers to maintain eye contact and emotional connection. For instance, when a patient expresses fear about pain, a peripheral display might quietly show: "Patient mentioned pain concerns 3 times—consider exploring what 'comfortable' means to them." Providers can glance at these suggestions when natural pauses occur, choosing whether to incorporate them based on their assessment of the conversation's emotional dynamics. Such ambient recording approaches are already being adopted in clinical practice, with EHR systems providing Microsoft's Dragon Ambient experience Copilot\footnote{https://www.microsoft.com/en-us/health-solutions/clinical-workflow/dragon-copilot}, enabling providers to record patient visits with consent directly through Epic's mobile application for direct EHR integration. {However, it is important to note that the feasibility of adding new AI systems in the clinical settings depends on organizational, legal, and data-governance conditions that shape what is feasible. Hospitals face strict regulations regarding consent, data retention, and access rights. Therefore, successful design must respect these boundaries and align with existing governance structures rather than introducing new liabilities that the ED is ill-equipped to handle.}

{Importantly, SIC support technologies in the ED cannot operate as generic conversational checklists. In this context, their value lies in functioning as “connection accelerators” and “emotional stabilizers”—tools that help providers quickly establish rapport and recover conversational flow without intruding on the human core of the interaction. For example, the system may surface lightweight, personalized cues (e.g., noting that a patient is a “Red Sox fan”) to reduce emotional distance in the opening moments of an SIC. In more challenging situations, ambient listening could flag conversational impasses—such as a sudden emotional outburst—and offer gentle, text-based phrasing options that help providers de-escalate or reorient the dialogue while maintaining empathy.}

%%{
%%It is crucial to note that there are existing solutions using ambient technologies for transcription (e.g., ~\cite{microsoft_2025_dragoncopilot}. However, the feasibility of adding new AI systems in the clinical settings depends on organizational, legal, and data-governance conditions that shape what is feasible. Hospitals face strict regulations regarding consent, data retention, and access rights. Therefore, successful design must respect these boundaries and align with existing governance structures rather than introducing new liabilities that the ED is ill-equipped to handle.}

{Prior HCI and CSCW work on Emotion AI has shown that emotion-detection AI systems often intensify emotional labor~\cite{boyd2023automated}, act as surveillance mechanisms~\cite{corvite2023data}, and leave users feeling judged or managed~\cite{roemmich2023emotion}. Our findings extend these critiques into the uniquely fragile setting of SICs, where even subtle forms of evaluative emotional inference could undermine trust at critical moments of decision-making. For this reason, we argue that any emotional support provided by AI in SICs must avoid scoring, classifying, or interpreting patient emotions. Instead, it should offer non-evaluative, context-sensitive scaffolds that protect provider psychological safety and patient privacy while enabling the conversation to continue with compassion and stability.}

{
\textbf{DG3: Provide Rationale-Driven and Provider-Curated AI \textbf{\color{teal}{Documentation}} to Support Longitudinal SIC Continuity and SIC Progress Tracking to Support Care Continuity}.} As mentioned in Section \ref{sec:documentation}, ED providers face substantial documentation burdens that force many to work unpaid hours after shifts, consuming significant time that could be redirected to direct patient care.
The challenge is compounded by the need to distill lengthy conversations into clinically relevant summaries that capture essential patient goals, values, and care preferences while filtering out less relevant content.
To address this challenge, AI systems should automatically generate structured SIC notes that highlight key information (e.g., patients' goals, values, and care preferences), provide suggested documentation prompts, and offer editable drafts that healthcare providers can review and modify. Recent research in LLMs \cite{Agaronnik_Davis_Manz_Tulsky_Lindvall_2025} and Natural Language Processing \cite{lindvall2022natural} have demonstrated their suitability for this task.
Implementation must prioritize clinical autonomy by ensuring transparency in AI-generated content, maintaining clear links to source conversation segments for verification, and providing intuitive editing interfaces that allow healthcare providers to correct, expand, or restructure summaries based on their clinical judgment and documentation requirements.

{While the general burden of clinical documentation is well-documented~\cite{zhang2022characteristics, chen2010documenting, saarinen2005does, davidson2004s}, our study identifies the specific information that must be emphasized during ED SIC documentation. Providers stressed that documenting SICs requires capturing the rationale behind decisions—such as the value that leads a patient to choose comfort-oriented care—rather than simply recording the final preference itself. At the same time, our findings reveal an important misalignment between what AI might naturally summarize from a conversation and what providers consider clinically meaningful. Patients often repeat emotionally charged statements that matter for rapport building but do not belong in clinical notes, while subtle value-laden remarks may be crucial for downstream decision-making. Because of this asymmetry, AI-generated SIC notes must function as drafts that providers curate, refine, and approve. Transparent links to the original conversation allow providers to elevate meaningful content, suppress irrelevant repetition, and ensure that the documentation reflects clinical significance.}

{Beyond producing notes, we identified a critical opportunity for AI-assisted SIC progress tracking, which reframes the SIC not as a standalone conversation but as a longitudinal trajectory. A complete SIC consists of multiple components, yet an ED provider under severe time pressure may complete only one or two (Section \ref{sec:documentation}). To address this, an AI system should track the patient’s progress across SIC components, create a dynamic record of what has been discussed, and highlight what remains. This externalizes the invisible work~\cite{Star_Strauss_1999} of remembering and coordinating SIC progress across teams. When the patient transitions to a new provider team, the AI system should clearly convey which SIC elements are outstanding so that future conversations do not restart from zero, thereby creating a connected SIC ecosystem across departments.}

{This design guideline extends prior HCI work on digital scribes~\cite{wendt2025deploying, olson2025use, li2021automating, han2024ascleai}. While these systems primarily focus on transcription accuracy and efficiency in outpatient settings, SIC documentation in the ED requires more than transcription or summarization. For SICs, AI systems must curate meaning, highlight rationales, reflect ED providers’ priorities, and support continuity of care. In doing so, documentation shifts from a static record to an active coordination tool for multidisciplinary teams.}

{
\textbf{DG4: Provide Immediate Post-SIC Feedback to Interrupt the Anxiety Loop and Build SIC Skills Through Reflection-on-Action}. }
ED providers mentioned in Section \ref{sec:preparation} that due to the lack of training and experience with SICs, they have difficulty with mental and emotional preparation, with many colleagues feeling uncomfortable initiating SICs. 
{
Such negative experience of SICs creates a "negative loop" to hinder future SICs, and causes a SIC skill-development gap that leads rapidly from one crisis to another without closure.
}
In response, ED providers expressed interest in receiving AI-generated feedback about their performance in SICs to improve their communication skills over time. 
To address their need, AI systems should analyze recorded SICs to provide personalized feedback and suggestions that help healthcare providers refine their communication approaches. 
Such systems could identify specific areas for improvement in conversation flow, question phrasing, and patient engagement techniques based on previous conversation recordings. 
{
Importantly, such feedback should not function as heavy educational platforms, but as immediate "Reflection-on-Action" engines.
Specifically, the system should provide concise feedback immediately after the end of SICs.
First, it provides objectively confirmed strengths to combat imposter syndrome and rebuild confidence.
Second, it should offer specific, actionable alternative phrasing for missed opportunities. 
Such feedback aligns with Schön’s theory of reflection-on-action~\cite{schon2017reflective}, transforming the SIC documentation moment into a micro-learning opportunity.
}

The AI-generated feedback approach builds on previous research highlighting AI's potential to analyze conversation content and provide feedback to healthcare providers~\cite{chua2022enhancing}, as well as demonstrated capabilities for evaluating communication skills in clinical settings~\cite{ryan2019using}. 
{
While prior research has explored feedback mechanisms for team communication~\cite{tausczik2013improving} and reflective tools for counselors~\cite{chen2023facilitating}, these interventions typically target non-urgent settings and operate on the assumption that users have dedicated time to reflect on their performance. In contrast, our design guidelines address the unique constraints of a highly time-pressured environment, such as EDs, where providers lack the capacity for post-shift study. 
By embedding feedback directly at the end of the workflow, our objective extends beyond mere skill enhancement; we aim to interrupt the psychological 'negative loop' and alleviate performance anxiety. 
As a result, ED providers could initiate future SICs with lower cognitive load and greater comfort and confidence.
}

%This approach builds on previous research proposing AI-human collaborative workflows for SICs~\cite{chua2022enhancing}, which highlighted AI's potential to analyze conversation content and provide feedback to healthcare providers, leveraging demonstrated capabilities for evaluating communication skills in clinical settings~\cite{ryan2019using}. Implementation should provide constructive feedback that builds healthcare provider confidence and comfort levels, ultimately improving their preparation and effectiveness in conducting future SICs in EDs.

%\textbf{DG5: SIC Progress Tracking Across Clinical Departments For Future SICs.}
%A final opportunity area identified in this study is AI-assisted progress tracking, which reframes the SIC as a longitudinal process rather than a singular event. As P5 noted that a complete SIC may consist of several components, such as treatment preferences, values, preferred lifestyle, and minimum acceptable lifestyle. However, an ED provider might only accomplish one or two under the time pressure (a challenge we mentioned in Section \ref{sec:documentation}). To address this challenge, an AI system that could track the patient's SIC progress along this route. The system would then notify the next provider team, potentially in a different department, of how far the patient progressed and prompt them to address the remaining parts. The purpose is to prepare and inform future SICs, creating a connected SIC ecosystem across departments so that care remains aligned with the patient’s core values throughout the entire treatment trajectory.

\subsection{Limitations and Future Work}

Our study has its limitations. The findings draw on experiences from eleven ED providers and nurses. {Recruiting participants for research in EDs required sustained effort, since participation depended on availability during unpredictable and high-pressure shifts. The findings, therefore, reflect the perspectives of the physicians and nurses who were able to participate within these constraints. Our aim was to capture the lived realities of conducting SICs in EDs and to describe how providers navigate these conversations under severe time pressure and shifting acuity.}

Second, our study reflects the specific constraints and workflows of the U.S. healthcare system. While the core challenges we identified, such as difficulty in navigating SICs in minutes, likely persist across different geographies, the specific manifestations of these issues may vary in other national contexts. We recommend that future research investigate the unique barriers in different national contexts and propose solutions tailored to these environments.
Moreover, our study focused on clinical staff, and perspectives from patients and family members were not represented. {Although the perspectives from patients and their family members are out of the scope of this study, future work that includes these stakeholders may reveal complementary challenges related to communication, family dynamics, cultural expectations, and long-term decision-making that ED providers alone may not identify.}

Furthermore, participants varied in their familiarity with AI and in the extent to which AI-enabled tools have been discussed or explored in their institutions. {Understanding these differences is important because ED providers' prior exposure may shape how they interpret the role, limits, and risks of AI. A more systematic characterization of existing digital infrastructures and AI-related initiatives across study sites would help contextualize how ED providers imagine AI-based support.}
Finally, our interviews may not fully capture tacit coordination, situated workarounds, or moment-to-moment pressures that arise during SICs. {In-situ observations could provide a more complete picture of how SICs unfold throughout a shift and how workflow pressures constrain decision-making.}

{Despite these limitations, the study provides an important and currently missing foundation for understanding SICs in emergency departments. Very little empirical work has documented how SICs are structured in the ED, how ED providers adapt them under severe time pressure, and where the most consequential breakdowns occur. 
Our identified end-to-end workflow of SICs in EDs and its challenges supply the baseline that is needed before effective AI support can be designed. Without this descriptive understanding, it would be difficult to define meaningful design requirements or evaluate the appropriateness of AI intervention.}
It is important to notice that while this work centers on technological solutions, such as AI, we acknowledge that there are some non-technical or organizational alternatives. 
For example, the challenges we identified, such as nighttime staffing shortages and fragmented points of contact, may be addressed in some level through increased headcount, optimized rostering, or updated directory systems. 
Future research should examine how these non-technical strategies might compare with or complement technological tools within clinical workflow.

{Building on this foundation, future work will focus on translating these opportunities into concrete system design. We plan to engage ED providers through participatory design to refine requirements and determine appropriate timing, modality, and granularity of support. We will then develop a functional prototype that incorporates high-priority features and evaluate it through usability testing and follow-up interviews. Long-term work will examine how AI-supported SICs operate across different types of EDs and how organizational structures influence the challenges and shape the efficiency–empathy balance identified in our study.}

%% file: sections/10conclusion.tex
\section{Conclusion}

This study examined how ED providers conduct SICs within minutes and identified both opportunities and concerns for AI support grounded in their workflows. Through interviews with eleven ED providers, we mapped a four-stage SIC workflow—\textcolor{cyan}{\textbf{Identification}}, \textbf{\color{magenta}{Preparation}}, \textbf{\color{brown}{Conduction}}, and \textbf{\color{teal}{Documentation}}—and surfaced the barriers in each stage. Fragmented EHR data and the absence of a structured SIC protocol make it difficult to access prior values-based information, disrupting continuity as patients move across departments. ED providers also conduct SICs in fast-paced, low-privacy environments while carrying significant emotional labor with little preparation time or support.

Despite these constraints, participants expressed a strong interest in AI that can synthesize patient information, offer light-touch conversational scaffolding, and streamline documentation. Meanwhile, they emphasized that such support must remain interpretable, protect emotional authenticity, and preserve the human connection central to SICs. In this context, AI must help ED providers balance the relentless demand for efficiency with the need to maintain empathy during high-stakes, time-pressured SICs. We argue for ambient and peripheral AI systems that can be engaged selectively, support continuity across fragmented workflows, and reinforce rather than replace providers’ clinical judgment and relational presence.

%% file: MAIN-sigconf-authordraft.bbl

\begin{thebibliography}{109}

%%% the Web address.

\ifx \showCODEN    \undefined \def \showCODEN     #1{\unskip}     \fi
\ifx \showISBNx    \undefined \def \showISBNx     #1{\unskip}     \fi
\ifx \showISBNxiii \undefined \def \showISBNxiii  #1{\unskip}     \fi
\ifx \showISSN     \undefined \def \showISSN      #1{\unskip}     \fi
\ifx \showLCCN     \undefined \def \showLCCN      #1{\unskip}     \fi
\ifx \shownote     \undefined \def \shownote      #1{#1}          \fi
\ifx \showarticletitle \undefined \def \showarticletitle #1{#1}   \fi
\ifx \showURL      \undefined \def \showURL       {\relax}        \fi
% invisible to TeX
\providecommand\bibfield[2]{#2}
\providecommand\bibinfo[2]{#2}
\providecommand\natexlab[1]{#1}
\providecommand\showeprint[2][]{arXiv:#2}

\bibitem[Ter(2013)]%
        {TertiaryCareCentersMeSH}
 \bibinfo{year}{2013}\natexlab{}.
\newblock \bibinfo{title}{Tertiary Care Centers [MeSH Heading]}.
\newblock \bibinfo{howpublished}{\url{https://www.ncbi.nlm.nih.gov/mesh?Db=mesh&term=Tertiary+Care+Centers}}.
\newblock
\newblock
\shownote{Accessed: 2025-11-26}.


\bibitem[Agaronnik et~al\mbox{.}(2025)]%
        {Agaronnik_Davis_Manz_Tulsky_Lindvall_2025}
\bibfield{author}{\bibinfo{person}{Nicole~D Agaronnik}, \bibinfo{person}{Joshua Davis}, \bibinfo{person}{Christopher~R Manz}, \bibinfo{person}{James~A Tulsky}, {and} \bibinfo{person}{Charlotta Lindvall}.} \bibinfo{year}{2025}\natexlab{}.
\newblock \showarticletitle{Large language models to identify advance care planning in patients with advanced cancer}.
\newblock \bibinfo{journal}{\emph{Journal of Pain and Symptom Management}} \bibinfo{volume}{69}, \bibinfo{number}{3} (\bibinfo{year}{2025}), \bibinfo{pages}{243--250}.
\newblock
\href{https://doi.org/10.1016/j.jpainsymman.2024.11.016}{doi:\nolinkurl{10.1016/j.jpainsymman.2024.11.016}}


\bibitem[Alam and Mueller(2023)]%
        {alam2023cognitive}
\bibfield{author}{\bibinfo{person}{Lamia Alam} {and} \bibinfo{person}{Shane~T Mueller}.} \bibinfo{year}{2023}\natexlab{}.
\newblock \showarticletitle{Cognitive Empathy within Patient-AI Communication for Diagnostic Reasoning}. In \bibinfo{booktitle}{\emph{Proceedings of the Human Factors and Ergonomics Society Annual Meeting}}, Vol.~\bibinfo{volume}{67}. SAGE Publications Sage CA: Los Angeles, CA, \bibinfo{pages}{1055--1062}.
\newblock
\href{https://doi.org/10.1177/21695067231193682}{doi:\nolinkurl{10.1177/21695067231193682}}


\bibitem[Avati et~al\mbox{.}(2017)]%
        {avati2017improving}
\bibfield{author}{\bibinfo{person}{Anand Avati}, \bibinfo{person}{Kenneth Jung}, \bibinfo{person}{Stephanie Harman}, \bibinfo{person}{Lance Downing}, \bibinfo{person}{Andrew Ng}, {and} \bibinfo{person}{Nigam~H Shah}.} \bibinfo{year}{2017}\natexlab{}.
\newblock \showarticletitle{Improving palliative care with deep learning}. In \bibinfo{booktitle}{\emph{2017 IEEE international conference on bioinformatics and biomedicine (BIBM)}}. IEEE, \bibinfo{pages}{311--316}.
\newblock
\href{https://doi.org/10.1109/BIBM.2017.8217669}{doi:\nolinkurl{10.1109/BIBM.2017.8217669}}


\bibitem[Bainbridge(1983)]%
        {Bainbridge_1983}
\bibfield{author}{\bibinfo{person}{Lisanne Bainbridge}.} \bibinfo{year}{1983}\natexlab{}.
\newblock \showarticletitle{Ironies of automation}.
\newblock \bibinfo{journal}{\emph{Automatica}} \bibinfo{volume}{19}, \bibinfo{number}{6} (\bibinfo{year}{1983}), \bibinfo{pages}{775–779}.
\newblock
\showISSN{0005-1098}
\href{https://doi.org/10.1016/0005-1098(83)90046-8}{doi:\nolinkurl{10.1016/0005-1098(83)90046-8}}


\bibitem[Bakker et~al\mbox{.}(2015)]%
        {Bakker_van_den_Hoven_Eggen_2015}
\bibfield{author}{\bibinfo{person}{Saskia Bakker}, \bibinfo{person}{Elise van~den Hoven}, {and} \bibinfo{person}{Berry Eggen}.} \bibinfo{year}{2015}\natexlab{}.
\newblock \showarticletitle{Peripheral interaction: characteristics and considerations}.
\newblock \bibinfo{journal}{\emph{Personal and Ubiquitous Computing}} \bibinfo{volume}{19}, \bibinfo{number}{1} (\bibinfo{date}{Jan.} \bibinfo{year}{2015}), \bibinfo{pages}{239–254}.
\newblock
\showISSN{1617-4917}
\href{https://doi.org/10.1007/s00779-014-0775-2}{doi:\nolinkurl{10.1007/s00779-014-0775-2}}


\bibitem[Bartle et~al\mbox{.}(2022)]%
        {bartle2022second}
\bibfield{author}{\bibinfo{person}{Vince Bartle}, \bibinfo{person}{Janice Lyu}, \bibinfo{person}{Freesoul El~Shabazz-Thompson}, \bibinfo{person}{Yunmin Oh}, \bibinfo{person}{Angela~Anqi Chen}, \bibinfo{person}{Yu-Jan Chang}, \bibinfo{person}{Kenneth Holstein}, {and} \bibinfo{person}{Nicola Dell}.} \bibinfo{year}{2022}\natexlab{}.
\newblock \showarticletitle{“a second voice”: Investigating opportunities and challenges for interactive voice assistants to support home health aides}. In \bibinfo{booktitle}{\emph{Proceedings of the 2022 CHI Conference on Human Factors in Computing Systems}}. \bibinfo{pages}{1--17}.
\newblock
\href{https://doi.org/10.1145/3491102.3517683}{doi:\nolinkurl{10.1145/3491102.3517683}}


\bibitem[Bascom et~al\mbox{.}(2024)]%
        {bascom2024designing}
\bibfield{author}{\bibinfo{person}{Emily Bascom}, \bibinfo{person}{Reggie Casanova-Perez}, \bibinfo{person}{Kelly Tobar}, \bibinfo{person}{Manas~Satish Bedmutha}, \bibinfo{person}{Harshini Ramaswamy}, \bibinfo{person}{Wanda Pratt}, \bibinfo{person}{Janice Sabin}, \bibinfo{person}{Brian Wood}, \bibinfo{person}{Nadir Weibel}, {and} \bibinfo{person}{Andrea Hartzler}.} \bibinfo{year}{2024}\natexlab{}.
\newblock \showarticletitle{Designing communication feedback systems to reduce healthcare providers’ implicit biases in patient encounters}. In \bibinfo{booktitle}{\emph{Proceedings of the 2024 CHI Conference on Human Factors in Computing Systems}}. \bibinfo{pages}{1--12}.
\newblock
\href{https://doi.org/10.1145/3613904.3642756}{doi:\nolinkurl{10.1145/3613904.3642756}}


\bibitem[Bennett et~al\mbox{.}(2020)]%
        {bennett2020national}
\bibfield{author}{\bibinfo{person}{Christopher~L Bennett}, \bibinfo{person}{Ashley~F Sullivan}, \bibinfo{person}{Adit~A Ginde}, \bibinfo{person}{John Rogers}, \bibinfo{person}{Janice~A Espinola}, \bibinfo{person}{Carson~E Clay}, {and} \bibinfo{person}{Carlos~A Camargo~Jr}.} \bibinfo{year}{2020}\natexlab{}.
\newblock \showarticletitle{National study of the emergency physician workforce, 2020}.
\newblock \bibinfo{journal}{\emph{Annals of emergency medicine}} \bibinfo{volume}{76}, \bibinfo{number}{6} (\bibinfo{year}{2020}), \bibinfo{pages}{695--708}.
\newblock
\href{https://doi.org/10.1016/j.annemergmed.2020.06.039}{doi:\nolinkurl{10.1016/j.annemergmed.2020.06.039}}


\bibitem[Bernacki et~al\mbox{.}(2015)]%
        {bernacki2015development}
\bibfield{author}{\bibinfo{person}{Rachelle Bernacki}, \bibinfo{person}{Mathilde Hutchings}, \bibinfo{person}{Judith Vick}, \bibinfo{person}{Grant Smith}, \bibinfo{person}{Joanna Paladino}, \bibinfo{person}{Stuart Lipsitz}, \bibinfo{person}{Atul~A Gawande}, {and} \bibinfo{person}{Susan~D Block}.} \bibinfo{year}{2015}\natexlab{}.
\newblock \showarticletitle{Development of the Serious Illness Care Program: a randomised controlled trial of a palliative care communication intervention}.
\newblock \bibinfo{journal}{\emph{BMJ open}} \bibinfo{volume}{5}, \bibinfo{number}{10} (\bibinfo{year}{2015}), \bibinfo{pages}{e009032}.
\newblock
\href{https://doi.org/10.1136/bmjopen-2015-009032}{doi:\nolinkurl{10.1136/bmjopen-2015-009032}}


\bibitem[Berry et~al\mbox{.}(2019)]%
        {berry2019supporting}
\bibfield{author}{\bibinfo{person}{Andrew~BL Berry}, \bibinfo{person}{Catherine~Y Lim}, \bibinfo{person}{Tad Hirsch}, \bibinfo{person}{Andrea~L Hartzler}, \bibinfo{person}{Linda~M Kiel}, \bibinfo{person}{Zo{\"e}~A Bermet}, {and} \bibinfo{person}{James~D Ralston}.} \bibinfo{year}{2019}\natexlab{}.
\newblock \showarticletitle{Supporting communication about values between people with multiple chronic conditions and their providers}. In \bibinfo{booktitle}{\emph{proceedings of the 2019 CHI Conference on Human Factors in Computing Systems}}. \bibinfo{pages}{1--14}.
\newblock
\href{https://doi.org/10.1145/3290605.3300700}{doi:\nolinkurl{10.1145/3290605.3300700}}


\bibitem[Berry et~al\mbox{.}(2021)]%
        {berry2021supporting}
\bibfield{author}{\bibinfo{person}{Andrew~BL Berry}, \bibinfo{person}{Catherine~Y Lim}, \bibinfo{person}{Calvin~A Liang}, \bibinfo{person}{Andrea~L Hartzler}, \bibinfo{person}{Tad Hirsch}, \bibinfo{person}{Dawn~M Ferguson}, \bibinfo{person}{Zo{\"e}~A Bermet}, {and} \bibinfo{person}{James~D Ralston}.} \bibinfo{year}{2021}\natexlab{}.
\newblock \showarticletitle{Supporting collaborative reflection on personal values and health}.
\newblock \bibinfo{journal}{\emph{Proceedings of the ACM on human-computer interaction}} \bibinfo{volume}{5}, \bibinfo{number}{CSCW2} (\bibinfo{year}{2021}), \bibinfo{pages}{1--39}.
\newblock
\href{https://doi.org/10.1145/3476040}{doi:\nolinkurl{10.1145/3476040}}


\bibitem[Boyd and Andalibi(2023)]%
        {boyd2023automated}
\bibfield{author}{\bibinfo{person}{Karen~L Boyd} {and} \bibinfo{person}{Nazanin Andalibi}.} \bibinfo{year}{2023}\natexlab{}.
\newblock \showarticletitle{Automated emotion recognition in the workplace: How proposed technologies reveal potential futures of work}.
\newblock \bibinfo{journal}{\emph{Proceedings of the ACM on human-computer interaction}} \bibinfo{volume}{7}, \bibinfo{number}{CSCW1} (\bibinfo{year}{2023}), \bibinfo{pages}{1--37}.
\newblock
\href{https://doi.org/10.1145/3579528}{doi:\nolinkurl{10.1145/3579528}}


\bibitem[Butz et~al\mbox{.}(2007)]%
        {butz2007shared}
\bibfield{author}{\bibinfo{person}{Arlene~M Butz}, \bibinfo{person}{Jennifer~M Walker}, \bibinfo{person}{Margaret Pulsifer}, {and} \bibinfo{person}{Marilyn Winkelstein}.} \bibinfo{year}{2007}\natexlab{}.
\newblock \showarticletitle{Shared decision making in school age children with asthma}.
\newblock \bibinfo{journal}{\emph{Pediatric nursing}} \bibinfo{volume}{33}, \bibinfo{number}{2} (\bibinfo{year}{2007}), \bibinfo{pages}{111}.
\newblock


\bibitem[Campbell et~al\mbox{.}(2013)]%
        {campbell2013coding}
\bibfield{author}{\bibinfo{person}{John~L Campbell}, \bibinfo{person}{Charles Quincy}, \bibinfo{person}{Jordan Osserman}, {and} \bibinfo{person}{Ove~K Pedersen}.} \bibinfo{year}{2013}\natexlab{}.
\newblock \showarticletitle{Coding in-depth semistructured interviews: Problems of unitization and intercoder reliability and agreement}.
\newblock \bibinfo{journal}{\emph{Sociological methods \& research}} \bibinfo{volume}{42}, \bibinfo{number}{3} (\bibinfo{year}{2013}), \bibinfo{pages}{294--320}.
\newblock
\href{https://doi.org/10.1177/0049124113500475}{doi:\nolinkurl{10.1177/0049124113500475}}


\bibitem[Chen et~al\mbox{.}(2025a)]%
        {chen2025patient}
\bibfield{author}{\bibinfo{person}{David Chen}, \bibinfo{person}{Kabir Chauhan}, \bibinfo{person}{Rod Parsa}, \bibinfo{person}{Zhihui~Amy Liu}, \bibinfo{person}{Fei-Fei Liu}, \bibinfo{person}{Ernie Mak}, \bibinfo{person}{Lawson Eng}, \bibinfo{person}{Breffni~Louise Hannon}, \bibinfo{person}{Jennifer Croke}, \bibinfo{person}{Andrew Hope}, {et~al\mbox{.}}} \bibinfo{year}{2025}\natexlab{a}.
\newblock \showarticletitle{Patient perceptions of empathy in physician and artificial intelligence chatbot responses to patient questions about cancer}.
\newblock \bibinfo{journal}{\emph{npj Digital Medicine}} \bibinfo{volume}{8}, \bibinfo{number}{1} (\bibinfo{year}{2025}), \bibinfo{pages}{275}.
\newblock
\href{https://doi.org/10.1038/s41746-025-01671-6}{doi:\nolinkurl{10.1038/s41746-025-01671-6}}


\bibitem[Chen et~al\mbox{.}(2023)]%
        {chen2023facilitating}
\bibfield{author}{\bibinfo{person}{Tianying Chen}, \bibinfo{person}{Michael~Xieyang Liu}, \bibinfo{person}{Emily Ding}, \bibinfo{person}{Emma O'Neil}, \bibinfo{person}{Mansi Agarwal}, \bibinfo{person}{Robert~E Kraut}, {and} \bibinfo{person}{Laura Dabbish}.} \bibinfo{year}{2023}\natexlab{}.
\newblock \showarticletitle{Facilitating counselor reflective learning with a real-time annotation tool}. In \bibinfo{booktitle}{\emph{Proceedings of the 2023 CHI Conference on Human Factors in Computing Systems}}. \bibinfo{pages}{1--17}.
\newblock
\href{https://doi.org/10.1145/3544548.3581551}{doi:\nolinkurl{10.1145/3544548.3581551}}


\bibitem[Chen(2010)]%
        {chen2010documenting}
\bibfield{author}{\bibinfo{person}{Yunan Chen}.} \bibinfo{year}{2010}\natexlab{}.
\newblock \showarticletitle{Documenting transitional information in EMR}. In \bibinfo{booktitle}{\emph{Proceedings of the SIGCHI Conference on Human Factors in Computing Systems}}. \bibinfo{pages}{1787--1796}.
\newblock
\href{https://doi.org/10.1145/1753326.1753594}{doi:\nolinkurl{10.1145/1753326.1753594}}


\bibitem[Chen et~al\mbox{.}(2025b)]%
        {chen2025designing}
\bibfield{author}{\bibinfo{person}{Zhanming Chen}, \bibinfo{person}{Alisha Ghaju}, \bibinfo{person}{May Hang}, \bibinfo{person}{Juan~Fernando Maestre}, {and} \bibinfo{person}{Ji~Youn Shin}.} \bibinfo{year}{2025}\natexlab{b}.
\newblock \showarticletitle{Designing Health Technologies for Immigrant Communities: Exploring Healthcare Providers' Communication Strategies with Patients}. In \bibinfo{booktitle}{\emph{Proceedings of the 2025 CHI Conference on Human Factors in Computing Systems}}. \bibinfo{pages}{1--19}.
\newblock
\href{https://doi.org/10.1145/3706598.3713782}{doi:\nolinkurl{10.1145/3706598.3713782}}


\bibitem[Chisholm et~al\mbox{.}(2001)]%
        {chisholm2001work}
\bibfield{author}{\bibinfo{person}{Carey~D Chisholm}, \bibinfo{person}{Amanda~M Dornfeld}, \bibinfo{person}{David~R Nelson}, {and} \bibinfo{person}{William~H Cordell}.} \bibinfo{year}{2001}\natexlab{}.
\newblock \showarticletitle{Work interrupted: a comparison of workplace interruptions in emergency departments and primary care offices}.
\newblock \bibinfo{journal}{\emph{Annals of emergency medicine}} \bibinfo{volume}{38}, \bibinfo{number}{2} (\bibinfo{year}{2001}), \bibinfo{pages}{146--151}.
\newblock
\href{https://doi.org/10.1067/mem.2001.115440}{doi:\nolinkurl{10.1067/mem.2001.115440}}


\bibitem[Chishtie et~al\mbox{.}(2023)]%
        {chishtie2023use}
\bibfield{author}{\bibinfo{person}{Jawad Chishtie}, \bibinfo{person}{Natalie Sapiro}, \bibinfo{person}{Natalie Wiebe}, \bibinfo{person}{Leora Rabatach}, \bibinfo{person}{Diane Lorenzetti}, \bibinfo{person}{Alexander~A Leung}, \bibinfo{person}{Doreen Rabi}, \bibinfo{person}{Hude Quan}, {and} \bibinfo{person}{Cathy~A Eastwood}.} \bibinfo{year}{2023}\natexlab{}.
\newblock \showarticletitle{Use of epic electronic health record system for health care research: scoping review}.
\newblock \bibinfo{journal}{\emph{Journal of medical Internet research}}  \bibinfo{volume}{25} (\bibinfo{year}{2023}), \bibinfo{pages}{e51003}.
\newblock
\href{https://doi.org/10.2196/51003}{doi:\nolinkurl{10.2196/51003}}


\bibitem[Chua et~al\mbox{.}(2022)]%
        {chua2022enhancing}
\bibfield{author}{\bibinfo{person}{Isaac~S Chua}, \bibinfo{person}{Christine~S Ritchie}, {and} \bibinfo{person}{David~W Bates}.} \bibinfo{year}{2022}\natexlab{}.
\newblock \showarticletitle{Enhancing serious illness communication using artificial intelligence}.
\newblock \bibinfo{journal}{\emph{NPJ digital medicine}} \bibinfo{volume}{5}, \bibinfo{number}{1} (\bibinfo{year}{2022}), \bibinfo{pages}{14}.
\newblock
\href{https://doi.org/10.1038/s41746-022-00556-2}{doi:\nolinkurl{10.1038/s41746-022-00556-2}}


\bibitem[Corvite et~al\mbox{.}(2023)]%
        {corvite2023data}
\bibfield{author}{\bibinfo{person}{Shanley Corvite}, \bibinfo{person}{Kat Roemmich}, \bibinfo{person}{Tillie~Ilana Rosenberg}, {and} \bibinfo{person}{Nazanin Andalibi}.} \bibinfo{year}{2023}\natexlab{}.
\newblock \showarticletitle{Data subjects' perspectives on emotion artificial intelligence use in the workplace: A relational ethics lens}.
\newblock \bibinfo{journal}{\emph{Proceedings of the ACM on Human-Computer Interaction}} \bibinfo{volume}{7}, \bibinfo{number}{CSCW1} (\bibinfo{year}{2023}), \bibinfo{pages}{1--38}.
\newblock
\href{https://doi.org/10.1145/3579600}{doi:\nolinkurl{10.1145/3579600}}


\bibitem[Cox et~al\mbox{.}(2025)]%
        {cox2025mobile}
\bibfield{author}{\bibinfo{person}{Christopher~E Cox}, \bibinfo{person}{Deepshikha~C Ashana}, \bibinfo{person}{Katelyn Dempsey}, \bibinfo{person}{Maren~K Olsen}, \bibinfo{person}{Alice Parish}, \bibinfo{person}{David Casarett}, \bibinfo{person}{Kimberly~S Johnson}, \bibinfo{person}{Krista~L Haines}, \bibinfo{person}{Colleen Naglee}, \bibinfo{person}{Jason~N Katz}, {et~al\mbox{.}}} \bibinfo{year}{2025}\natexlab{}.
\newblock \showarticletitle{Mobile app--facilitated collaborative palliative care intervention for critically ill older adults: a randomized clinical trial}.
\newblock \bibinfo{journal}{\emph{JAMA Internal Medicine}} \bibinfo{volume}{185}, \bibinfo{number}{2} (\bibinfo{year}{2025}), \bibinfo{pages}{173--183}.
\newblock
\href{https://doi.org/10.1001/jamainternmed.2024.6838}{doi:\nolinkurl{10.1001/jamainternmed.2024.6838}}


\bibitem[Cuadra et~al\mbox{.}(2024)]%
        {cuadra2024illusion}
\bibfield{author}{\bibinfo{person}{Andrea Cuadra}, \bibinfo{person}{Maria Wang}, \bibinfo{person}{Lynn~Andrea Stein}, \bibinfo{person}{Malte~F Jung}, \bibinfo{person}{Nicola Dell}, \bibinfo{person}{Deborah Estrin}, {and} \bibinfo{person}{James~A Landay}.} \bibinfo{year}{2024}\natexlab{}.
\newblock \showarticletitle{The illusion of empathy? notes on displays of emotion in human-computer interaction}. In \bibinfo{booktitle}{\emph{Proceedings of the 2024 CHI Conference on Human Factors in Computing Systems}}. \bibinfo{pages}{1--18}.
\newblock
\href{https://doi.org/10.1145/3613904.3642336}{doi:\nolinkurl{10.1145/3613904.3642336}}


\bibitem[Curtis et~al\mbox{.}(2018)]%
        {curtis2018effect}
\bibfield{author}{\bibinfo{person}{J~Randall Curtis}, \bibinfo{person}{Lois Downey}, \bibinfo{person}{Anthony~L Back}, \bibinfo{person}{Elizabeth~L Nielsen}, \bibinfo{person}{Sudiptho Paul}, \bibinfo{person}{Alexandria~Z Lahdya}, \bibinfo{person}{Patsy~D Treece}, \bibinfo{person}{Priscilla Armstrong}, \bibinfo{person}{Ronald Peck}, {and} \bibinfo{person}{Ruth~A Engelberg}.} \bibinfo{year}{2018}\natexlab{}.
\newblock \showarticletitle{Effect of a patient and clinician communication-priming intervention on patient-reported goals-of-care discussions between patients with serious illness and clinicians: a randomized clinical trial}.
\newblock \bibinfo{journal}{\emph{JAMA Internal Medicine}} \bibinfo{volume}{178}, \bibinfo{number}{7} (\bibinfo{year}{2018}), \bibinfo{pages}{930--940}.
\newblock
\href{https://doi.org/10.1001/jamainternmed.2018.2317}{doi:\nolinkurl{10.1001/jamainternmed.2018.2317}}


\bibitem[Curtis et~al\mbox{.}(2023)]%
        {curtis2023intervention}
\bibfield{author}{\bibinfo{person}{J~Randall Curtis}, \bibinfo{person}{Robert~Y Lee}, \bibinfo{person}{Lyndia~C Brumback}, \bibinfo{person}{Erin~K Kross}, \bibinfo{person}{Lois Downey}, \bibinfo{person}{Janaki Torrence}, \bibinfo{person}{Nicole LeDuc}, \bibinfo{person}{Kasey~Mallon Andrews}, \bibinfo{person}{Jennifer Im}, \bibinfo{person}{Joanna Heywood}, {et~al\mbox{.}}} \bibinfo{year}{2023}\natexlab{}.
\newblock \bibinfo{title}{Intervention to promote communication about goals of care for hospitalized patients with serious illness: A randomized clinical trial}.
\newblock \bibinfo{numpages}{2028--2037}~pages.
\newblock
\href{https://doi.org/10.1001/jama.2023.8812}{doi:\nolinkurl{10.1001/jama.2023.8812}}


\bibitem[Davidson et~al\mbox{.}(2004)]%
        {davidson2004s}
\bibfield{author}{\bibinfo{person}{Steven~J Davidson}, \bibinfo{person}{Frank~L Zwemer~Jr}, \bibinfo{person}{Larry~A Nathanson}, \bibinfo{person}{Kenneth~N Sable}, {and} \bibinfo{person}{Abu~NGA Khan}.} \bibinfo{year}{2004}\natexlab{}.
\newblock \showarticletitle{Where's the beef? The promise and the reality of clinical documentation}.
\newblock \bibinfo{journal}{\emph{Academic Emergency Medicine}} \bibinfo{volume}{11}, \bibinfo{number}{11} (\bibinfo{year}{2004}), \bibinfo{pages}{1127--1134}.
\newblock
\href{https://doi.org/10.1197/j.aem.2004.08.004}{doi:\nolinkurl{10.1197/j.aem.2004.08.004}}


\bibitem[Desai et~al\mbox{.}(2023)]%
        {desai2023painless}
\bibfield{author}{\bibinfo{person}{Smit Desai}, \bibinfo{person}{Morgan Lundy}, {and} \bibinfo{person}{Jessie Chin}.} \bibinfo{year}{2023}\natexlab{}.
\newblock \showarticletitle{“A painless way to learn:” designing an interactive storytelling voice user interface to engage older adults in informal health information learning}. In \bibinfo{booktitle}{\emph{Proceedings of the 5th International Conference on Conversational User Interfaces}}. \bibinfo{pages}{1--16}.
\newblock
\href{https://doi.org/10.1145/3571884.3597141}{doi:\nolinkurl{10.1145/3571884.3597141}}


\bibitem[Fereday and Muir-Cochrane(2006)]%
        {fereday2006demonstrating}
\bibfield{author}{\bibinfo{person}{Jennifer Fereday} {and} \bibinfo{person}{Eimear Muir-Cochrane}.} \bibinfo{year}{2006}\natexlab{}.
\newblock \showarticletitle{Demonstrating rigor using thematic analysis: A hybrid approach of inductive and deductive coding and theme development}.
\newblock \bibinfo{journal}{\emph{International journal of qualitative methods}} \bibinfo{volume}{5}, \bibinfo{number}{1} (\bibinfo{year}{2006}), \bibinfo{pages}{80--92}.
\newblock
\href{https://doi.org/10.1177/160940690600500107}{doi:\nolinkurl{10.1177/160940690600500107}}


\bibitem[Foong et~al\mbox{.}(2024)]%
        {foong2024designing}
\bibfield{author}{\bibinfo{person}{Pin~Sym Foong}, \bibinfo{person}{Natasha Ureyang}, \bibinfo{person}{Charisse Foo}, \bibinfo{person}{Sajeban Antonyrex}, {and} \bibinfo{person}{Gerald~CH Koh}.} \bibinfo{year}{2024}\natexlab{}.
\newblock \showarticletitle{Designing for caregiver-facing values elicitation tools}. In \bibinfo{booktitle}{\emph{Proceedings of the 2024 CHI Conference on Human Factors in Computing Systems}}. \bibinfo{pages}{1--20}.
\newblock
\href{https://doi.org/10.1145/3613904.3642214}{doi:\nolinkurl{10.1145/3613904.3642214}}


\bibitem[Fox et~al\mbox{.}(2023)]%
        {Fox_Shorey_Kang_Montiel_Valle_Rodriguez_2023}
\bibfield{author}{\bibinfo{person}{Sarah~E. Fox}, \bibinfo{person}{Samantha Shorey}, \bibinfo{person}{Esther~Y. Kang}, \bibinfo{person}{Dominique Montiel~Valle}, {and} \bibinfo{person}{Estefania Rodriguez}.} \bibinfo{year}{2023}\natexlab{}.
\newblock \showarticletitle{Patchwork: The Hidden, Human Labor of AI Integration within Essential Work}.
\newblock \bibinfo{journal}{\emph{Proc. ACM Hum.-Comput. Interact.}} \bibinfo{volume}{7}, \bibinfo{number}{CSCW1} (\bibinfo{date}{April} \bibinfo{year}{2023}), \bibinfo{pages}{81:1--81:20}.
\newblock
\href{https://doi.org/10.1145/3579514}{doi:\nolinkurl{10.1145/3579514}}


\bibitem[Galvin et~al\mbox{.}(2017)]%
        {galvin2017adverse}
\bibfield{author}{\bibinfo{person}{Rose Galvin}, \bibinfo{person}{Yannick Gilleit}, \bibinfo{person}{Emma Wallace}, \bibinfo{person}{Grainne Cousins}, \bibinfo{person}{Manon Bolmer}, \bibinfo{person}{Timothy Rainer}, \bibinfo{person}{Susan~M Smith}, {and} \bibinfo{person}{Tom Fahey}.} \bibinfo{year}{2017}\natexlab{}.
\newblock \showarticletitle{Adverse outcomes in older adults attending emergency departments: a systematic review and meta-analysis of the Identification of Seniors At Risk (ISAR) screening tool}.
\newblock \bibinfo{journal}{\emph{Age and ageing}} \bibinfo{volume}{46}, \bibinfo{number}{2} (\bibinfo{year}{2017}), \bibinfo{pages}{179--186}.
\newblock
\href{https://doi.org/10.1093/ageing/afw233}{doi:\nolinkurl{10.1093/ageing/afw233}}


\bibitem[Garg(2021)]%
        {garg2021understanding}
\bibfield{author}{\bibinfo{person}{Radhika Garg}.} \bibinfo{year}{2021}\natexlab{}.
\newblock \showarticletitle{Understanding Tensions and Resilient Practices that Emerged from Technology Use in Asian Indian Families in the US During the COVID-19 Pandemic}.
\newblock \bibinfo{journal}{\emph{Proc. ACM Hum.-Comput. Interact}}  \bibinfo{volume}{5} (\bibinfo{year}{2021}).
\newblock
\href{https://doi.org/10.1145/3479558}{doi:\nolinkurl{10.1145/3479558}}


\bibitem[Geerse et~al\mbox{.}(2019)]%
        {geerse2019qualitative}
\bibfield{author}{\bibinfo{person}{Olaf~P Geerse}, \bibinfo{person}{Daniela~J Lamas}, \bibinfo{person}{Justin~J Sanders}, \bibinfo{person}{Joanna Paladino}, \bibinfo{person}{Jane Kavanagh}, \bibinfo{person}{Natalie~J Henrich}, \bibinfo{person}{Annette~J Berendsen}, \bibinfo{person}{Thijo~JN Hiltermann}, \bibinfo{person}{Erik~K Fromme}, \bibinfo{person}{Rachelle~E Bernacki}, {et~al\mbox{.}}} \bibinfo{year}{2019}\natexlab{}.
\newblock \showarticletitle{A qualitative study of serious illness conversations in patients with advanced cancer}.
\newblock \bibinfo{journal}{\emph{Journal of palliative medicine}} \bibinfo{volume}{22}, \bibinfo{number}{7} (\bibinfo{year}{2019}), \bibinfo{pages}{773--781}.
\newblock
\href{https://doi.org/10.1089/jpm.2018.0487}{doi:\nolinkurl{10.1089/jpm.2018.0487}}


\bibitem[Goodman(1961)]%
        {goodman1961snowball}
\bibfield{author}{\bibinfo{person}{Leo~A Goodman}.} \bibinfo{year}{1961}\natexlab{}.
\newblock \showarticletitle{Snowball sampling}.
\newblock \bibinfo{journal}{\emph{The annals of mathematical statistics}} (\bibinfo{year}{1961}), \bibinfo{pages}{148--170}.
\newblock
\href{https://doi.org/10.4135/9781526421036831710}{doi:\nolinkurl{10.4135/9781526421036831710}}


\bibitem[Guest et~al\mbox{.}(2006)]%
        {guest2006many}
\bibfield{author}{\bibinfo{person}{Greg Guest}, \bibinfo{person}{Arwen Bunce}, {and} \bibinfo{person}{Laura Johnson}.} \bibinfo{year}{2006}\natexlab{}.
\newblock \showarticletitle{How many interviews are enough? An experiment with data saturation and variability}.
\newblock \bibinfo{journal}{\emph{Field methods}} \bibinfo{volume}{18}, \bibinfo{number}{1} (\bibinfo{year}{2006}), \bibinfo{pages}{59--82}.
\newblock
\href{https://doi.org/10.1177/1525822X05279903}{doi:\nolinkurl{10.1177/1525822X05279903}}


\bibitem[Hachem et~al\mbox{.}(2025)]%
        {hachem2025electronic}
\bibfield{author}{\bibinfo{person}{Yasmina Hachem}, \bibinfo{person}{Joshua Lakin}, \bibinfo{person}{Winifred Teuteberg}, \bibinfo{person}{Amelia Cullinan}, \bibinfo{person}{Matthew~J Gonzales}, \bibinfo{person}{Charlotta Lindvall}, \bibinfo{person}{Pallavi Kumar}, \bibinfo{person}{Laura Dingfield}, \bibinfo{person}{Laurel Kilpatrick}, \bibinfo{person}{Jeff Greenwald}, {et~al\mbox{.}}} \bibinfo{year}{2025}\natexlab{}.
\newblock \showarticletitle{Electronic Health Record Serious Illness Conversation Dashboards: An Implementation Case Series}.
\newblock \bibinfo{journal}{\emph{Journal of pain and symptom management}} \bibinfo{volume}{69}, \bibinfo{number}{2} (\bibinfo{year}{2025}), \bibinfo{pages}{e139--e146}.
\newblock
\href{https://doi.org/10.1016/j.jpainsymman.2024.10.032}{doi:\nolinkurl{10.1016/j.jpainsymman.2024.10.032}}


\bibitem[Han et~al\mbox{.}(2024)]%
        {han2024ascleai}
\bibfield{author}{\bibinfo{person}{Jiyeon Han}, \bibinfo{person}{Jimin Park}, \bibinfo{person}{Jinyoung Huh}, \bibinfo{person}{Uran Oh}, \bibinfo{person}{Jaeyoung Do}, {and} \bibinfo{person}{Daehee Kim}.} \bibinfo{year}{2024}\natexlab{}.
\newblock \showarticletitle{AscleAI: A LLM-based clinical note management system for enhancing clinician productivity}. In \bibinfo{booktitle}{\emph{Extended Abstracts of the CHI Conference on Human Factors in Computing Systems}}. \bibinfo{pages}{1--7}.
\newblock
\href{https://doi.org/10.1145/3613905.3650784}{doi:\nolinkurl{10.1145/3613905.3650784}}


\bibitem[Hao et~al\mbox{.}(2024)]%
        {hao2024advancing}
\bibfield{author}{\bibinfo{person}{Yuexing Hao}, \bibinfo{person}{Zeyu Liu}, \bibinfo{person}{Robert~N Riter}, {and} \bibinfo{person}{Saleh Kalantari}.} \bibinfo{year}{2024}\natexlab{}.
\newblock \showarticletitle{Advancing patient-centered shared decision-making with ai systems for older adult cancer patients}. In \bibinfo{booktitle}{\emph{Proceedings of the 2024 CHI Conference on Human Factors in Computing Systems}}. \bibinfo{pages}{1--20}.
\newblock
\href{https://doi.org/10.1145/3613904.3642353}{doi:\nolinkurl{10.1145/3613904.3642353}}


\bibitem[Hartson and Pyla(2012)]%
        {hartson2012ux}
\bibfield{author}{\bibinfo{person}{Rex Hartson} {and} \bibinfo{person}{Pardha~S Pyla}.} \bibinfo{year}{2012}\natexlab{}.
\newblock \bibinfo{booktitle}{\emph{The UX Book: Process and guidelines for ensuring a quality user experience}}.
\newblock \bibinfo{publisher}{Elsevier}.
\newblock


\bibitem[Jacobs et~al\mbox{.}(2021)]%
        {jacobs2021designing}
\bibfield{author}{\bibinfo{person}{Maia Jacobs}, \bibinfo{person}{Jeffrey He}, \bibinfo{person}{Melanie F.~Pradier}, \bibinfo{person}{Barbara Lam}, \bibinfo{person}{Andrew~C Ahn}, \bibinfo{person}{Thomas~H McCoy}, \bibinfo{person}{Roy~H Perlis}, \bibinfo{person}{Finale Doshi-Velez}, {and} \bibinfo{person}{Krzysztof~Z Gajos}.} \bibinfo{year}{2021}\natexlab{}.
\newblock \showarticletitle{Designing AI for trust and collaboration in time-constrained medical decisions: a sociotechnical lens}. In \bibinfo{booktitle}{\emph{Proceedings of the 2021 chi conference on human factors in computing systems}}. \bibinfo{pages}{1--14}.
\newblock
\href{https://doi.org/10.1145/3411764.3445385}{doi:\nolinkurl{10.1145/3411764.3445385}}


\bibitem[Jang et~al\mbox{.}(2014)]%
        {jang2014bodydiagrams}
\bibfield{author}{\bibinfo{person}{Amy Jang}, \bibinfo{person}{Diana~L MacLean}, {and} \bibinfo{person}{Jeffrey Heer}.} \bibinfo{year}{2014}\natexlab{}.
\newblock \showarticletitle{BodyDiagrams: improving communication of pain symptoms through drawing}. In \bibinfo{booktitle}{\emph{Proceedings of the SIGCHI Conference on Human Factors in Computing Systems}}. \bibinfo{pages}{1153--1162}.
\newblock
\href{https://doi.org/10.1145/2556288.2557223}{doi:\nolinkurl{10.1145/2556288.2557223}}


\bibitem[Jo et~al\mbox{.}(2024)]%
        {jo2024exploring}
\bibfield{author}{\bibinfo{person}{Eunkyung Jo}, \bibinfo{person}{Rachael Zehrung}, \bibinfo{person}{Katherine Genuario}, \bibinfo{person}{Alexandra Papoutsaki}, {and} \bibinfo{person}{Daniel~A Epstein}.} \bibinfo{year}{2024}\natexlab{}.
\newblock \showarticletitle{Exploring Patient-Generated Annotations to Digital Clinical Symptom Measures for Patient-Centered Communication}.
\newblock \bibinfo{journal}{\emph{Proceedings of the ACM on Human-Computer Interaction}} \bibinfo{volume}{8}, \bibinfo{number}{CSCW2} (\bibinfo{year}{2024}), \bibinfo{pages}{1--26}.
\newblock
\href{https://doi.org/10.1145/3686997}{doi:\nolinkurl{10.1145/3686997}}


\bibitem[Johnson et~al\mbox{.}(2025)]%
        {johnson2025benefits}
\bibfield{author}{\bibinfo{person}{Jennifer Johnson}, \bibinfo{person}{Timmy Li}, \bibinfo{person}{Megan Mandile}, \bibinfo{person}{Santiago Lopez}, \bibinfo{person}{Molly McCann-Pineo}, \bibinfo{person}{Landon Witz}, {and} \bibinfo{person}{Payal Sud}.} \bibinfo{year}{2025}\natexlab{}.
\newblock \showarticletitle{Benefits of Emergency Department-Initiated Goals of Care Conversations and Palliative Care Consultations Among Older Adults with Chronic or Serious Life-Limiting Illnesses}.
\newblock \bibinfo{journal}{\emph{JACEP Open}} \bibinfo{volume}{6}, \bibinfo{number}{4} (\bibinfo{year}{2025}), \bibinfo{pages}{100165}.
\newblock
\href{https://doi.org/10.1016/j.acepjo.2025.100165}{doi:\nolinkurl{10.1016/j.acepjo.2025.100165}}


\bibitem[Kelley and Bollens-Lund(2018)]%
        {kelley2018identifying}
\bibfield{author}{\bibinfo{person}{Amy~S Kelley} {and} \bibinfo{person}{Evan Bollens-Lund}.} \bibinfo{year}{2018}\natexlab{}.
\newblock \showarticletitle{Identifying the population with serious illness: the “denominator” challenge}.
\newblock \bibinfo{journal}{\emph{Journal of palliative medicine}} \bibinfo{volume}{21}, \bibinfo{number}{S2} (\bibinfo{year}{2018}), \bibinfo{pages}{S--7}.
\newblock
\href{https://doi.org/10.1089/jpm.2017.0548}{doi:\nolinkurl{10.1089/jpm.2017.0548}}


\bibitem[Kim et~al\mbox{.}(2016)]%
        {kim2016natural}
\bibfield{author}{\bibinfo{person}{Yan~S Kim}, \bibinfo{person}{Gabriel~J Escobar}, \bibinfo{person}{Scott~D Halpern}, \bibinfo{person}{John~D Greene}, \bibinfo{person}{Patricia Kipnis}, {and} \bibinfo{person}{Vincent Liu}.} \bibinfo{year}{2016}\natexlab{}.
\newblock \showarticletitle{The natural history of changes in preferences for life-sustaining treatments and implications for inpatient mortality in younger and older hospitalized adults}.
\newblock \bibinfo{journal}{\emph{Journal of the American Geriatrics Society}} \bibinfo{volume}{64}, \bibinfo{number}{5} (\bibinfo{year}{2016}), \bibinfo{pages}{981--989}.
\newblock
\href{https://doi.org/10.1111/jgs.14048}{doi:\nolinkurl{10.1111/jgs.14048}}


\bibitem[Kong et~al\mbox{.}(2024)]%
        {kong2024envisioning}
\bibfield{author}{\bibinfo{person}{Elaine Kong}, \bibinfo{person}{Kuo-Ting Huang}, {and} \bibinfo{person}{Aakash Gautam}.} \bibinfo{year}{2024}\natexlab{}.
\newblock \showarticletitle{Envisioning Possibilities and Challenges of AI for Personalized Cancer Care}. In \bibinfo{booktitle}{\emph{Companion Publication of the 2024 Conference on Computer-Supported Cooperative Work and Social Computing}}. \bibinfo{pages}{415--421}.
\newblock
\href{https://doi.org/10.1145/3678884.3681885}{doi:\nolinkurl{10.1145/3678884.3681885}}


\bibitem[Kumar et~al\mbox{.}(2023)]%
        {kumar2023exploring}
\bibfield{author}{\bibinfo{person}{Harsh Kumar}, \bibinfo{person}{Yiyi Wang}, \bibinfo{person}{Jiakai Shi}, \bibinfo{person}{Ilya Musabirov}, \bibinfo{person}{Norman~AS Farb}, {and} \bibinfo{person}{Joseph~Jay Williams}.} \bibinfo{year}{2023}\natexlab{}.
\newblock \showarticletitle{Exploring the use of large language models for improving the awareness of mindfulness}. In \bibinfo{booktitle}{\emph{Extended Abstracts of the 2023 CHI Conference on Human Factors in Computing Systems}}. \bibinfo{pages}{1--7}.
\newblock
\href{https://doi.org/10.1145/3544549.3585614}{doi:\nolinkurl{10.1145/3544549.3585614}}


\bibitem[Kwon et~al\mbox{.}(2019)]%
        {kwon2019artificial}
\bibfield{author}{\bibinfo{person}{Joon-myoung Kwon}, \bibinfo{person}{Kyung-Hee Kim}, \bibinfo{person}{Ki-Hyun Jeon}, \bibinfo{person}{Sang~Eun Lee}, \bibinfo{person}{Hae-Young Lee}, \bibinfo{person}{Hyun-Jai Cho}, \bibinfo{person}{Jin~Oh Choi}, \bibinfo{person}{Eun-Seok Jeon}, \bibinfo{person}{Min-Seok Kim}, \bibinfo{person}{Jae-Joong Kim}, {et~al\mbox{.}}} \bibinfo{year}{2019}\natexlab{}.
\newblock \showarticletitle{Artificial intelligence algorithm for predicting mortality of patients with acute heart failure}.
\newblock \bibinfo{journal}{\emph{PloS one}} \bibinfo{volume}{14}, \bibinfo{number}{7} (\bibinfo{year}{2019}), \bibinfo{pages}{e0219302}.
\newblock
\href{https://doi.org/10.1371/journal.pone.0219302}{doi:\nolinkurl{10.1371/journal.pone.0219302}}


\bibitem[Lakin et~al\mbox{.}(2017)]%
        {lakin2017systematic}
\bibfield{author}{\bibinfo{person}{Joshua~R Lakin}, \bibinfo{person}{Luca~A Koritsanszky}, \bibinfo{person}{Rebecca Cunningham}, \bibinfo{person}{Francine~L Maloney}, \bibinfo{person}{Brandon~J Neal}, \bibinfo{person}{Joanna Paladino}, \bibinfo{person}{Marissa~C Palmor}, \bibinfo{person}{Christine Vogeli}, \bibinfo{person}{Timothy~G Ferris}, \bibinfo{person}{Susan~D Block}, {et~al\mbox{.}}} \bibinfo{year}{2017}\natexlab{}.
\newblock \showarticletitle{A systematic intervention to improve serious illness communication in primary care}.
\newblock \bibinfo{journal}{\emph{Health Affairs}} \bibinfo{volume}{36}, \bibinfo{number}{7} (\bibinfo{year}{2017}), \bibinfo{pages}{1258--1264}.
\newblock
\href{https://doi.org/10.1377/hlthaff.2017.0219}{doi:\nolinkurl{10.1377/hlthaff.2017.0219}}


\bibitem[Lee et~al\mbox{.}(2021b)]%
        {lee2021human}
\bibfield{author}{\bibinfo{person}{Min~Hun Lee}, \bibinfo{person}{Daniel~P Siewiorek}, \bibinfo{person}{Asim Smailagic}, \bibinfo{person}{Alexandre Bernardino}, {and} \bibinfo{person}{Sergi~Berm{\'u}dez Berm{\'u}dez~i Badia}.} \bibinfo{year}{2021}\natexlab{b}.
\newblock \showarticletitle{A human-ai collaborative approach for clinical decision making on rehabilitation assessment}. In \bibinfo{booktitle}{\emph{Proceedings of the 2021 CHI conference on human factors in computing systems}}. \bibinfo{pages}{1--14}.
\newblock
\href{https://doi.org/10.1145/3411764.3445472}{doi:\nolinkurl{10.1145/3411764.3445472}}


\bibitem[Lee et~al\mbox{.}(2021a)]%
        {lee2021identifying}
\bibfield{author}{\bibinfo{person}{Robert~Y Lee}, \bibinfo{person}{Lyndia~C Brumback}, \bibinfo{person}{William~B Lober}, \bibinfo{person}{James Sibley}, \bibinfo{person}{Elizabeth~L Nielsen}, \bibinfo{person}{Patsy~D Treece}, \bibinfo{person}{Erin~K Kross}, \bibinfo{person}{Elizabeth~T Loggers}, \bibinfo{person}{James~A Fausto}, \bibinfo{person}{Charlotta Lindvall}, {et~al\mbox{.}}} \bibinfo{year}{2021}\natexlab{a}.
\newblock \showarticletitle{Identifying goals of care conversations in the electronic health record using natural language processing and machine learning}.
\newblock \bibinfo{journal}{\emph{Journal of pain and symptom management}} \bibinfo{volume}{61}, \bibinfo{number}{1} (\bibinfo{year}{2021}), \bibinfo{pages}{136--142}.
\newblock
\href{https://doi.org/10.1016/j.jpainsymman.2020.08.024}{doi:\nolinkurl{10.1016/j.jpainsymman.2020.08.024}}


\bibitem[Li et~al\mbox{.}(2021)]%
        {li2021automating}
\bibfield{author}{\bibinfo{person}{Brenna Li}, \bibinfo{person}{Noah Crampton}, \bibinfo{person}{Thomas Yeates}, \bibinfo{person}{Yu Xia}, \bibinfo{person}{Xirong Tian}, {and} \bibinfo{person}{Khai Truong}.} \bibinfo{year}{2021}\natexlab{}.
\newblock \showarticletitle{Automating clinical documentation with digital scribes: Understanding the impact on physicians}. In \bibinfo{booktitle}{\emph{Proceedings of the 2021 CHI Conference on Human Factors in Computing Systems}}. \bibinfo{pages}{1--12}.
\newblock
\href{https://doi.org/10.1145/3411764.3445172}{doi:\nolinkurl{10.1145/3411764.3445172}}


\bibitem[Li et~al\mbox{.}(2024)]%
        {li2024beyond}
\bibfield{author}{\bibinfo{person}{Brenna Li}, \bibinfo{person}{Ofek Gross}, \bibinfo{person}{Noah Crampton}, \bibinfo{person}{Mamta Kapoor}, \bibinfo{person}{Saba Tauseef}, \bibinfo{person}{Mohit Jain}, \bibinfo{person}{Khai~N Truong}, {and} \bibinfo{person}{Alex Mariakakis}.} \bibinfo{year}{2024}\natexlab{}.
\newblock \showarticletitle{Beyond the Waiting Room: Patient's Perspectives on the Conversational Nuances of Pre-Consultation Chatbots}. In \bibinfo{booktitle}{\emph{Proceedings of the 2024 CHI Conference on Human Factors in Computing Systems}}. \bibinfo{pages}{1--24}.
\newblock
\href{https://doi.org/10.1145/3613904.3641913}{doi:\nolinkurl{10.1145/3613904.3641913}}


\bibitem[Liaqat et~al\mbox{.}(2024)]%
        {liaqat2024promoting}
\bibfield{author}{\bibinfo{person}{Salaar Liaqat}, \bibinfo{person}{Daniyal Liaqat}, \bibinfo{person}{Tatiana Son}, \bibinfo{person}{Tiago Falk}, \bibinfo{person}{Robert Wu}, \bibinfo{person}{Andrea~S Gershon}, \bibinfo{person}{Eyal De~Lara}, {and} \bibinfo{person}{Alex Mariakakis}.} \bibinfo{year}{2024}\natexlab{}.
\newblock \showarticletitle{Promoting Engagement in Remote Patient Monitoring Using Asynchronous Messaging}. In \bibinfo{booktitle}{\emph{Proceedings of the 2024 CHI Conference on Human Factors in Computing Systems}}. \bibinfo{pages}{1--18}.
\newblock
\href{https://doi.org/10.1145/3613904.3642630}{doi:\nolinkurl{10.1145/3613904.3642630}}


\bibitem[Lindvall et~al\mbox{.}(2022)]%
        {lindvall2022natural}
\bibfield{author}{\bibinfo{person}{Charlotta Lindvall}, \bibinfo{person}{Chih-Ying Deng}, \bibinfo{person}{Edward Moseley}, \bibinfo{person}{Nicole Agaronnik}, \bibinfo{person}{Areej El-Jawahri}, \bibinfo{person}{Michael~K Paasche-Orlow}, \bibinfo{person}{Joshua~R Lakin}, \bibinfo{person}{Angelo Volandes}, \bibinfo{person}{James~A Tulsky}, \bibinfo{person}{ACP-PEACE Investigators}, {et~al\mbox{.}}} \bibinfo{year}{2022}\natexlab{}.
\newblock \showarticletitle{Natural language processing to identify advance care planning documentation in a multisite pragmatic clinical trial}.
\newblock \bibinfo{journal}{\emph{Journal of pain and symptom management}} \bibinfo{volume}{63}, \bibinfo{number}{1} (\bibinfo{year}{2022}), \bibinfo{pages}{e29--e36}.
\newblock
\href{https://doi.org/10.1016/j.jpainsymman.2021.06.025}{doi:\nolinkurl{10.1016/j.jpainsymman.2021.06.025}}


\bibitem[Maity and Deroy(2024)]%
        {maity2024future}
\bibfield{author}{\bibinfo{person}{Subhankar Maity} {and} \bibinfo{person}{Aniket Deroy}.} \bibinfo{year}{2024}\natexlab{}.
\newblock \showarticletitle{The future of learning in the age of generative ai: Automated question generation and assessment with large language models}.
\newblock \bibinfo{journal}{\emph{arXiv preprint arXiv:2410.09576}} (\bibinfo{year}{2024}).
\newblock
\href{https://doi.org/10.48550/arXiv.2410.09576}{doi:\nolinkurl{10.48550/arXiv.2410.09576}}


\bibitem[Mandel et~al\mbox{.}(2023)]%
        {mandel2023pilot}
\bibfield{author}{\bibinfo{person}{Ernest~I Mandel}, \bibinfo{person}{Francine~L Maloney}, \bibinfo{person}{Nathan~J Pertsch}, \bibinfo{person}{Jonathon~D Gass}, \bibinfo{person}{Justin~J Sanders}, \bibinfo{person}{Rachelle~E Bernacki}, {and} \bibinfo{person}{Susan~D Block}.} \bibinfo{year}{2023}\natexlab{}.
\newblock \showarticletitle{A pilot study of the serious illness conversation guide in a dialysis clinic}.
\newblock \bibinfo{journal}{\emph{American Journal of Hospice and Palliative Medicine{\textregistered}}} \bibinfo{volume}{40}, \bibinfo{number}{10} (\bibinfo{year}{2023}), \bibinfo{pages}{1106--1113}.
\newblock
\href{https://doi.org/10.1177/10499091221147303}{doi:\nolinkurl{10.1177/10499091221147303}}


\bibitem[Manz et~al\mbox{.}(2020)]%
        {manz2020effect}
\bibfield{author}{\bibinfo{person}{Christopher~R Manz}, \bibinfo{person}{Ravi~B Parikh}, \bibinfo{person}{Dylan~S Small}, \bibinfo{person}{Chalanda~N Evans}, \bibinfo{person}{Corey Chivers}, \bibinfo{person}{Susan~H Regli}, \bibinfo{person}{C~William Hanson}, \bibinfo{person}{Justin~E Bekelman}, \bibinfo{person}{Charles~AL Rareshide}, \bibinfo{person}{Nina O’Connor}, {et~al\mbox{.}}} \bibinfo{year}{2020}\natexlab{}.
\newblock \showarticletitle{Effect of integrating machine learning mortality estimates with behavioral nudges to clinicians on serious illness conversations among patients with cancer: a stepped-wedge cluster randomized clinical trial}.
\newblock \bibinfo{journal}{\emph{JAMA oncology}} \bibinfo{volume}{6}, \bibinfo{number}{12} (\bibinfo{year}{2020}), \bibinfo{pages}{e204759--e204759}.
\newblock
\href{https://doi.org/10.1001/jamaoncol.2020.4759}{doi:\nolinkurl{10.1001/jamaoncol.2020.4759}}


\bibitem[Mazmanian et~al\mbox{.}(2013)]%
        {Mazmanian_Orlikowski_Yates_2013}
\bibfield{author}{\bibinfo{person}{Melissa Mazmanian}, \bibinfo{person}{Wanda~J. Orlikowski}, {and} \bibinfo{person}{JoAnne Yates}.} \bibinfo{year}{2013}\natexlab{}.
\newblock \showarticletitle{The Autonomy Paradox: The Implications of Mobile Email Devices for Knowledge Professionals}.
\newblock \bibinfo{journal}{\emph{Organization Science}} \bibinfo{volume}{24}, \bibinfo{number}{5} (\bibinfo{date}{Oct.} \bibinfo{year}{2013}), \bibinfo{pages}{1337–1357}.
\newblock
\showISSN{1047-7039}
\href{https://doi.org/10.1287/orsc.1120.0806}{doi:\nolinkurl{10.1287/orsc.1120.0806}}


\bibitem[Morley et~al\mbox{.}(2018)]%
        {morley2018emergency}
\bibfield{author}{\bibinfo{person}{Claire Morley}, \bibinfo{person}{Maria Unwin}, \bibinfo{person}{Gregory~M Peterson}, \bibinfo{person}{Jim Stankovich}, {and} \bibinfo{person}{Leigh Kinsman}.} \bibinfo{year}{2018}\natexlab{}.
\newblock \showarticletitle{Emergency department crowding: a systematic review of causes, consequences and solutions}.
\newblock \bibinfo{journal}{\emph{PloS one}} \bibinfo{volume}{13}, \bibinfo{number}{8} (\bibinfo{year}{2018}), \bibinfo{pages}{e0203316}.
\newblock
\href{https://doi.org/10.1371/journal.pone.0203316}{doi:\nolinkurl{10.1371/journal.pone.0203316}}


\bibitem[Murray et~al\mbox{.}(2025)]%
        {murray2025goc}
\bibfield{author}{\bibinfo{person}{Julia Murray}, \bibinfo{person}{Zacharia Grami}, \bibinfo{person}{Katherine Benson}, \bibinfo{person}{Christopher Hritz}, \bibinfo{person}{Samantha Lawson}, \bibinfo{person}{Corita~Reilley Grudzen}, \bibinfo{person}{Allison Cuthel}, {and} \bibinfo{person}{Lauren Talanda-Fath Southerland}.} \bibinfo{year}{2025}\natexlab{}.
\newblock \showarticletitle{Effect of a multi-component palliative care intervention on goals of care discussions for critical patients in the emergency department}.
\newblock \bibinfo{journal}{\emph{Internal and Emergency Medicine}} (\bibinfo{year}{2025}).
\newblock
\href{https://doi.org/10.1007/s11739-025-04048-5}{doi:\nolinkurl{10.1007/s11739-025-04048-5}}
\newblock
\shownote{Advance online publication}.


\bibitem[Naik et~al\mbox{.}(2016)]%
        {naik2016health}
\bibfield{author}{\bibinfo{person}{Aanand~D Naik}, \bibinfo{person}{Lindsey~A Martin}, \bibinfo{person}{Jennifer Moye}, {and} \bibinfo{person}{Michele~J Karel}.} \bibinfo{year}{2016}\natexlab{}.
\newblock \showarticletitle{Health values and treatment goals of older, multimorbid adults facing life-threatening illness}.
\newblock \bibinfo{journal}{\emph{Journal of the American Geriatrics Society}} \bibinfo{volume}{64}, \bibinfo{number}{3} (\bibinfo{year}{2016}), \bibinfo{pages}{625--631}.
\newblock
\href{https://doi.org/10.1111/jgs.14027}{doi:\nolinkurl{10.1111/jgs.14027}}


\bibitem[Ni et~al\mbox{.}(2011)]%
        {ni2011anatonme}
\bibfield{author}{\bibinfo{person}{Tao Ni}, \bibinfo{person}{Amy~K Karlson}, {and} \bibinfo{person}{Daniel Wigdor}.} \bibinfo{year}{2011}\natexlab{}.
\newblock \showarticletitle{AnatOnMe: facilitating doctor-patient communication using a projection-based handheld device}. In \bibinfo{booktitle}{\emph{Proceedings of the SIGCHI conference on human factors in computing systems}}. \bibinfo{pages}{3333--3342}.
\newblock
\href{https://doi.org/10.1145/1978942.1979437}{doi:\nolinkurl{10.1145/1978942.1979437}}


\bibitem[Nowell et~al\mbox{.}(2017)]%
        {Nowell_Norris_White_Moules_2017}
\bibfield{author}{\bibinfo{person}{Lorelli~S. Nowell}, \bibinfo{person}{Jill~M. Norris}, \bibinfo{person}{Deborah~E. White}, {and} \bibinfo{person}{Nancy~J. Moules}.} \bibinfo{year}{2017}\natexlab{}.
\newblock \showarticletitle{Thematic Analysis: Striving to Meet the Trustworthiness Criteria}.
\newblock \bibinfo{journal}{\emph{International Journal of Qualitative Methods}} \bibinfo{volume}{16}, \bibinfo{number}{1} (\bibinfo{date}{Dec.} \bibinfo{year}{2017}), \bibinfo{pages}{1609406917733847}.
\newblock
\showISSN{1609-4069}
\href{https://doi.org/10.1177/1609406917733847}{doi:\nolinkurl{10.1177/1609406917733847}}


\bibitem[Olson et~al\mbox{.}(2025)]%
        {olson2025use}
\bibfield{author}{\bibinfo{person}{Kristine~D Olson}, \bibinfo{person}{Daniella Meeker}, \bibinfo{person}{Matt Troup}, \bibinfo{person}{Timothy~D Barker}, \bibinfo{person}{Vinh~H Nguyen}, \bibinfo{person}{Jennifer~B Manders}, \bibinfo{person}{Cheryl~D Stults}, \bibinfo{person}{Veena~G Jones}, \bibinfo{person}{Sachin~D Shah}, \bibinfo{person}{Tina Shah}, {et~al\mbox{.}}} \bibinfo{year}{2025}\natexlab{}.
\newblock \showarticletitle{Use of ambient AI scribes to reduce administrative burden and professional burnout}.
\newblock \bibinfo{journal}{\emph{JAMA Network Open}} \bibinfo{volume}{8}, \bibinfo{number}{10} (\bibinfo{year}{2025}), \bibinfo{pages}{e2534976--e2534976}.
\newblock
\href{https://doi.org/10.1001/jamanetworkopen.2025.34976}{doi:\nolinkurl{10.1001/jamanetworkopen.2025.34976}}


\bibitem[Ouchi et~al\mbox{.}(2019)]%
        {ouchi2019goals}
\bibfield{author}{\bibinfo{person}{Kei Ouchi}, \bibinfo{person}{Naomi George}, \bibinfo{person}{Jeremiah~D Schuur}, \bibinfo{person}{Emily~L Aaronson}, \bibinfo{person}{Charlotta Lindvall}, \bibinfo{person}{Edward Bernstein}, \bibinfo{person}{Rebecca~L Sudore}, \bibinfo{person}{Mara~A Schonberg}, \bibinfo{person}{Susan~D Block}, {and} \bibinfo{person}{James~A Tulsky}.} \bibinfo{year}{2019}\natexlab{}.
\newblock \showarticletitle{Goals-of-care conversations for older adults with serious illness in the emergency department: challenges and opportunities}.
\newblock \bibinfo{journal}{\emph{Annals of emergency medicine}} \bibinfo{volume}{74}, \bibinfo{number}{2} (\bibinfo{year}{2019}), \bibinfo{pages}{276--284}.
\newblock
\href{https://doi.org/10.1016/j.annemergmed.2019.01.003}{doi:\nolinkurl{10.1016/j.annemergmed.2019.01.003}}


\bibitem[Ouchi et~al\mbox{.}(2017)]%
        {ouchi2017preparing}
\bibfield{author}{\bibinfo{person}{Kei Ouchi}, \bibinfo{person}{Vinicius Knabben}, \bibinfo{person}{Laura Rivera-Reyes}, \bibinfo{person}{Niharika Ganta}, \bibinfo{person}{Laura~P Gelfman}, \bibinfo{person}{Rebecca Sudore}, \bibinfo{person}{Ula Hwang}, {and} \bibinfo{person}{GEDI~WISE Investigators}.} \bibinfo{year}{2017}\natexlab{}.
\newblock \showarticletitle{Preparing older adults with serious illness to formulate their goals for medical care in the emergency department}.
\newblock \bibinfo{journal}{\emph{Journal of palliative medicine}} \bibinfo{volume}{20}, \bibinfo{number}{4} (\bibinfo{year}{2017}), \bibinfo{pages}{404--408}.
\newblock
\href{https://doi.org/10.1089/jpm.2016.0109}{doi:\nolinkurl{10.1089/jpm.2016.0109}}


\bibitem[Paladino et~al\mbox{.}(2020)]%
        {paladino2020training}
\bibfield{author}{\bibinfo{person}{Joanna Paladino}, \bibinfo{person}{Laurel Kilpatrick}, \bibinfo{person}{Nina O'Connor}, \bibinfo{person}{Ramya Prabhakar}, \bibinfo{person}{Anna Kennedy}, \bibinfo{person}{Brandon~J Neal}, \bibinfo{person}{Jane Kavanagh}, \bibinfo{person}{Justin Sanders}, \bibinfo{person}{Susan Block}, {and} \bibinfo{person}{Erik Fromme}.} \bibinfo{year}{2020}\natexlab{}.
\newblock \showarticletitle{Training clinicians in serious illness communication using a structured guide: evaluation of a training program in three health systems}.
\newblock \bibinfo{journal}{\emph{Journal of Palliative Medicine}} \bibinfo{volume}{23}, \bibinfo{number}{3} (\bibinfo{year}{2020}), \bibinfo{pages}{337--345}.
\newblock
\href{https://doi.org/10.1089/jpm.2019.0334}{doi:\nolinkurl{10.1089/jpm.2019.0334}}


\bibitem[Paladino et~al\mbox{.}(2023)]%
        {paladino2023improving}
\bibfield{author}{\bibinfo{person}{Joanna Paladino}, \bibinfo{person}{Justin~J Sanders}, \bibinfo{person}{Erik~K Fromme}, \bibinfo{person}{Susan Block}, \bibinfo{person}{Juliet~C Jacobsen}, \bibinfo{person}{Vicki~A Jackson}, \bibinfo{person}{Christine~S Ritchie}, {and} \bibinfo{person}{Suzanne Mitchell}.} \bibinfo{year}{2023}\natexlab{}.
\newblock \showarticletitle{Improving serious illness communication: a qualitative study of clinical culture}.
\newblock \bibinfo{journal}{\emph{BMC palliative care}} \bibinfo{volume}{22}, \bibinfo{number}{1} (\bibinfo{year}{2023}), \bibinfo{pages}{104}.
\newblock
\href{https://doi.org/10.1186/s12904-023-01229-x}{doi:\nolinkurl{10.1186/s12904-023-01229-x}}


\bibitem[Patel et~al\mbox{.}(2013)]%
        {patel2013visual}
\bibfield{author}{\bibinfo{person}{Rupa~A Patel}, \bibinfo{person}{Andrea Hartzler}, \bibinfo{person}{Wanda Pratt}, \bibinfo{person}{Anthony Back}, \bibinfo{person}{Mary Czerwinski}, {and} \bibinfo{person}{Asta Roseway}.} \bibinfo{year}{2013}\natexlab{}.
\newblock \showarticletitle{Visual feedback on nonverbal communication: a design exploration with healthcare professionals}. In \bibinfo{booktitle}{\emph{2013 7th International Conference on Pervasive Computing Technologies for Healthcare and Workshops}}. IEEE, \bibinfo{pages}{105--112}.
\newblock
\href{https://doi.org/10.4108/icst.pervasivehealth.2013.252024}{doi:\nolinkurl{10.4108/icst.pervasivehealth.2013.252024}}


\bibitem[P{\'e}loquin et~al\mbox{.}(2025)]%
        {peloquin2025discussions}
\bibfield{author}{\bibinfo{person}{Fannie P{\'e}loquin}, \bibinfo{person}{{\'E}mile Marmen}, \bibinfo{person}{V{\'e}ronique G{\'e}linas}, \bibinfo{person}{Ariane Plaisance}, \bibinfo{person}{Maude Linteau}, \bibinfo{person}{Audrey Nolet}, \bibinfo{person}{Nathalie Germain}, {and} \bibinfo{person}{Patrick~M Archambault}.} \bibinfo{year}{2025}\natexlab{}.
\newblock \showarticletitle{Discussions about goals of care in the emergency department: a qualitative study of emergency physicians’ opinions using the normalization process theory}.
\newblock \bibinfo{journal}{\emph{Canadian Journal of Emergency Medicine}} (\bibinfo{year}{2025}), \bibinfo{pages}{1--9}.
\newblock
\href{https://doi.org/10.1007/s43678-025-00911-8}{doi:\nolinkurl{10.1007/s43678-025-00911-8}}


\bibitem[Pourhomayoun and Shakibi(2020)]%
        {pourhomayoun2020predicting}
\bibfield{author}{\bibinfo{person}{Mohammad Pourhomayoun} {and} \bibinfo{person}{Mahdi Shakibi}.} \bibinfo{year}{2020}\natexlab{}.
\newblock \showarticletitle{Predicting mortality risk in patients with COVID-19 using artificial intelligence to help medical decision-making}.
\newblock \bibinfo{journal}{\emph{MedRxiv}} (\bibinfo{year}{2020}), \bibinfo{pages}{2020--03}.
\newblock
\href{https://doi.org/10.1101/2020.03.30.20047308}{doi:\nolinkurl{10.1101/2020.03.30.20047308}}


\bibitem[Prachanukool et~al\mbox{.}(2022)]%
        {prachanukool2022emergency}
\bibfield{author}{\bibinfo{person}{Thidathit Prachanukool}, \bibinfo{person}{Susan~D Block}, \bibinfo{person}{Donna Berry}, \bibinfo{person}{Rachel~S Lee}, \bibinfo{person}{Sarah Rossmassler}, \bibinfo{person}{Mohammad~A Hasdianda}, \bibinfo{person}{Wei Wang}, \bibinfo{person}{Rebecca Sudore}, \bibinfo{person}{Mara~A Schonberg}, \bibinfo{person}{James~A Tulsky}, {et~al\mbox{.}}} \bibinfo{year}{2022}\natexlab{}.
\newblock \showarticletitle{Emergency department-based, nurse-initiated, serious illness conversation intervention for older adults: a protocol for a randomized controlled trial}.
\newblock \bibinfo{journal}{\emph{Trials}} \bibinfo{volume}{23}, \bibinfo{number}{1} (\bibinfo{year}{2022}), \bibinfo{pages}{866}.
\newblock
\href{https://doi.org/10.1186/s13063-022-06797-6}{doi:\nolinkurl{10.1186/s13063-022-06797-6}}


\bibitem[Ramesh et~al\mbox{.}(2024)]%
        {ramesh2024data}
\bibfield{author}{\bibinfo{person}{Shri~Harini Ramesh}, \bibinfo{person}{Alicia Ouskine}, \bibinfo{person}{Elahe Khorasani}, \bibinfo{person}{Mona Ebrahimipour}, \bibinfo{person}{Hillel Finestone}, \bibinfo{person}{Adrian~DC Chan}, {and} \bibinfo{person}{Fateme Rajabiyazdi}.} \bibinfo{year}{2024}\natexlab{}.
\newblock \showarticletitle{A Data Visualization Tool for Patients and Healthcare Providers to Communicate during Inpatient Stroke Rehabilitation}. In \bibinfo{booktitle}{\emph{Proceedings of the 50th Graphics Interface Conference}}. \bibinfo{pages}{1--14}.
\newblock
\href{https://doi.org/10.1145/3670947.3670978}{doi:\nolinkurl{10.1145/3670947.3670978}}


\bibitem[Roemmich et~al\mbox{.}(2023)]%
        {roemmich2023emotion}
\bibfield{author}{\bibinfo{person}{Kat Roemmich}, \bibinfo{person}{Florian Schaub}, {and} \bibinfo{person}{Nazanin Andalibi}.} \bibinfo{year}{2023}\natexlab{}.
\newblock \showarticletitle{Emotion AI at work: Implications for workplace surveillance, emotional labor, and emotional privacy}. In \bibinfo{booktitle}{\emph{Proceedings of the 2023 CHI conference on human factors in computing systems}}. \bibinfo{pages}{1--20}.
\newblock
\href{https://doi.org/10.1145/3544548.3580950}{doi:\nolinkurl{10.1145/3544548.3580950}}


\bibitem[Ryan et~al\mbox{.}(2019)]%
        {ryan2019using}
\bibfield{author}{\bibinfo{person}{Padhraig Ryan}, \bibinfo{person}{Saturnino Luz}, \bibinfo{person}{Pierre Albert}, \bibinfo{person}{Carl Vogel}, \bibinfo{person}{Charles Normand}, {and} \bibinfo{person}{Glyn Elwyn}.} \bibinfo{year}{2019}\natexlab{}.
\newblock \showarticletitle{Using artificial intelligence to assess clinicians’ communication skills}.
\newblock \bibinfo{journal}{\emph{Bmj}}  \bibinfo{volume}{364} (\bibinfo{year}{2019}).
\newblock
\href{https://doi.org/10.1136/bmj.l161}{doi:\nolinkurl{10.1136/bmj.l161}}


\bibitem[Ryu et~al\mbox{.}(2023)]%
        {ryu2023you}
\bibfield{author}{\bibinfo{person}{Hyeyoung Ryu}, \bibinfo{person}{Andrew~BL Berry}, \bibinfo{person}{Catherine~Y Lim}, \bibinfo{person}{Andrea Hartzler}, \bibinfo{person}{Tad Hirsch}, \bibinfo{person}{Juanita~I Trejo}, \bibinfo{person}{Zo{\"e}~Abigail Bermet}, \bibinfo{person}{Brandi Crawford-Gallagher}, \bibinfo{person}{Vi Tran}, \bibinfo{person}{Dawn Ferguson}, {et~al\mbox{.}}} \bibinfo{year}{2023}\natexlab{}.
\newblock \showarticletitle{“You Can See the Connections”: Facilitating Visualization of Care Priorities in People Living with Multiple Chronic Health Conditions}. In \bibinfo{booktitle}{\emph{Proceedings of the 2023 CHI Conference on Human Factors in Computing Systems}}. \bibinfo{pages}{1--17}.
\newblock
\href{https://doi.org/10.1145/3544548.3580908}{doi:\nolinkurl{10.1145/3544548.3580908}}


\bibitem[Saarinen and Aho(2005)]%
        {saarinen2005does}
\bibfield{author}{\bibinfo{person}{K Saarinen} {and} \bibinfo{person}{M Aho}.} \bibinfo{year}{2005}\natexlab{}.
\newblock \showarticletitle{Does the implementation of a clinical information system decrease the time intensive care nurses spend on documentation of care?}
\newblock \bibinfo{journal}{\emph{Acta anaesthesiologica scandinavica}} \bibinfo{volume}{49}, \bibinfo{number}{1} (\bibinfo{year}{2005}), \bibinfo{pages}{62--65}.
\newblock
\href{https://doi.org/10.1111/j.1399-6576.2005.00546.x}{doi:\nolinkurl{10.1111/j.1399-6576.2005.00546.x}}


\bibitem[Samaras et~al\mbox{.}(2010)]%
        {samaras2010older}
\bibfield{author}{\bibinfo{person}{Nikolaos Samaras}, \bibinfo{person}{Thierry Chevalley}, \bibinfo{person}{Dimitrios Samaras}, {and} \bibinfo{person}{Gabriel Gold}.} \bibinfo{year}{2010}\natexlab{}.
\newblock \showarticletitle{Older patients in the emergency department: a review}.
\newblock \bibinfo{journal}{\emph{Annals of emergency medicine}} \bibinfo{volume}{56}, \bibinfo{number}{3} (\bibinfo{year}{2010}), \bibinfo{pages}{261--269}.
\newblock
\href{https://doi.org/10.1016/j.annemergmed.2010.04.015}{doi:\nolinkurl{10.1016/j.annemergmed.2010.04.015}}


\bibitem[Sch{\"o}n(2017)]%
        {schon2017reflective}
\bibfield{author}{\bibinfo{person}{Donald~A Sch{\"o}n}.} \bibinfo{year}{2017}\natexlab{}.
\newblock \bibinfo{booktitle}{\emph{The reflective practitioner: How professionals think in action}}.
\newblock \bibinfo{publisher}{Routledge}.
\newblock
\href{https://doi.org/10.4324/9781315237473}{doi:\nolinkurl{10.4324/9781315237473}}


\bibitem[Sedgwick(2013)]%
        {sedgwick2013convenience}
\bibfield{author}{\bibinfo{person}{Philip Sedgwick}.} \bibinfo{year}{2013}\natexlab{}.
\newblock \showarticletitle{Convenience sampling}.
\newblock \bibinfo{journal}{\emph{Bmj}}  \bibinfo{volume}{347} (\bibinfo{year}{2013}).
\newblock
\href{https://doi.org/10.1136/bmj.f6304}{doi:\nolinkurl{10.1136/bmj.f6304}}


\bibitem[Seitz(2024)]%
        {seitz2024artificial}
\bibfield{author}{\bibinfo{person}{Lennart Seitz}.} \bibinfo{year}{2024}\natexlab{}.
\newblock \showarticletitle{Artificial empathy in healthcare chatbots: Does it feel authentic?}
\newblock \bibinfo{journal}{\emph{Computers in Human Behavior: Artificial Humans}} \bibinfo{volume}{2}, \bibinfo{number}{1} (\bibinfo{year}{2024}), \bibinfo{pages}{100067}.
\newblock
\href{https://doi.org/10.1016/j.chbah.2024.100067}{doi:\nolinkurl{10.1016/j.chbah.2024.100067}}


\bibitem[Seljelid et~al\mbox{.}(2021)]%
        {seljelid2021digital}
\bibfield{author}{\bibinfo{person}{Berit Seljelid}, \bibinfo{person}{Cecilie Varsi}, \bibinfo{person}{Lise Solberg~Nes}, \bibinfo{person}{Kristin~Astrid {\O}ystese}, {and} \bibinfo{person}{Elin B{\o}r{\o}sund}.} \bibinfo{year}{2021}\natexlab{}.
\newblock \showarticletitle{A digital patient-provider communication intervention (InvolveMe): qualitative study on the implementation preparation based on identified facilitators and barriers}.
\newblock \bibinfo{journal}{\emph{Journal of Medical Internet Research}} \bibinfo{volume}{23}, \bibinfo{number}{4} (\bibinfo{year}{2021}), \bibinfo{pages}{e22399}.
\newblock
\href{https://doi.org/10.2196/22399}{doi:\nolinkurl{10.2196/22399}}


\bibitem[Seo et~al\mbox{.}(2025)]%
        {seo2025enhancing}
\bibfield{author}{\bibinfo{person}{Woosuk Seo}, \bibinfo{person}{Young-Ho Kim}, \bibinfo{person}{Ji~Eun Kim}, \bibinfo{person}{Megan~Tao Fan}, \bibinfo{person}{Mark~S Ackerman}, \bibinfo{person}{Sung~Won Choi}, {and} \bibinfo{person}{Sun~Young Park}.} \bibinfo{year}{2025}\natexlab{}.
\newblock \showarticletitle{Enhancing Pediatric Communication: The Role of an AI-Driven Chatbot in Facilitating Child-Parent-Provider Interaction}. In \bibinfo{booktitle}{\emph{Proceedings of the 2025 CHI Conference on Human Factors in Computing Systems}}. \bibinfo{pages}{1--16}.
\newblock
\href{https://doi.org/10.1145/3706598.3713134}{doi:\nolinkurl{10.1145/3706598.3713134}}


\bibitem[Shilling et~al\mbox{.}(2024)]%
        {shilling2024let}
\bibfield{author}{\bibinfo{person}{Danielle~M Shilling}, \bibinfo{person}{Christopher~R Manz}, \bibinfo{person}{Jacob~J Strand}, {and} \bibinfo{person}{Manali~I Patel}.} \bibinfo{year}{2024}\natexlab{}.
\newblock \showarticletitle{Let Us Have the conversation: serious illness communication in oncology: definitions, barriers, and successful approaches}.
\newblock \bibinfo{journal}{\emph{American Society of Clinical Oncology Educational Book}} \bibinfo{volume}{44}, \bibinfo{number}{3} (\bibinfo{year}{2024}), \bibinfo{pages}{e431352}.
\newblock
\href{https://doi.org/10.1200/EDBK_431352}{doi:\nolinkurl{10.1200/EDBK_431352}}


\bibitem[Smith et~al\mbox{.}(2012)]%
        {smith2012half}
\bibfield{author}{\bibinfo{person}{Alexander~K Smith}, \bibinfo{person}{Ellen McCarthy}, \bibinfo{person}{Ellen Weber}, \bibinfo{person}{Irena~Stijacic Cenzer}, \bibinfo{person}{John Boscardin}, \bibinfo{person}{Jonathan Fisher}, {and} \bibinfo{person}{Kenneth Covinsky}.} \bibinfo{year}{2012}\natexlab{}.
\newblock \showarticletitle{Half of older Americans seen in emergency department in last month of life; most admitted to hospital, and many die there}.
\newblock \bibinfo{journal}{\emph{Health Affairs}} \bibinfo{volume}{31}, \bibinfo{number}{6} (\bibinfo{year}{2012}), \bibinfo{pages}{1277--1285}.
\newblock
\href{https://doi.org/10.1377/hlthaff.2011.0922}{doi:\nolinkurl{10.1377/hlthaff.2011.0922}}


\bibitem[Smriti et~al\mbox{.}(2024)]%
        {Smriti_Wang_Huh-Yoo_2024}
\bibfield{author}{\bibinfo{person}{Diva Smriti}, \bibinfo{person}{Lu Wang}, {and} \bibinfo{person}{Jina Huh-Yoo}.} \bibinfo{year}{2024}\natexlab{}.
\newblock \showarticletitle{Emotion Work in Caregiving: The Role of Technology to Support Informal Caregivers of Persons Living With Dementia}.
\newblock \bibinfo{journal}{\emph{Proc. ACM Hum.-Comput. Interact.}} \bibinfo{volume}{8}, \bibinfo{number}{CSCW1} (\bibinfo{date}{April} \bibinfo{year}{2024}), \bibinfo{pages}{48:1--48:34}.
\newblock
\href{https://doi.org/10.1145/3637325}{doi:\nolinkurl{10.1145/3637325}}


\bibitem[Star and Strauss(1999)]%
        {Star_Strauss_1999}
\bibfield{author}{\bibinfo{person}{Susan~Leigh Star} {and} \bibinfo{person}{Anselm Strauss}.} \bibinfo{year}{1999}\natexlab{}.
\newblock \showarticletitle{Layers of Silence, Arenas of Voice: The Ecology of Visible and Invisible Work}.
\newblock \bibinfo{journal}{\emph{Computer Supported Cooperative Work (CSCW)}} \bibinfo{volume}{8}, \bibinfo{number}{1} (\bibinfo{date}{March} \bibinfo{year}{1999}), \bibinfo{pages}{9–30}.
\newblock
\showISSN{1573-7551}
\href{https://doi.org/10.1023/A:1008651105359}{doi:\nolinkurl{10.1023/A:1008651105359}}


\bibitem[Steinhauser et~al\mbox{.}(2000)]%
        {steinhauser2000factors}
\bibfield{author}{\bibinfo{person}{Karen~E Steinhauser}, \bibinfo{person}{Nicholas~A Christakis}, \bibinfo{person}{Elizabeth~C Clipp}, \bibinfo{person}{Maya McNeilly}, \bibinfo{person}{Lauren McIntyre}, {and} \bibinfo{person}{James~A Tulsky}.} \bibinfo{year}{2000}\natexlab{}.
\newblock \showarticletitle{Factors considered important at the end of life by patients, family, physicians, and other care providers}.
\newblock \bibinfo{journal}{\emph{Jama}} \bibinfo{volume}{284}, \bibinfo{number}{19} (\bibinfo{year}{2000}), \bibinfo{pages}{2476--2482}.
\newblock
\href{https://doi.org/10.1001/jama.284.19.2476}{doi:\nolinkurl{10.1001/jama.284.19.2476}}


\bibitem[Sudore et~al\mbox{.}(2017)]%
        {sudore2017effect}
\bibfield{author}{\bibinfo{person}{Rebecca~L Sudore}, \bibinfo{person}{John Boscardin}, \bibinfo{person}{Mariko~A Feuz}, \bibinfo{person}{Ryan~D McMahan}, \bibinfo{person}{Mary~T Katen}, {and} \bibinfo{person}{Deborah~E Barnes}.} \bibinfo{year}{2017}\natexlab{}.
\newblock \showarticletitle{Effect of the PREPARE website vs an easy-to-read advance directive on advance care planning documentation and engagement among veterans: a randomized clinical trial}.
\newblock \bibinfo{journal}{\emph{JAMA internal medicine}} \bibinfo{volume}{177}, \bibinfo{number}{8} (\bibinfo{year}{2017}), \bibinfo{pages}{1102--1109}.
\newblock
\href{https://doi.org/10.1001/jamainternmed.2017.1607}{doi:\nolinkurl{10.1001/jamainternmed.2017.1607}}


\bibitem[Sunnerhagen and Francisco(2013)]%
        {sunnerhagen2013enhancing}
\bibfield{author}{\bibinfo{person}{K~Stibrant Sunnerhagen} {and} \bibinfo{person}{GE Francisco}.} \bibinfo{year}{2013}\natexlab{}.
\newblock \showarticletitle{Enhancing patient--provider communication for long-term post-stroke spasticity management}.
\newblock \bibinfo{journal}{\emph{Acta Neurologica Scandinavica}} \bibinfo{volume}{128}, \bibinfo{number}{5} (\bibinfo{year}{2013}), \bibinfo{pages}{305--310}.
\newblock


\bibitem[Tausczik and Pennebaker(2013)]%
        {tausczik2013improving}
\bibfield{author}{\bibinfo{person}{Yla~R Tausczik} {and} \bibinfo{person}{James~W Pennebaker}.} \bibinfo{year}{2013}\natexlab{}.
\newblock \showarticletitle{Improving teamwork using real-time language feedback}. In \bibinfo{booktitle}{\emph{Proceedings of the SIGCHI conference on human factors in computing systems}}. \bibinfo{pages}{459--468}.
\newblock
\href{https://doi.org/10.1145/2470654.2470720}{doi:\nolinkurl{10.1145/2470654.2470720}}


\bibitem[Verdezoto et~al\mbox{.}(2021)]%
        {Verdezoto_Bagalkot_Akbar_Sharma_Mackintosh_Harrington_Griffiths_2021}
\bibfield{author}{\bibinfo{person}{Nervo Verdezoto}, \bibinfo{person}{Naveen Bagalkot}, \bibinfo{person}{Syeda~Zainab Akbar}, \bibinfo{person}{Swati Sharma}, \bibinfo{person}{Nicola Mackintosh}, \bibinfo{person}{Deirdre Harrington}, {and} \bibinfo{person}{Paula Griffiths}.} \bibinfo{year}{2021}\natexlab{}.
\newblock \showarticletitle{The Invisible Work of Maintenance in Community Health: Challenges and Opportunities for Digital Health to Support Frontline Health Workers in Karnataka, South India}.
\newblock \bibinfo{journal}{\emph{Proc. ACM Hum.-Comput. Interact.}} \bibinfo{volume}{5}, \bibinfo{number}{CSCW1} (\bibinfo{date}{April} \bibinfo{year}{2021}), \bibinfo{pages}{91:1--91:31}.
\newblock
\href{https://doi.org/10.1145/3449165}{doi:\nolinkurl{10.1145/3449165}}


\bibitem[{Vital Decisions}(2024)]%
        {vitaldecisions_mylivingvoice}
\bibfield{author}{\bibinfo{person}{{Vital Decisions}}.} \bibinfo{year}{2024}\natexlab{}.
\newblock \bibinfo{title}{{My Living Voice}}.
\newblock
\newblock
\shownote{Accessed: 2025-04-16}.


\bibitem[Wang et~al\mbox{.}(2019)]%
        {wang2019development}
\bibfield{author}{\bibinfo{person}{Liqin Wang}, \bibinfo{person}{Long Sha}, \bibinfo{person}{Joshua~R Lakin}, \bibinfo{person}{Julie Bynum}, \bibinfo{person}{David~W Bates}, \bibinfo{person}{Pengyu Hong}, {and} \bibinfo{person}{Li Zhou}.} \bibinfo{year}{2019}\natexlab{}.
\newblock \showarticletitle{Development and validation of a deep learning algorithm for mortality prediction in selecting patients with dementia for earlier palliative care interventions}.
\newblock \bibinfo{journal}{\emph{JAMA network open}} \bibinfo{volume}{2}, \bibinfo{number}{7} (\bibinfo{year}{2019}), \bibinfo{pages}{e196972--e196972}.
\newblock
\href{https://doi.org/10.1001/jamanetworkopen.2019.6972}{doi:\nolinkurl{10.1001/jamanetworkopen.2019.6972}}


\bibitem[Wang et~al\mbox{.}(2024)]%
        {wang2024commsense}
\bibfield{author}{\bibinfo{person}{Zhiyuan Wang}, \bibinfo{person}{Nusayer Hassan}, \bibinfo{person}{Virginia LeBaron}, \bibinfo{person}{Tabor Flickinger}, \bibinfo{person}{David Ling}, \bibinfo{person}{James Edwards}, \bibinfo{person}{Congyu Wu}, \bibinfo{person}{Mehdi Boukhechba}, {and} \bibinfo{person}{Laura~E Barnes}.} \bibinfo{year}{2024}\natexlab{}.
\newblock \showarticletitle{CommSense: A Wearable Sensing Computational Framework for Evaluating Patient-Clinician Interactions}.
\newblock \bibinfo{journal}{\emph{Proceedings of the ACM on Human-Computer Interaction}} \bibinfo{volume}{8}, \bibinfo{number}{CSCW2} (\bibinfo{year}{2024}), \bibinfo{pages}{1--31}.
\newblock
\href{https://doi.org/10.1145/3686952}{doi:\nolinkurl{10.1145/3686952}}


\bibitem[Weiser and Brown(1996)]%
        {Weiser1996THECA}
\bibfield{author}{\bibinfo{person}{Mark~D. Weiser} {and} \bibinfo{person}{John~Seely Brown}.} \bibinfo{year}{1996}\natexlab{}.
\newblock \showarticletitle{THE COMING AGE OF CALM TECHNOLOGY[1]}.
\newblock
\href{https://doi.org/10.1007/978-1-4612-0685-9_6}{doi:\nolinkurl{10.1007/978-1-4612-0685-9_6}}


\bibitem[Wendt et~al\mbox{.}(2025)]%
        {wendt2025deploying}
\bibfield{author}{\bibinfo{person}{Staci~J Wendt}, \bibinfo{person}{Catherine~T Dinh}, \bibinfo{person}{Michael Sutcliffe}, \bibinfo{person}{Kyle Jones}, \bibinfo{person}{James~M Scanlan}, {and} \bibinfo{person}{J~Scott Smitherman}.} \bibinfo{year}{2025}\natexlab{}.
\newblock \showarticletitle{Deploying ambient clinical intelligence to improve care: a research article assessing the impact of nuance DAX on documentation burden and burnout}.
\newblock \bibinfo{journal}{\emph{Future Healthcare Journal}} (\bibinfo{year}{2025}), \bibinfo{pages}{100450}.
\newblock
\href{https://doi.org/10.1016/j.fhj.2025.100450}{doi:\nolinkurl{10.1016/j.fhj.2025.100450}}


\bibitem[Wilson et~al\mbox{.}(2020)]%
        {wilson2020end}
\bibfield{author}{\bibinfo{person}{Jennifer~G Wilson}, \bibinfo{person}{Diana~P English}, \bibinfo{person}{Clark~G Owyang}, \bibinfo{person}{Erica~A Chimelski}, \bibinfo{person}{Corita~R Grudzen}, \bibinfo{person}{Hong-nei Wong}, \bibinfo{person}{Rebecca~A Aslakson}, \bibinfo{person}{Rebecca Aslakson}, \bibinfo{person}{Katherine Ast}, \bibinfo{person}{Thomas Carroll}, {et~al\mbox{.}}} \bibinfo{year}{2020}\natexlab{}.
\newblock \showarticletitle{End-of-life care, palliative care consultation, and palliative care referral in the emergency department: a systematic review}.
\newblock \bibinfo{journal}{\emph{Journal of pain and symptom management}} \bibinfo{volume}{59}, \bibinfo{number}{2} (\bibinfo{year}{2020}), \bibinfo{pages}{372--383}.
\newblock
\href{https://doi.org/10.1016/j.jpainsymman.2019.09.020}{doi:\nolinkurl{10.1016/j.jpainsymman.2019.09.020}}


\bibitem[Wong et~al\mbox{.}(2019)]%
        {wong2019associations}
\bibfield{author}{\bibinfo{person}{Eunice~C Wong}, \bibinfo{person}{Rebecca~L Collins}, \bibinfo{person}{Joshua Breslau}, \bibinfo{person}{M~Audrey Burnam}, \bibinfo{person}{Matthew~S Cefalu}, {and} \bibinfo{person}{Elizabeth Roth}.} \bibinfo{year}{2019}\natexlab{}.
\newblock \showarticletitle{Associations between provider communication and personal recovery outcomes}.
\newblock \bibinfo{journal}{\emph{BMC psychiatry}}  \bibinfo{volume}{19} (\bibinfo{year}{2019}), \bibinfo{pages}{1--8}.
\newblock
\href{https://doi.org/10.1186/s12888-019-2084-9}{doi:\nolinkurl{10.1186/s12888-019-2084-9}}


\bibitem[Wright et~al\mbox{.}(2008)]%
        {wright2008associations}
\bibfield{author}{\bibinfo{person}{Alexi~A Wright}, \bibinfo{person}{Baohui Zhang}, \bibinfo{person}{Alaka Ray}, \bibinfo{person}{Jennifer~W Mack}, \bibinfo{person}{Elizabeth Trice}, \bibinfo{person}{Tracy Balboni}, \bibinfo{person}{Susan~L Mitchell}, \bibinfo{person}{Vicki~A Jackson}, \bibinfo{person}{Susan~D Block}, \bibinfo{person}{Paul~K Maciejewski}, {et~al\mbox{.}}} \bibinfo{year}{2008}\natexlab{}.
\newblock \showarticletitle{Associations between end-of-life discussions, patient mental health, medical care near death, and caregiver bereavement adjustment}.
\newblock \bibinfo{journal}{\emph{Jama}} \bibinfo{volume}{300}, \bibinfo{number}{14} (\bibinfo{year}{2008}), \bibinfo{pages}{1665--1673}.
\newblock
\href{https://doi.org/10.1001/jama.300.14.1665}{doi:\nolinkurl{10.1001/jama.300.14.1665}}


\bibitem[Wright et~al\mbox{.}(2004)]%
        {wright2004doctors}
\bibfield{author}{\bibinfo{person}{Emma~Burkitt Wright}, \bibinfo{person}{Christopher Holcombe}, {and} \bibinfo{person}{Peter Salmon}.} \bibinfo{year}{2004}\natexlab{}.
\newblock \showarticletitle{Doctors' communication of trust, care, and respect in breast cancer: qualitative study}.
\newblock \bibinfo{journal}{\emph{Bmj}} \bibinfo{volume}{328}, \bibinfo{number}{7444} (\bibinfo{year}{2004}), \bibinfo{pages}{864}.
\newblock
\href{https://doi.org/10.1136/bmj.38046.771308.7C}{doi:\nolinkurl{10.1136/bmj.38046.771308.7C}}


\bibitem[Yang et~al\mbox{.}(2024)]%
        {yang2024talk2care}
\bibfield{author}{\bibinfo{person}{Ziqi Yang}, \bibinfo{person}{Xuhai Xu}, \bibinfo{person}{Bingsheng Yao}, \bibinfo{person}{Ethan Rogers}, \bibinfo{person}{Shao Zhang}, \bibinfo{person}{Stephen Intille}, \bibinfo{person}{Nawar Shara}, \bibinfo{person}{Guodong~Gordon Gao}, {and} \bibinfo{person}{Dakuo Wang}.} \bibinfo{year}{2024}\natexlab{}.
\newblock \showarticletitle{Talk2care: An llm-based voice assistant for communication between healthcare providers and older adults}.
\newblock \bibinfo{journal}{\emph{Proceedings of the ACM on Interactive, Mobile, Wearable and Ubiquitous Technologies}} \bibinfo{volume}{8}, \bibinfo{number}{2} (\bibinfo{year}{2024}), \bibinfo{pages}{1--35}.
\newblock
\href{https://doi.org/10.1145/3659625}{doi:\nolinkurl{10.1145/3659625}}


\bibitem[You et~al\mbox{.}(2017)]%
        {you2017barriers}
\bibfield{author}{\bibinfo{person}{John~J You}, \bibinfo{person}{Natasha Aleksova}, \bibinfo{person}{Anique Ducharme}, \bibinfo{person}{Jane MacIver}, \bibinfo{person}{Lisa Mielniczuk}, \bibinfo{person}{Robert~A Fowler}, \bibinfo{person}{Catherine Demers}, \bibinfo{person}{Brian Clarke}, \bibinfo{person}{Marie-Claude Parent}, \bibinfo{person}{Mustafa Toma}, {et~al\mbox{.}}} \bibinfo{year}{2017}\natexlab{}.
\newblock \showarticletitle{Barriers to goals of care discussions with patients who have advanced heart failure: results of a multicenter survey of hospital-based cardiology clinicians}.
\newblock \bibinfo{journal}{\emph{Journal of cardiac failure}} \bibinfo{volume}{23}, \bibinfo{number}{11} (\bibinfo{year}{2017}), \bibinfo{pages}{786--793}.
\newblock
\href{https://doi.org/10.1016/j.cardfail.2017.06.003}{doi:\nolinkurl{10.1016/j.cardfail.2017.06.003}}


\bibitem[Zhang et~al\mbox{.}(2024)]%
        {zhang2024rethinking}
\bibfield{author}{\bibinfo{person}{Shao Zhang}, \bibinfo{person}{Jianing Yu}, \bibinfo{person}{Xuhai Xu}, \bibinfo{person}{Changchang Yin}, \bibinfo{person}{Yuxuan Lu}, \bibinfo{person}{Bingsheng Yao}, \bibinfo{person}{Melanie Tory}, \bibinfo{person}{Lace~M Padilla}, \bibinfo{person}{Jeffrey Caterino}, \bibinfo{person}{Ping Zhang}, {et~al\mbox{.}}} \bibinfo{year}{2024}\natexlab{}.
\newblock \showarticletitle{Rethinking human-AI collaboration in complex medical decision making: a case study in sepsis diagnosis}. In \bibinfo{booktitle}{\emph{Proceedings of the 2024 CHI Conference on Human Factors in Computing Systems}}. \bibinfo{pages}{1--18}.
\newblock
\href{https://doi.org/10.1145/3613904.3642343}{doi:\nolinkurl{10.1145/3613904.3642343}}


\bibitem[Zhang et~al\mbox{.}(2022)]%
        {zhang2022characteristics}
\bibfield{author}{\bibinfo{person}{Zhan Zhang}, \bibinfo{person}{Karen Joy}, \bibinfo{person}{Richard Harris}, {and} \bibinfo{person}{Sun~Young Park}.} \bibinfo{year}{2022}\natexlab{}.
\newblock \showarticletitle{Characteristics and challenges of clinical documentation in self-organized fast-paced medical work}.
\newblock \bibinfo{journal}{\emph{Proceedings of the ACM on human-computer interaction}} \bibinfo{volume}{6}, \bibinfo{number}{CSCW2} (\bibinfo{year}{2022}), \bibinfo{pages}{1--21}.
\newblock
\href{https://doi.org/10.1145/3555111}{doi:\nolinkurl{10.1145/3555111}}


\bibitem[Zheng et~al\mbox{.}(2025)]%
        {zheng2025customizing}
\bibfield{author}{\bibinfo{person}{Xi Zheng}, \bibinfo{person}{Zhuoyang Li}, \bibinfo{person}{Xinning Gui}, {and} \bibinfo{person}{Yuhan Luo}.} \bibinfo{year}{2025}\natexlab{}.
\newblock \showarticletitle{Customizing emotional support: How do individuals construct and interact with LLM-powered chatbots}. In \bibinfo{booktitle}{\emph{Proceedings of the 2025 CHI Conference on Human Factors in Computing Systems}}. \bibinfo{pages}{1--20}.
\newblock
\href{https://doi.org/10.1145/3706598.3713453}{doi:\nolinkurl{10.1145/3706598.3713453}}


\end{thebibliography}
